\documentclass[12pt]{article}

\input epsf
\usepackage{amssymb,amsmath,wrapfig}
\usepackage{cite}
\usepackage{graphicx}
\usepackage{slashed}
\usepackage{multirow}
\usepackage{float}
\usepackage{textcomp}
\usepackage{fancyhdr}

\restylefloat{table}

\makeatletter
\@addtoreset{equation}{section}
\makeatother

\begin{document}

        \begin{titlepage}

        \begin{center}

        \vskip .3in \noindent

        {\Large \bf{Hunting composite vector resonances at the LHC: naturalness facing data}}

         \vspace{1.5cm}

        Davide Greco$^1$ and Da Liu$^{1,2}$\\

          \vspace{1cm}

       {\small $^1$ Institut de Th\'{e}orie des Ph\'{e}nom\`{e}nes Physique,
        EPFL, Lausanne, Switzerland\\
       \vspace{.1cm}

        $^2$ State Key Laboratory of Theoretical Physics, Institute of Theoretical Physics, Chinese Academy of Sciences, Beijing, People's Republic of China
        \vspace{.1cm}}

        \vskip .5in
        {\bf Abstract }
        \vskip .1in
        
          \end{center}
We introduce a simplified low-energy effective Lagrangian description of the 					phenomenology of heavy vector resonances in the minimal composite Higgs model, based on the coset $SO(5)/SO(4)$, analysing in detail their interaction with lighter top partners. Our construction is based on robust assumptions on the symmetry structure of the theory and on plausible natural assumptions on its dynamics. We apply our simplified approach to triplets in the representations $(\textbf{3}, \textbf{1})$ and $(\textbf{1}, \textbf{3})$ and to singlets in the representation $(\textbf{1},\textbf{1})$ of $SO(4)$. Our model captures the basic features of their phenomenology in terms of a minimal set of free parameters and can be efficiently used as a benchmark in the search for heavy spin-1 states at the LHC and at future colliders. We devise an efficient semi-analytic method to convert experimental limits on $\sigma \times BR$ into bounds on the free parameters of the theory and we recast the presently available 8 TeV LHC data on experimental searches of spin-1 resonances as exclusion regions in the parameter space of the models. These latter are conveniently interpreted as a test of the notion of naturalness.

        \noindent

        \vfill
        \eject


        \end{titlepage}

\section{Introduction} 
\label{sec:intro}

The discovery of a new scalar resonance at the LHC marked an important step towards our comprehension of the dynamics hiding behind electroweak symmetry breaking (EWSB). The remarkable compatibility of its properties with those of the Standard Model (SM) Higgs boson and the absence of any new physics predicted by many beyond-the-Standard-Model (BSM) scenarios are forcing us to deeply reconsider the role of naturalness in the dynamics of this particle. A concrete realization of naturalness is offered by the composite Higgs scenario: a new strongly coupled sector confining at the TeV scale and inducing the spontaneous breaking of a global symmetry can produce a light pseudo Nambu-Goldstone boson (pNGB) Higgs at 125 GeV, \cite{VacuumMisal}. Probing the compositeness of the newly discovered scalar is therefore a crucial task for understanding how natural its features are. This is indeed the main question we would like to address in this paper: assuming naturalness as a good guiding principle for the existence of a new strongly coupled physics at the TeV scale, how can the presently available LHC data be used to test the validity of our notion of naturalness? 

A possible way to answer this question is to study the phenomenological properties and the possibility of a direct discovery of other composite resonances generated by the strong sector. In particular, one of the robust predictions of this class of theories is the existence of spin-1 resonances excited from the vacuum by the conserved currents of the strong dynamics. They form multiplets of the unbroken global symmetry and can behave rather differently from the heavy $Z^\prime$ states in weakly coupled extensions of the SM. These vectors, in fact, interact strongly with the longitudinally polarized $W$ and $Z$ bosons and the Higgs and thus tend to be broader than the weakly coupled ones. The strength of their interactions with the SM fermions depends on whether these latter participate to the strong dynamics or are purely elementary states. A simple possibility is that SM fermions couple to the EWSB dynamics according to their masses, so that the lightest ones are the most weakly coupled. This idea has an elegant implementation in the framework of partial compositeness \cite{WarpedComposite} and can give a qualitative understanding of the hierarchies in the Yukawa matrices of the SM fermions in terms of RG flows \cite{HologFerm, MinimComp}. A second robust characteristic of composite Higgs models is the existence of spin-1/2 resonances, the top partners. In the most natural realizations, these fermionic states are lighter then the heavy vector particles, \cite{LightFermions, LightFermions2, LightFermions3, LightFermions4, LightFermions5, LightFermions6}. In a natural scenario we therefore expect the phenomenology of spin-1 states to be significantly affected by the presence of lighter composite fermions.

In this work, we study the phenomenology of spin-1 resonances in composite Higgs theories by means of a simplified description based on an effective Lagrangian, focussing on their interaction with lighter top partners. This is aimed at capturing the main features relevant for the production and decay of the heavy vectors at high-energy colliders and their effects in low-energy experiments, avoiding the complications of a full model. Although simplified, our procedure will be sophisticated enough to properly include those aspects which are distinctive predictions of the class of theories under consideration, such as for example the pNGB nature of the Higgs boson. We will focus on the minimal $SO(5)\times U(1)_X /SO(4)\times U(1)_X$ composite Higgs model and consider vector triplets transforming as a $(\textbf{3}, \textbf{1})$ and $(\textbf{1}, \textbf{3})$ of $SO(4)\sim SU(2)_L \times SU(2)_R$ and vector singlets transforming only under the unbroken $U(1)_X$. We will study in detail the interactions of these bosonic states with top partners and include the effects implied by the partial compositeness of SM fermions. The importance of lighter composite fermions on the phenomenology of vector resonances has been pointed out also in \cite{Prod5} and in \cite{Vignaroli}; this latter considered the case of a $SU(2)_L$ charged heavy spin-1 state. Our approach, however, differs for the method used in deriving the effective Lagrangian and for taking into account all the spin-1 resonances in the simplest representations of $H$.

Our construction provides a benchmark model to be used in searches for heavy spin-1 states at the LHC and at future colliders. A simple kinematic model based on the width and the production cross section times decay branching ratio ($\sigma \times BR$) is sufficient to guide searches for narrow resonances in individual channels and to set limits, see the discussion in \cite{Bridge}. However, combining the results obtained in different final states as well as interpreting the limits on $\sigma \times BR$ in explicit models of BSM physics and developing a detailed analysis of the interaction with lighter fermionic states requires an underlying dynamical description, such as the one given by a simplified Lagrangian. Here we provide such a dynamical description for spin-1 resonances coupled to lighter top partners appearing in a natural and sufficiently large class of composite Higgs theories. Our simplified Lagrangian fully takes into account the non-linear effects due to multiple Higgs vev insertions and does not rely on an expansion in $v/f$, where $v$ is the electroweak scale and $f$ is the decay constant of the pNGB Higgs. In the limit $v/f \ll 1$, it can be matched onto the more general one of \cite{Bridge}, which covers a more ample spectrum of possibilities in terms of a larger number of free parameters. In this sense, the main virtue of our model is that of describing the phenomenology of spin-1 resonances in composite Higgs theories in terms of a minimal set of physical quantities: one mass and one coupling strength for each heavy vector. Expressing the experimental results in such a restricted parameter space is thus extremely simple and gives an immediate understanding of the reach of current searches in the framework of strongly interacting models for EWSB. It also provides an immediate way to test how natural the Higgs sector is expected to be.

This paper is organized as follows. In Section \ref{sec:Assumptions}, we review the most important characteristics of the minimal composite Higgs model that are relevant for our construction and we analyse the dynamical assumptions that justify our effective Lagrangian approach. In Section \ref{sec:Models}, we introduce the models for the three vector resonances under consideration and we discuss their mass spectrum and physical interactions.\footnote{Part of the results appearing in this section has already been presented in \cite{Proceeding}.}~The main production mechanisms and decay modes are discussed in Section \ref{sec:ProdDec}, where we describe the most important channels that can be relevant for a future discovery at the LHC. The presently available 8 TeV LHC data are used to derive exclusion limits on the parameter space of our models in Section \ref{sec:LHCBounds}. Our conclusions are finally summarized in Section \ref{sec:Concl}. 


\section{Behind the models} 
\label{sec:Assumptions}

Our main purpose is to introduce an effective Lagrangian description of the interactions between heavy vectors and top partners in the minimal composite Higgs scenario. We aim at deriving a simplified model, based on a minimal set of free parameters, which is suitable for studying the production and decay of these new heavy states at colliders, but still capable of capturing the most important features of the underlying strong dynamics. We will indeed make some robust assumptions on the symmetry structure of the theory, dictated by the pNGB nature of the Higgs, and some plausible dynamical assumptions on its spectrum, dictated by naturalness arguments, that can provide enough information to determine the most prominent phenomenological aspects of these constructions. 

\subsection{The symmetry structure and the degrees of freedom}

We start analysing the basic features of the minimal composite Higgs model that will have relevant consequences for the phenomenology of the heavy resonances. We assume the existence of a new strongly interacting sector with an approximate global symmetry  in the UV, $G=SO(5) \times U(1)_X$, spontaneously broken to $H = SO(4)\times U(1)_X \sim SU(2)_L \times SU(2)_R \times U(1)_X$ at an energy scale $f$.~\footnote{The abelian group $U(1)_X$ must be included in order to reproduce the correct hypercharge of the fermion fields, which is given by $Y = T_{R}^3+X$, $T_{R}^3$ being the third generator of $SU(2)_R$}~The four Goldstone bosons, $\Pi^{\widehat{a}}$, resulting from the spontaneous breaking of the global symmetry transform as a $\mathbf{(2,2)_0}$ under the linearly-realized unbroken subgroup, $H$; in the absence of an explicit breaking of $SO(5)$ they are exactly massless. The SM electroweak bosons gauge the $SU(2)_L \times U(1)_Y$ subgroup of the global group, thus introducing a preferred orientation in the coset space $SO(5)/SO(4)$ with respect to the global $SO(4)$. The misalignment between the direction fixed by the local group and the vacuum where the theory is realized can be conveniently parametrized by an angle $\theta$, which serves as an order parameter for EWSB, \cite{EffectRes}. The interaction between the Goldstone bosons and the SM fields explicitly breaks the global symmetry and generates a potential for the Higgs at loop level resulting in a non-vanishing vev for its modulus. As a consequence, three Goldstone bosons are eaten to give mass to the SM gauge bosons and a massive Higgs field, $h(x)$, remains in the spectrum. The misalignment angle can be identified as $\theta = \left \langle h \right \rangle / f$ and the electroweak scale is dynamically generated at $v=f \sin \theta$. It is convenient to introduce the parameter 
\begin{equation}\label{Csi}
\xi = \sin^2 \theta = \left({v \over f}\right)^2
\end{equation}
characterising the separation between the electroweak and the strong scale; in a natural theory, we expect $\xi \sim 1$, but it is conceivable that a small amount of tuning can give rise to $\xi \ll 1$. In particular, compatibility with the constraints coming from electroweak precision tests and Higgs coupling measurements generically implies $\xi \lesssim 0.2$, \cite{Bridge, EWFit1, CLIC}.

In this framework, we will construct effective Lagrangians respecting the non-linearly realized $SO(5)$ global group using the standard CCWZ formalism, as developed in \cite{CCWZ1} and \cite{CCWZ2}. According to this procedure, a Lagrangian invariant under the global $SO(5)$ can be written following the rules of a local $SO(4)$ symmetry; the basic building blocks are given by the Goldstone boson matrix, $U(\Pi)$, and the $d_\mu$ and $E_\mu$ symbols, resulting from the Maurer-Cartan form $U^\dagger D_{\mu} U$, which are reviewed in Appendix \ref{AppCCWZ}. 

Considering now the degrees of freedom, they comprise elementary states, which include the gauge bosons $W_\mu$ and $B_\mu$ and the SM fermions, and composite states, which, besides the pNGB Higgs and the longitudinally polarized $W$ and $Z$ bosons, include particles with specific transformation properties under the unbroken $SO(4)$. As regards the interactions between these two sectors, the gauge bosons couple through the gauging of the SM subgroup of $G$, whereas the elementary fermions couple linearly to the composite dynamics, according to the paradigm of partial compositeness, \cite{PartComp}. Since this linear interaction is responsible for generating the masses of leptons and quarks, we expect the heaviest SM fermions to be more strongly coupled to the new sector and to have the strongest interactions with the composite resonances. At the energy scale that can be probed at the LHC, it is therefore a well justified approximation to consider all leptons and quarks, except for the heaviest doublet $q_L = ( t_L, b_L )$ and the right-handed top quark $t_R$, to be fully elementary and massless, so that we can neglect their linear coupling to the strong dynamics. On the other hand, the top-bottom doublet is taken to have a direct linear interaction with an operator $\mathcal{O_R}$, transforming in a representation $r_{\mathcal{O}}$ of $SO(5)\times U(1)_X$, so that in the UV the Lagrangian is:
\begin{equation}\label{PartialComp}
\mathcal{L} = y_L \bar{q}_L^{\alpha}\Delta_{\alpha, I_{\mathcal{O}}} \mathcal{O_R}^{I_ \mathcal{O}} + \text{h.c.} = y_L (\bar{Q}_L)_{I_\mathcal{O}} \mathcal{O_R}^{I_\mathcal{O}} + \text{h.c.},
\end{equation}
where $I_{\mathcal{O}}$ denotes the indices of the operator $\mathcal{O_R}$ and $ (\bar{Q}_L)_{I_\mathcal{O}} = \bar{q}_L^{\alpha}\Delta_{\alpha, I_{\mathcal{O}}}$ indicates the embedding of $q_L$ into a full representation of $SO(5)$, as discussed in \cite{Tasi}. This kind of mixing explicitly breaks the global symmetry of the strong dynamics, $y_L \Delta$ being a spurion under $G$, generating a contribution to the Higgs potential via loop effects. In order to obtain a sufficiently light Higgs, we therefore expect $y_L$ to be a relatively small parameter. The choice of the representation $r_\mathcal{O}$ does not depend on the details of the low-energy physics and it is to some extent free. Many possibilities have been studied in the literature, \cite{Hunters, Top9}; for simplicity, we will only consider the minimal case where $r_{\mathcal{O}}= \textbf{5}_{\textbf{2/3}}$, so that the form of the embedding will be unambiguously fixed:
\begin{equation}\label{Embeddings}
(Q_L^5)_I ={1\over \sqrt{2}} \left(ib_L \quad b_L \quad it_L \quad -t_L \quad 0\right)^T,
\end{equation}
which formally transforms under $g\in SO(5)$ as $(Q_L^5)_I \rightarrow g_I^J (Q_L^5)_J$ and has $X$-charge equal to 2/3. As regards the $t_R$, we will consider two different scenarios. First, we will assume that this particle arises as a composite resonance of the strong sector, transforming like a singlet under $SO(4)$ and with hypercharge $2/3$. Then, similarly to what happens to the heaviest doublet, we will be interested in studying the phenomenological implications of a partially composite $t_R$, for reasons that will become clear in the following. In this particular case, the $t_R$ is assumed to be linearly coupled to an operator $\mathcal{O}_L$ of the strong sector transforming as a $\mathbf{5_{2/3}}$, with the UV lagrangian
\begin{equation}\label{tRPartComp}
\mathcal{L}= y_R \bar{t}_R \Delta_{I} \mathcal{O}_L^{I}+ \text{h.c.} = y_R (\bar{Q}_{R}^5)_{I}\mathcal{O}_L^{I}+\text{h.c.},
\end{equation}  
where the embedding is in this case fixed by the standard model quantum numbers to be:
\begin{equation}\label{EmbedtR}
(Q_R^5)_I= \left(0 \quad 0 \quad 0 \quad 0 \quad t_R\right)^T.
\end{equation}
$(Q_R^5)_I$ formally transforms under $SO(5)$ like $(Q_L^5)_I$ and has $X$-charge $2/3$. The parameter $y_R$ is expected to be of the order of the corresponding $y_L$ in order to accommodate a reasonably tuned light Higgs in the spectrum.  

We have discussed all the basic ingredients of the model, concerning both the new symmetries and the particles we have to deal with. In this work, as highlighted in the Introduction, we will be mainly interested in studying the phenomenology of composite spin-1 states, $\rho_\mu$, focusing on triplets transforming as a $(\mathbf{3,1})_0$ and $(\mathbf{1,3})_0$ under the unbroken $SO(4)\times U(1)_X$ and on vector singlets, which are left invariant by $SO(4)$ and transform only under the abelian group $U(1)_X$, analysing in detail their interplay with lighter spin-1/2 heavy states.

\subsection{Dynamical assumptions}

Since we aim at building a simplified description of the interactions between vectors and top partners, we need to make some generic assumptions on the dynamics of the strong sector that can guide us in the construction of an effective Lagrangian and can give a basic understanding of its regime of validity. Following the SILH approach, \cite{SILH}, we can broadly parametrize the new confining dynamics with a mass scale $m_*$ and a coupling $g_*$, which are related by the NDA estimate
\begin{equation}\label{NDA}
m_* \sim g_* f,
\end{equation}  
reproducing the usual relation between the Goldstone boson decay constant and the mass of the composite states. We will however generalize this simple approximation, taking into account both the theoretical implications of naturalness and the constraints coming from electroweak precision tests. On the theoretical level, in fact, we naturally expect the fermionic resonances to be light, since they are directly responsible for cutting off the quadratically divergent contributions to the Higgs mass coming from the SM top quark loops, as explained in \cite{LightFermions, LightFermions2, LightFermions3, LightFermions4, LightFermions5, LightFermions6}. In particular, a reasonably tuned pNGB Higgs generically requires top partners to have a mass around 1 TeV. On the other hand, as described also in Appendix \ref{AppSTU}, vector resonances contribute at tree level to the $\hat{S}$ parameter, thus implying their mass to be generically bigger than 2 TeV. 

These considerations are the main reason for parametrizing the confining dynamics with two different scales, a lighter one for the spin-1/2 and a heavier one for the spin-1 resonances, pointing towards a natural scenario where the phenomenology of vector particles can be considerably affected by the presence of a lower-lying layer of fermionic states. We therefore introduce a mass scale, $m_\psi$, and a coupling, $g_\psi$, for the top partners, such that
\begin{equation}\label{FermionNDA}
m_\psi = a_\psi g_\psi f,
\end{equation}     
and a mass scale, $m_\rho$, and a coupling, $g_\rho$, for the vector resonances, with the analogous relation
\begin{equation}\label{VectorNDA}
m_\rho = a_\rho g_\rho f,
\end{equation}
where $a_\psi$ and $a_\rho$ are $O(1)$ parameters, as implied by NDA. Supposing the fermionic scale to be smaller than the vector scale therefore implies the obvious relation between the two couplings of the new dynamics:
\begin{equation}\label{TwoCouplings}
g_\psi < {a_\rho \over a_\psi} g_\rho .
\end{equation}  
In particular, a naturally light composite Higgs generically requires the fermionic coupling constant to be favoured in the range $1 \lesssim g_{\psi} \lesssim 3$. We will be mainly interested in studying how these assumptions on the strong sector can be tested in the context of a phenomenological model for the production of heavy spin-1 states and their decay to top partners and SM particles.

We have some other considerations to make on the two scales in order to justify our effective Lagrangian approach. Following the criterion of $\textit{partial}$ $\textit{UV}$ $\textit{completion}$, firstly introduced in \cite{EffectRes}, we assume that the bosonic resonances we want to study have a mass $M_\rho$ much lower than the vector scale and bigger than the fermionic scale, $m_\psi < M_\rho \ll m_\rho$, so that we can integrate out all the heavier states and write a Lagrangian in an expansion of $M_\rho/m_\rho$. This approximation obviously starts loosing its validity as soon as the mass separation becomes smaller, $m_\psi \ll M_\rho \sim m_\rho$, in which case the interference effects with other resonances become non-negligible and our analysis is only a qualitative description of the underlying dynamics. We apply this point of view to the triplets in the representation $(\textbf{3},\textbf{1})_0$, $\rho_\mu^L$, and $(\textbf{1},\textbf{3})_0$, $\rho_\mu^R$, and to the singlet, $\rho_\mu^X$, building one model for each of them. In every case we will suppose that the other two vectors have a mass $M_\rho \sim m_\rho$, so that they belong to the tower of heavier resonances that are being integrated out, resulting in a great simplification of the phenomenology. This assumption is dictated mainly by the need of building the simplest description of the interplay between heavy vectors and top partners and we have no deep reasons for excluding the opposite case, namely that the spin-1 resonances are almost degenerate in mass. We will however make some comments about this possibility in Appendix \ref{AppIntEff}, showing under which conditions the mutual interaction between the vectors can be safely neglected even when their spectrum is degenerate. 

Finally, we must discuss the role of the fermionic scale in our effective expansion. In fact, since we are about to derive a phenomenological Lagrangian which is valid up to the first vector resonance, we should in principle include its interactions with all the fermions at the scale $m_\psi$ and falling into various representations of the unbroken $SO(4)$. In order to avoid the complications arising from such a full model, we will only take into account the lightest heavy fermions, assuming that their mass satisfies the condition $M_\Psi < m_\psi$, so that the decay channel of the vectors to these fermionic states is the most favoured one among the decays to other resonances. Under this conditions, we can more safely neglect the remaining tower of spin-1/2 states. For our construction to be fully meaningful, we need a criterion to understand under which representation of $SO(4)$ the lightest heavy fermions should transform. This is easily found by noticing that in explicit models the lightest fermionic resonances that must be present in the spectrum are the top partners falling into the representations of $H$ that can be excited from the vacuum by the operators $\mathcal{O}_R$ and $\mathcal{O}_L$ linearly coupled to the $q_L$ doublet and the $t_R$, when this latter is partially composite, \cite{LightFermions2}. Since we chose $r_{\mathcal{O}} = 5_{2/3}$ for both cases, we can decompose $\mathcal{O}_R$ and $\mathcal{O}_L$ under $SO(4)$, obtaining $\mathbf{5_{2/3}}=\mathbf{4_{2/3}}+\mathbf{1}_{2/3} $, therefore justifying the introduction of top partners in the fourplet and in the singlet of the unbroken group. Moreover, we must notice that limiting our analysis to the lightest fermionic resonances becomes a very crude approximation when $M_\Psi \sim m_\psi$, requiring a more complete construction; we leave this study to future work, with the aim to provide in the present analysis a simplified model with a few degrees of freedom and parameters that can be more thoroughly used to guide searches of new physics at the LHC.

We now have all the elements to derive a phenomenological Lagrangian describing the interplay between vector and fermion resonances, based on symmetry principles and general reasonable assumptions on the nature of the strong dynamics. In conclusion, we will write three models, one for a $\rho_\mu^L$ and top partners in the fourplet, one for a $\rho_\mu^R$ and again top partners in the fourplet, and a last one for a $\rho^\mu_X$ and top partners in the singlet.

\section{The models}
\label{sec:Models}

After the clarification of the symmetries and the dynamical assumptions behind our approach, we are now in a good position for explicitly introducing the Lagrangians for the three vector resonances. We will devote this section to describe the three models and some of their basic phenomenological characteristics.

\subsection{A Lagrangian for $\rho_\mu^L$}

We start considering a theory for the $(\textbf{3},\textbf{1})_0$ triplet and top partners in the fourplet, introducing therefore the fermionic field
\begin{equation}\label{Fourplet}
\Psi = {1 \over \sqrt{2}} \left(
\begin{array}{l}
i B - i X_{5/3} \\
B + X_{5/3} \\
i T + i X_{2/3} \\
-T + X_{2/3}
\end{array}
\right),
\end{equation}
which has X-charge 2/3. The vector resonance transforms non-homogeneously under the unbroken $SO(4)$,
\begin{equation}\label{RhoTrans}
\rho_\mu^L \rightarrow h(\Pi, g) \rho_\mu^L h^\dagger(\Pi, g) - i h(\Pi, g) \partial_\mu h^\dagger (\Pi, g),
\end{equation}
where $h(\Pi,g) \in SO(4)$, as described in Appendix A. The partner field transforms instead linearly, so that
\begin{equation}\label{TopPartTransf}
\Psi \rightarrow h(\Pi, g) \Psi,
\end{equation}
and it decomposes into two doublets under $SU(2)_L\times U(1)_Y$, the $(T, B)$ doublet with the same quantum numbers of top and bottom quarks and the $(X_{5/3}, X_{2/3})$ doublet with an exotic particle of charge $5/3$ and a second top-like resonance, $X_{2/3}$.

Following now the CCWZ prescription and considering the $t_R$ a full composite condensate of the strong sector, at leading order in the derivative expansion the Lagrangian is:
\begin{equation}\label{LagLeftBref}
\mathcal{L}_{L} = \mathcal{L}_{light}+\mathcal{L}_{\Psi}+\mathcal{L}_{\rho_L},
\end{equation}
where the three different contributions stand for:
\begin{equation}\label{LagLeft}
\begin{array}{ll}
\displaystyle \mathcal{L}_{light}  =  & \displaystyle {f^2 \over 4}(d^{\widehat{a}}_\mu)^2 -{1 \over 4 } W_{\mu \nu }^a W ^{a \mu \nu} - {1 \over 4 }B_{\mu \nu}B^{\mu \nu} + \bar{\psi} \gamma^\mu (i \partial_\mu + g_{el}{\sigma^a \over 2}W_\mu^a P_L + g_{el}^{\prime}Y B_\mu) \psi \\

& \displaystyle + i \bar{q}_L \slashed{D} q_L + i \bar{t}_R \slashed{D}t_R,  \\[0,2cm]

\displaystyle \mathcal{L}_{\Psi}  = & \bar{\Psi}\gamma^\mu (i {\nabla}_\mu + X g_{el}^\prime B_\mu -M_\Psi ) \Psi + \left[i c_1 \bar{\Psi}_R^i \slashed{d}_i t_R + y_L f (\bar{Q}_L^5)^I U_{I i}\Psi_R^i \right. \\

& \displaystyle \left. + y_L c_2 f (\bar{Q}_L^5)^I U_{I5}t_R + \text{h.c.} \right],\\[0,2cm]

\displaystyle  \mathcal{L}_{\rho_L} = & \displaystyle - {1\over 4 } \rho_{\mu \nu }^{a_L}\rho^{a_L \mu \nu} + {m^2_{\rho_L} \over 2 g_{\rho_L}^2}( g_{\rho_L} \rho_\mu^{a_L}-E_\mu^{a_L})^2 + c_3 \bar{\Psi}^i \gamma^\mu (g_{\rho_L} \rho_\mu^{a_L}-E_\mu^{a_L})T^{a_L}_{ij} \Psi^j.
\end{array}
\end{equation}
In the first Lagrangian, containing the kinetic terms of the elementary sector, the composite Goldstone bosons and third family quarks, we have collectively indicated with $\psi$ all the massless fermions, namely the leptons and the first two quark families, so that the $\psi$ field has to be understood as a sum over these different species. The second Lagrangian, $\mathcal{L}_{\Psi}$, on the other hand, describes the kinetic term of the top partners and their interactions with third family quarks, which are generated in the IR by the UV Lagrangian (\ref{PartialComp}). We have used the notation of Appendix A to indicate the CCWZ covariant derivative, $\nabla_\mu$, which is necessary to respect the non-linearly realised $SO(5)$, and we have added the contribution of the $B_\mu$ field in order to preserve the SM gauge invariance. Finally, the last Lagrangian, $\mathcal{L}_{\rho_L}$, introduces the kinetic and mass terms of the vector resonance and its interaction with the top partners. In particular, since $\rho_L$ transforms non-homogeneously under the unbroken $SO(4)$, the field strength must be
\begin{equation}\label{RhoKin}
\rho_{\mu \nu}^{a_L} = \partial_\mu \rho_\nu ^{a_L}-\partial_\nu \rho_\mu ^{a_L} + g_{\rho_L} \epsilon^{a_L b_L c_L}\rho_\mu ^{b_L} \rho_\nu ^{c_L}.
\end{equation}
We note that additional higher derivative operators can in general be included in the previous Lagrangian and they can play a relevant role at energies of order of the resonances mass, as discussed for example in \cite{EffectRes}. We will omit them for simplicity, referring to \cite{Bridge} for a more complete discussion of the effects of these additional terms on the phenomenology of vector resonances.

From Eq.~(\ref{LagLeft}), we immediately see that the only source of interactions among the composite $\rho_L$ and the elementary gauge fields is the $\rho_L - W$ and the $\rho_L - B$ mass mixings that follow from the mass term in $\mathcal{L}_{\rho_L}$. Given the expression of the CCWZ connections, the global mass matrix of spin-1 fields $(W, B, \rho_L)$ is non-diagonal and must be diagonalised by a proper field rotation, in order to obtain the couplings and the Lorentz structure of the vertices in the mass eigenstate basis. Similarly, the mass matrix of these spin-1/2 fields arising from the Lagrangian $\mathcal{L}_{\Psi}$ is in general non-diagonal and we need another rotation, on the fermionic sector, in order to describe the particle spectrum.

Before discussing the two rotations, let us first count how many parameters appear in our Lagrangian. There are eight couplings, $(g_{el}, g_{el}^{\prime}, g_{\rho_L}, c_1, c_2, c_3, y_L, f)$, two mass scales, $(m_{\rho_L}, M_\psi)$, and the misalignment angle, that can be conveniently traded for the variable $\xi$, for a total of eleven free parameters. Notice that we have listed the NG decay constant $f$ as a coupling, since it controls the strength of the NG boson interactions. The couplings $g_{el}$ and $g_{el}^{\prime}$ arise as a result of the weak gauging of the SM subgroup of $H$, $g_{\rho_L}$ instead sets the strength of the interactions between the vectors and other composite states, including the Higgs and the longitudinally polarized $W$ and $Z$ bosons, whereas $c_1$, $c_2$ and $c_3$ are $O(1)$ parameters, as suggested by power counting. All the Lagrangian input parameters can be re-expressed in terms of physical quantities in the mass eigenstate basis. Three of them must be fixed in order to reproduce the basic electroweak observables, which we conveniently choose to be $G_F$, $\alpha_{em}$ and $m_Z$. Of the remaining eight input parameters, $\xi$ controls the modifications of the Higgs couplings from the SM values and is thus an observable, $c_2$ will be fixed in order to reproduce the physical top mass and the other six can be traded for the following physical quantities: the masses of two top partners, for instance $m_{X_{5/3}}$ and $m_B$, the mass of the charged heavy vector and its couplings to elementary fermions and to the top-bottom pair, and finally the coupling of one heavy fermion to a gauge boson and top quark.

In order to fix three of the input parameters in terms of $G_F$, $\alpha_{em}$ and $m_Z$, we need the expression of the latter in terms of the former. It turns out that $G_F$ and $\alpha_{em}$ are very simple to compute and read:
\begin{equation}\label{AlphaGF}
\begin{array}{ll}
\displaystyle G_F = {1 \over \sqrt{2} f^2 \xi}, \qquad \displaystyle {1 \over 4 \pi \alpha_{em}} = {1\over g_{el}^2}+{1\over g_{\rho_L}^2}+{1\over g_{el}^{\prime 2}} = {1 \over g^2}+{1\over g^{\prime 2}},
\end{array}
\end{equation} 
where we have conveniently defined the SM coupling $g$ and $g^{\prime}$ as
\begin{equation}\label{SMCoup}
{1\over g^2} \equiv {1\over g_{el}^2}+{1\over g_{\rho_L}^2}, \qquad g^{\prime} \equiv g_{el}^{\prime}.
\end{equation}
It is important to notice that $\alpha_{em}$ does not get corrections after EWSB at any order in $\xi$, due to the surviving electromagnetic gauge invariance. The formula for $G_F$ can be most easily derived by integrating out first the composite $\rho$ using the equations of motion at leading order in the derivative expansion, $\rho_\mu^{a_L} = E_\mu^{a_L} + O(p^3)$. From equation (\ref{LagLeft}), one can then see that the low-energy Lagrangian for the elementary fields contains one extra operator, $(E_{\mu \nu}^L)^2$, which however does not contribute to $G_F$. This means that the expression of $G_F$ in terms of the elementary parameters does not receive any tree-level contribution from the composite $\rho$, hence the simple formula in (\ref{AlphaGF}). Finally, the expression for $m_Z$ is in general quite complicated and can be obtained only after the rotation to the mass eigenstate basis; we will not report it here, but we will discuss its approximation while describing the physical spectrum of our theory. By making use of such a formula and of equation (\ref{AlphaGF}), for given values of the other input parameters, we can fix $g_{el}$, $g_{el}^{\prime}$ and $f$ so as to reproduce the experimental values of $G_F$, $\alpha_{em}$ and $m_Z$.

We now discuss the rotation to the mass eigenstate basis and the physical spectrum of the model. As regards the fermionic mass matrix, it has already been extensively analysed in \cite{Hunters} and we will not examine here the details, limiting ourselves to report the basic results. After the diagonalization, it is straightforward to derive the masses of the top quark and of the four top partners; they are found to be:
\begin{equation}\label{MassFerm}
\begin{array}{ll}
\displaystyle m_{top} = \frac{{c_2} y_L f}{\sqrt{2}}  \frac{ M_{\Psi}}{\sqrt{M_{\Psi }^2+ y_L^2 f^2}}\sqrt{\xi }\left[1+ O(\xi)\right], \qquad \displaystyle m_{X_{5/3}} = m_{X_{2/3}} = M_\Psi, \\ [0,4cm]

\displaystyle m_T = \sqrt{{M_{\Psi}^2}+y_L^2f^2}- \frac{ y_L^2 f^2\left({M_{\Psi}^2} -\left({c_2^2}- 1 \right) y_L^2 f^2\right)}{4 \left({M_{\Psi}^2}+ y_L^2f^2\right)^{3/2}}\xi + O(\xi ^2), \qquad \displaystyle m_B = \sqrt{M_\Psi ^2 + y_L ^2 f^2},
\end{array}
\end{equation} 
where we have listed the expressions at leading order in $\xi$. The lightest top partners are $X_{5/3}$ and $X_{2/3}$, whose mass is exactly equal to the Lagrangian parameter $M_{\Psi}$ and does not receive any correction after EWSB; in particular the $X_{5/3}$ particle cannot mix because of its exotic charge and it is left invariant by the rotation. The $B$ fermion is the heaviest particle and also in this case its mass is not altered after EWSB. The $T$ partner, on the other hand, is relatively lighter than $B$, due to $O(\xi)$ corrections, whereas the bottom quark remains massless, since we are not including the linear coupling of $b_R$ to the strong sector. This latter interactions will in general induce small corrections to the above relations of order $O(m_b/m_{top})$. In order to obtain the correct order of magnitude for the top mass, we expect $y_L \sim y_t$, where $y_t$ is the top Yukawa coupling. We will use the above expression for $m_{top}$ in the following in order to fix the parameter $c_2$ to reproduce the top quark mass. Finally, neglecting EWSB effects, we can find very simple expressions for the rotation angles; the mass matrix is in fact diagonalised by the following field rotation:
\begin{equation}\label{FermRot}
t_L \rightarrow \frac{{M}_{\Psi }}{\sqrt{y_L^2 f^2+{M}_{\Psi }^2}}t_L-\frac{  y_L f}{\sqrt{ y_L^2 f^2 +{M}_{\Psi }^2}}T_L ,\quad b_L \rightarrow \frac{ {M}_{\Psi }}{\sqrt{y_L^2 f^2+{M}_{\Psi }^2}}b_L-\frac{y_L f }{\sqrt{y_L^2 f^2+{M}_{\Psi }^2}}B_L,
\end{equation}
with the $T_L$ and $B_L$ particles transforming orthogonally. The right-handed $t_R$, $T_R$ and $B_R$ and the top partner $X_{2\over 3}$ are instead left unchanged.

Let us now focus on the spin-1 sector of the theory. The mass term of the Lagrangian can be written as
\begin{equation}\label{MassTerm}
\mathcal{L}_{mass} = X^{+} M_{\pm}^2 X^{-} + {1 \over 2} X^0 M_0^2 X^0,
\end{equation}
where $X^{\pm} = (X^1 \pm i X^2)/ \sqrt{2}$, with $X^{1,2} = \{ W^{1,2}, \rho_L^{1,2} \}$, and $X^0 = \{W^3, \rho_L, B \}$. The mass matrix therefore decomposes in a $2\times 2$ charged block, $M_{\pm}^2$, and a $3 \times 3$ neutral block, $M_0^2$. The expression for the charged sector is
\begin{equation}\label{ChargedBlock}
M_{\pm}^2=\left(
\begin{array}{cc}
\displaystyle  \frac{g_{\text{el}}^2 }{4 g_{\rho _L}^2}\left(g_{\rho _L}^2 f^2 \xi +A(\xi) m_{\rho _L}^2\right) & \displaystyle -\frac{ g_{\text{el}}}{2 g_{\rho _L}}B(\xi) m_{\rho _L}^2 \\
\displaystyle  -\frac{g_{\text{el}}}{2 g_{\rho _L}}B(\xi) m_{\rho _L}^2 & m_{\rho _L}^2 \\
\end{array}
\right),
\end{equation} 
while the neutral block can be easily found to be
\begin{equation}\label{NeutralBlock}
M_{0}^2=\left(
\begin{array}{ccc}
\displaystyle \frac{g_{\text{el}}^2 }{4 g_{\rho _L}^2}\left( g_{\rho _L}^2 f^2 \xi +A(\xi) m_{\rho _L}^2\right) & \displaystyle -\frac{ g_{\text{el}}}{2 g_{\rho _L}} B(\xi) m_{\rho _L}^2 & \displaystyle \frac{ g_{\text{el}} g_{\text{el}}'}{4 g_{\rho _L}^2} \left(m_{\rho _L}^2-f^2 g_{\rho _L}^2\right)\xi \\

\displaystyle -\frac{g_{\text{el}} }{2 g_{\rho _L}}B(\xi) m_{\rho _L}^2 &\displaystyle m_{\rho _L}^2 & \displaystyle -\frac{ g_{\text{el}}'}{2 g_{\rho _L}}C(\xi) m_{\rho _L}^2 \\

\displaystyle \frac{g_{\text{el}}  g_{\text{el}}'}{4 g_{\rho _L}^2}\left(m_{\rho _L}^2-f^2 g_{\rho _L}^2\right)\xi & \displaystyle -\frac{ g_{\text{el}}'}{2 g_{\rho _L}}C(\xi) m_{\rho _L}^2 & \displaystyle  \frac{ \left(g_{\text{el}}'\right){}^2}{4 g_{\rho _L}^2}\left(g_{\rho _L}^2 f^2 \xi -D(\xi) m_{\rho _L}^2\right) \\
\end{array}
\right),
\end{equation}
where we have expressed the misalignment angle $\theta$  as a function of $\xi$, according to equation (\ref{Csi}), and we have defined the functions
\begin{equation}\label{xiFunctions}
\begin{array}{llll}
\displaystyle A(\xi) = \left(2 \sqrt{1-\xi }+2-\xi \right) , & \quad B(\xi) = \left(1+\sqrt{1-\xi }\right),  \\
\displaystyle C(\xi) = \left(1-\sqrt{1-\xi }\right), & \quad D(\xi) =  \left(2 \sqrt{1-\xi }-2+\xi \right).
\end{array}
\end{equation}
It is now straightforward to analytically diagonalise the two matrices, but in general the expressions for the eigenvalues and the eigenvectors are quite complicated. It is thus more convenient to perform a numerical diagonalization, unless specific limits are considered in which expressions simplify. We will provide in Appendix \ref{AppNumDiag} a $\mathtt{Mathematica}$ code which makes such a numerical diagonalization for given values of the input parameters and generates all the relevant couplings and masses. In the rest of our study, however, we will work in the limit $\xi \ll 1$, which, besides being experimentally favoured, can also lead to simple analytical formulae for the physical couplings between the heavy triplet and the other particles in our theory. We will therefore expand the mass matrix and its eigenvectors and eigenvalues at leading order in $\xi$ so that our approximation will break down when $\xi \gtrsim 0.4$, in which case the corrections coming from subsequent powers in the expansion become non-negligible.  

The spectrum of the spin-1 sector is easily found once the mass matrix is diagonalised at linear order in $\xi$; after EWSB, the only massless state is the photon, since it is the gauge field associated with the unbroken $U(1)_{em}$, whereas for the remaining massive bosons we get:\footnote{Here and in the following we will generically indicate with $m_\rho$ the lagrangian parameters corresponding to the mass of one of the vector resonances and with $M_\rho$ the corresponding physical masses obtained by inverting the expressions of the latter in terms of the former.}
\begin{equation}\label{Spin-1Masses}
\begin{array}{ll}
\displaystyle m_W^2 = {g^2 \over 4}f^2 \xi, \qquad \displaystyle m_Z^2 = {g^2 + g^{\prime 2} \over 4} f^2 \xi, \\

\displaystyle M_{\rho_L^{\pm}}^2=M_{\rho_L^0}^2= \frac{g_{\rho _L}^2 }{g_{\rho _L}^2-g^2}m_{\rho _L}^2-{g^2 \xi \over 4}\left(\frac{ f^2 g^2-2  m_{\rho _L}^2}{ g^2- g_{\rho _L}^2}\right),
\end{array}
\end{equation}
where we have used the SM couplings $g$ and $g^{\prime}$ introduced in equation (\ref{SMCoup}). As it is clear from the previous expression, the masses of the $W$ and $Z$ bosons originate only after EWSB; if we now define the electroweak scale as $v = \sqrt{\xi} f$, through equation (\ref{AlphaGF}), then $m_W$ and $m_Z$ have formally the same expression as in the SM.~\footnote{With this choice, the $O(\xi^2)$ corrections appear in $m_W$ and $m_Z$, but not in $v$. One could equivalently define $v$ through the formula $m_W = {g v \over 2}$, so that $G_F$ in equation (\ref{AlphaGF}) deviates from its SM expression at  $O(\xi^2)$, once rewritten in terms of $v$.}~The masses of the heavy triplet arise instead at zeroth order in $\xi$ and get corrections after EWSB; at leading order in $\xi$, these corrections are equal for the two charged and the neutral resonances, since they do not depend on $g^{\prime}$, which is the only parameter in the bosonic sector to break the custodial symmetry. This degeneracy will be in general removed by $O(\xi^2)$ contributions. 

Once the form of the rotation to the mass eigenstate basis is derived, it is straightforward to obtain the physical interactions between the vector resonances, the SM fields and the top partners. We will focus in the following on trilinear vertices, which are the most relevant ones for studying the production and decay of heavy spin-1 states at the LHC, and we will refer to Appendix \ref{AppCoup} for the expression of the Lagrangian and the couplings in the mass eigenstate basis. 

We start analysing some qualitative features of the interactions among the vector resonances, the gauge bosons and the Higgs field. We notice first of all that the Lorentz structure of the vertices involving the heavy spin-1 states and two gauge bosons is the same as the one for triple gauge vertices in the SM. This is because the kinetic terms for both composite and elementary fields in Eq.~(\ref{LagLeft}) imply interactions of the SM type, since $\mathcal{L}_L$ has been truncated to two derivatives interactions, and rotating to the mass eigenbasis does not obviously change their Lorentz structure. Moreover, the values of the $g_{\rho^+_LWZ}$, $g_{\rho^+_LWH}$, $g_{\rho^0_LWW}$ and $g_{\rho^0_LZH}$ couplings can be easily extracted by using the Equivalence Theorem for $M_{\rho_L} \gg m_{Z/W}$; in this limit, the leading contribution to the interaction comes from the longitudinal polarizations of the SM vector fields and the overall strength equals that of the coupling of one $\rho_\mu^L$ to two NG bosons, $\rho_\mu^L \pi \pi$, up to small corrections of order $O(m_{Z/W}^2/M_{\rho_L}^2)$. As it can be directly seen from equation (\ref{LagLeft}), the $\rho_\mu^L \pi \pi$ coupling is proportional to $g_{\rho_L} a_{\rho_L}^2$, where the $O(1)$ parameter $a_{\rho_L}=m_{\rho_L}/(g_{\rho_L}f)$ is introduced analogously to Eq.~(\ref{NDA}) in order to enforce the NDA relation between the mass and coupling of the resonance. The free parameter $g_{\rho_L}$ plays therefore a dominant role in setting the strength of the interaction between the vectors and the SM gauge fields and Higgs. 

The interactions of the heavy vectors with the SM leptons and first two quark families, on the other hand, follow entirely from the universal composite-elementary mixing, that is from the elementary component of the heavy spin-1 mass eigenstate. As a consequence, the three couplings $g_{\rho_L^+ ffL}$, $g_{\rho_L^0 ffL}$ and $g_{\rho_L^0 ffY}$ do not depend on the fermion species and are therefore universal. After rotation to the mass eigenstate basis, the first two couplings scale like $\sim g^2 / g_{\rho_L}$, whereas the last one is of order $\sim g^{\prime 2}/g_{\rho_L}$. Moreover, since the $\rho_\mu^L$ triplet mixes with the elementary $W_\mu$ before EWSB and with the gauge field $B_\mu$ only after EWSB, the functions $g_{\rho_L^+ ffL}$ and $g_{\rho_L^0 ffL}$ arise at zeroth order in $\xi$ and they are equal up to $O(\xi)$ terms, since the breaking of the custodial symmetry due to the hypercharge $g^{\prime}$ enters only through EWSB effects. The coupling $g_{\rho_L^0 ffY}$ is instead generated only by the $\rho_\mu^L - B_\mu$ mixing and is therefore proportional to $\xi$, so that its contribution to the interaction between the neutral vector and massless fermions is sub-leading. From the above discussions it obviously follows that, in the limit $g_{\rho_L} \gg g$, the heavy resonances are most strongly coupled to composite states, namely the longitudinal $W$ and $Z$ bosons and the Higgs, whereas their coupling strength to lighter fermions is extremely weak.

Let us now consider the interactions among the heavy triplet and the partially composite top-bottom pair and the $t_R$. Besides the universal terms in the functions $g_{\rho_L^+ tb}$, $g_{\rho_L^0 t_L t_L}$ and $g_{\rho_L^0 b_L b_L}$ coming from the vector elementary-composite mixing, these couplings also receive an additional contribution before EWSB, due to the fermionic mixing, from the direct interaction of the vector resonances with top partners proportional to the $O(1)$ parameter $c_3$. The heaviest SM quarks are thus effectively more strongly coupled to the resonances than the lighter ones. After rotation to the mass eigenstate basis, all the previous functions scale in the same way and are of order 
\begin{equation}\label{EnhanCoup}
g_{{\rho_L^+}tb} \sim {g^2 \over g_{\rho_L}} + c_3 g_{\rho_L} {y_L^2 f^2 \over y_L^2 f^2 + M_\Psi^2}.
\end{equation}
As regards the $t_R$, the additional contributions to the function $g_{\rho_L^0 t_R t_R}$  must arise only after EWSB, because this particle is a singlet under the unbroken group $H$, whereas the $\rho_\mu^L$ resonance has isospin 1 under the $SU(2)_L$ subgroup of $SO(4)$. Isospin conservation therefore forbids any new interaction coming both from the term proportional to the parameter $c_1$ in $\mathcal{L}_{\Psi}$ and from the term proportional to $c_3$ in $\mathcal{L}_{\rho_L}$ before EWSB, so that this coupling does not receive a relevant enhancement for small values of the misalignment angle. 

The last set of interactions that has a prominent role in the phenomenology of composite vectors is that involving the top partners; we start considering how the spin-1 resonances couple with a heavy fermion and one third family quark. Before EWSB, the only couplings allowed by isospin conservation are $g_{\rho_L^+ T_L b_L}$, $g_{\rho_L^+ B_L t_L}$, $g_{\rho_L^0 T_L t_L}$, $g_{\rho_L^0 B_L b_L}$; they are generated by the last term in $\mathcal{L}_{\rho_L}$, since the kinetic terms are invariant under the rotation in the fermionic sector and the interaction $i c_1 \bar{\Psi}_i \slashed{d}^i t_R$ in $\mathcal{L}_\Psi$ can only contribute after EWSB. Once the rotation to the mass eigenstate basis is performed, all the previous couplings scale obviously like
\begin{equation}\label{TopPartQuarkCoup}
g_{\rho_L^+ T_L b_L} \sim c_3 g_{\rho_L} \frac{y_L f M_\Psi}{y_L^2 f^2 + M_\Psi^2},
\end{equation} 
and will receive further $O(\xi)$ corrections for non-zero values of the misalignment angle. We thus expect the decay channels to $T \bar{b}$, $B \bar{t}$, $T \bar{t}$ and $B \bar{b}$ to play an important role in the decay of the heavy vectors, especially for large values of the strong coupling constant $g_{\rho_L}$ and for high degrees of quark compositeness. All the remaining couplings between a spin-1 resonance, a top partner and a third family quark must originate after EWSB, since at least an insertion of the Higgs vev is needed to conserve the isospin, so that they will in general give a sub-dominant contribution to the phenomenology of vector resonances. 

We now consider the couplings between two heavy fermions and one heavy boson. The same analysis made for the previous situation is valid also in this case and we still expect the dominant interaction to be given by the term proportional to $c_3$ in $\mathcal{L}_{\rho_L}$. The universal contribution due to the elementary-composite mixing in the top partners kinetic term scales indeed like $g^2/g_{\rho_L}$ and the direct interaction between spin-1 and spin-1/2 resonances induces an additional contribution proportional to $g_{\rho_L}$. For large values of the strong coupling constant, the universal piece will therefore be suppressed whereas the second will be enhanced, analogously to what happens for the partially composite quarks. The functions generated before EWSB are those allowed by isospin conservation, namely $g_{\rho_L^+ T_L B_L}$, $g_{\rho_L^0 T_L T_L}$, $g_{\rho_L^0 B_L B_L}$, which all scale like
\begin{equation}\label{TopPartCoup1}
g_{\rho_L^+ T_L B_L} \sim {g^2 \over g_{\rho_L}} + c_3 g_{\rho_L}\frac{M_\Psi^2}{y_L^2f^2+M_\Psi^2},
\end{equation}  
and $g_{\rho_L^+ X_{{2\over 3}} X_{{5\over 3}}}$, $g_{\rho_L^+ T_R B_R}$, $g_{\rho_L^0 X_{{5\over 3}} X_{{5\over 3}}}$, $g_{\rho_L^0 X_{{2\over 3}} X_{{2\over 3}}}$, $g_{\rho_L^0 T_R T_R}$ and $g_{\rho_L^0 B_R B_R}$, which instead are all of order
\begin{equation}\label{TopPartCoup2}
g_{\rho_L^+ X_{{2\over 3}} X_{{5\over 3}}} \sim {g^2 \over g_{\rho_L}} + c_3 g_{\rho_L}.
\end{equation}
These second set of couplings does not receive any contribution from the rotation angles in Eq. (\ref{FermRot}) because the $X_{2/3}$, $T_R$ and $B_R$ fields are left invariant by the rotation in the fermionic sector before EWSB. We therefore expect the decay channel of vectors to $T\bar{B}$, $T\bar{T}$, $B\bar{B}$, $X_{2 \over 3} \bar{X_{5 \over 3}}$, $X_{5 \over 3} \bar{X_{5 \over 3}}$ and $X_{2 \over 3}\bar{X_{2 \over 3}}$ to be the most important one, when kinematically allowed, among the decays to two top partners. The other possible decay channels will instead be suppressed by the small value of $\xi$ since they must originate only after EWSB. 

We have finally summarized these results in Table 1, where we have listed all the relevant couplings arising before EWSB, neglecting the $O(\xi)$ corrections.  
\begin{table}[ht!]\label{Table1}
\begin{tabular}{|l|l|}
  \hline
   Couplings & Scaling \\[0,3cm]
  \hline
   $g_{\rho_L^+ W_L Z_L}$, $g_{\rho_L^+ W_L H}$, $g_{\rho_L^0 W_L W_L}$, $g_{\rho_L^0 Z_LH}$ & $a_{\rho_L}^2 g_{\rho_L}$ \\ [0,3cm]
 
  $g_{\rho_L^+ ffL}$, $g_{\rho_L^0 ffL}$ &  $\displaystyle {g^2 \over g_{\rho_L}}$ \\ [0,3cm]
   
  $g_{\rho_L^+ tb}$, $g_{\rho_L^0 t_L t_L}$, $g_{\rho_L^0 b_L b_L}$ & ${\displaystyle g^2 \over \displaystyle g_{\rho_L}} + c_3 g_{\rho_L} \displaystyle{y_L^2 f^2 \over y_L^2 f^2 + M_\Psi^2}$ \\ [0,3cm]
  
  $g_{\rho_L^+ T_L b_L}$, $g_{\rho_L^+ B_L t_L}$, $g_{\rho_L^0 T_L t_L}$, $g_{\rho_L^0 B_L b_L}$ & $\displaystyle c_3 g_{\rho_L} \frac{y_L f M_\Psi}{y_L^2 f^2 + M_\Psi^2}$ \\ [0,3cm]
 
  $g_{\rho_L^+ T_L B_L}$, $g_{\rho_L^0 T_L T_L}$, $g_{\rho_L^0 B_L B_L}$ & $\displaystyle {g^2 \over g_{\rho_L}} + c_3 g_{\rho_L}\frac{M_\Psi^2}{y_L^2f^2+M_\Psi^2}$ \\ [0,3cm]

  $g_{\rho_L^+ X_{{2\over 3}} X_{{5\over 3}}}$, $g_{\rho_L^+ T_R B_R}$, $g_{\rho_L^0 X_{{5\over 3}} X_{{5\over 3}}}$, $g_{\rho_L^0 X_{{2\over 3}} X_{{2\over 3}}}$, $g_{\rho_L^0 T_R T_R}$, $g_{\rho_L^0 B_R B_R}$ & $ \displaystyle{g^2 \over g_{\rho_L}} + c_3 g_{\rho_L}$\\ [0,3cm]
  \hline
\end{tabular}
\caption{\small List of the couplings arising before EWSB and their scaling with the strong coupling constant $g_{\rho_L}$ in the mass eigenstate basis, for the $\rho^\mu_L$ resonance coupled to top partners.}
\end{table}

\subsection{A Lagrangian for $\rho_\mu^R$}

We now introduce the Lagrangian for the $(\textbf{1},\textbf{3})_0$ vector resonance coupled to top partners in the fourplet, with fully composite $t_R$; it is given by:
\begin{equation}\label{LagRightBref}
\mathcal{L}_R = \mathcal{L}_{light} + \mathcal{L}_\Psi + \mathcal{L}_{\rho_R},
\end{equation}
where $ \mathcal{L}_{light}$ and $\mathcal{L}_\Psi$ have the same expression as in Eq.~(\ref{LagLeft}), whereas $ \mathcal{L}_{\rho_R}$ is
\begin{equation}\label{LagRight}
\mathcal{L}_{\rho_R}= - {1\over 4 } \rho_{\mu \nu }^{a_R}\rho^{a_R \mu \nu} + {m^2_{\rho_R} \over 2 g_{\rho_R}^2}( g_{\rho_R} \rho_\mu^{a_R}-E_\mu^{a_R})^2 + c_4 \bar{\Psi}^i \gamma^\mu (g_{\rho_R} \rho_\mu^{a_R}-E_\mu^{a_R})T^{a_R}_{ij} \Psi^j.
\end{equation}
The theory possesses again eleven parameters with $m_{\rho_R}$, $g_{\rho_R}$ and $c_4$ indicating respectively the mass and strong coupling constant of the $\rho^\mu_R$ resonance and the $O(1)$ parameter which plays the analogous role of $c_3$. As in the previous case, we can re-express all the Lagrangian input parameters in terms of physical quantities and fix $g_{el}$, $g_{el}^\prime$ and $f$ in order to reproduce the experimental values of $\alpha$, $G_F$ and $m_Z$, as described in Eq.~(\ref{AlphaGF}). We can define the SM $g$ and $g^\prime$ weak couplings as
\begin{equation}\label{SMCoupR}
g \equiv g_{el} \qquad {1\over {g^\prime}^2} \equiv {1\over {g_{el}^\prime}^2} + {1\over g_{\rho_R}^2},
\end{equation}
so that, differently to the $\rho_\mu^L$ case, we can now identify $g$ as the elementary gauge coupling constant.

Due to the interaction between the composite $\rho_R$ and the elementary gauge fields induced by the $\rho_R-W$ and $\rho_R-B$ mixings, the mass matrix of the bosonic sector of the theory is again non-diagonal. Analogously to Eq.~(\ref{MassTerm}), we can introduce the $2\times2$ charged block
\begin{equation}\label{RightChargedMat}
M_{\pm}^2 = \left(
\begin{array}{cc}
 \displaystyle \frac{g_{\text{el}}^2 }{4 g_{\rho _R}^2} \left( g_{\rho _R}^2 f^2 \xi  -D(\xi) m_{\rho _R}^2\right)& \displaystyle -\frac{ g_{\text{el}}}{2 g_{\rho _R}}C(\xi) m_{\rho _R}^2 \\
\displaystyle - \frac{g_{\text{el}}}{2 g_{\rho _R}}C(\xi) m_{\rho _R}^2 & \displaystyle m_{\rho _R}^2 \\
\end{array}
\right)
\end{equation} 
and the $3\times 3$ neutral block
\begin{equation}\label{RightNeutMat}
M_0^2 = \left(
\begin{array}{ccc}
\displaystyle \frac{\left(g_{\text{el}}'\right){}^2}{4 g_{\rho _R}^2}\left(g_{\rho _R}^2 f^2 \xi  -D(\xi) m_{\rho _R}^2\right) & \displaystyle  -\frac{g_{\text{el}}'}{2 g_{\rho _R}}C(\xi) m_{\rho _R}^2& \displaystyle \frac{ g_{\text{el}}  g_{\text{el}}'}{4 g_{\rho _R}^2}\left(m_{\rho _R}^2-f^2 g_{\rho _R}^2\right)\xi \\
\displaystyle -\frac{g_{\text{el}}'}{2 g_{\rho _R}}C(\xi) m_{\rho _R}^2 & \displaystyle m_{\rho _R}^2 & \displaystyle -\frac{g_{\text{el}}}{2 g_{\rho _R}}B(\xi) m_{\rho _R}^2 \\
\displaystyle \frac{ g_{\text{el}}  g_{\text{el}}'}{4 g_{\rho _R}^2}\left(m_{\rho _R}^2-f^2 g_{\rho _R}^2\right)\xi & \displaystyle -\frac{g_{\text{el}}}{2 g_{\rho _R}}B(\xi) m_{\rho _R}^2 & \displaystyle \frac{g_{\text{el}}^2}{4 g_{\rho _R}^2}\left(g_{\rho _R}^2 f^2 \xi  +A(\xi) m_{\rho _R}^2\right) \\
\end{array}
\right),
\end{equation}
that can be diagonalized numerically with the code provided in Appendix \ref{AppNumDiag}. The spectrum contains the massless photon, the $W$ and $Z$ boson, whose masses, at linear order in $\xi$, get the same expression as in Eq.~(\ref{Spin-1Masses}), and the right-handed triplet with masses
\begin{equation}\label{RightMasses}
\begin{array}{ll}
M_{\rho_R^\pm}^2 = m_{\rho_R}^2 + O(\xi^2), \qquad \displaystyle M_{\rho_R^{0}}= \frac{g_{\rho _R}^2 }{g_{\rho _R}^2-g'^2}m_{\rho _R}^2-\frac{g'^2 \xi}{4}\left(\frac{f^2 g'^2-2  m_{\rho _R}^2}{ g'^2- g_{\rho _R}^2}\right)+ O(\xi^2).
\end{array}
\end{equation} 
We see that the mass of the charged heavy vector coincides with the Lagrangian parameter $m_{\rho_R}$, up to $O(\xi^2)$ corrections, and that the spectrum is degenerate even at zeroth order in $\xi$ due to the dependence on $g^\prime$ which explicitly breaks the custodial symmetry.

We can easily derive the couplings of the spin-1 resonance to SM particles and top partners in the mass eigenstate basis once the rotation is performed; we will briefly describe their most important features, stressing the main differences from the left-handed vector.

Following the same reasoning of the previous analysis, we can verify that the functions $g_{\rho_R^+ ZW}$, $g_{\rho_R^+ WH}$, $g_{\rho_R^0 WW}$, $g_{\rho_R^0 ZH}$ scale all like $a_{\rho_R}^2 g_{\rho_R}$, in the limit when the Equivalence Theorem is a very good approximation, namely $M_{\rho_R^{\pm/0}} \gg m_{W/Z}$. As regards the fully elementary fermions, the universal composite-elementary mixing is such that also the couplings $g_{\rho_R^+ ffL}$, $g_{\rho_R^0 ffL}$ and $g_{\rho_R^0 ffY}$ scale in the same way as in left-handed case. However, since the $\rho_\mu^R$ triplet mixes with the elementary $W_\mu$ field after EWSB and with the gauge boson $B_\mu$ before EWSB, the couplings $g_{\rho_R^+ ffL}$ and $g_{\rho_R^0 ffL}$ arise at linear order in $\xi$ and are no longer equal due to the effects of the hypercharge $g^\prime$, whereas the $g_{\rho_R^0 ffY}$ function, induced only by the $\rho_\mu^R-B_\mu$ mixing, is generated at zeroth order in $\xi$ and gives the most relevant contribution. As a consequence, the charged heavy vectors couple very weakly to the lightest SM fermions, contrary to the $\rho_\mu^L$ resonance. Finally, the couplings to the partially composite $t_L$ and $b_L$ are enhanced by the interaction proportional to $c_4$. However, being $\rho_R$ an $SU(2)_L$ singlet, before EWSB it can couple only to the $SU(2)_L$ singlet current ($t\bar{t} + b\bar{b}$), so that the enhancement in $g_{\rho_R^+ tb}$ is proportional to $\xi$ and therefore suppressed by the small value of the misalignment angle. On the other hand, the couplings $g_{\rho_R^0 t_L t_L}$ and $g_{\rho_R^0 b_L b_L}$ are allowed by isospin conservation even at zeroth order in $\xi$ and they scale like their left-handed counterparts.

Considering now the couplings to one top partner and one third family quark, the functions arising before EWSB are $g_{\rho_R^+ X_{{2\over 3}L}b_L}$, $g_{\rho_R^+ X_{{5\over 3}L}t_L}$, $g_{\rho_R^0 T_L t_L}$ and $g_{\rho_R^0 B_L b_L}$ and again they are generated by the interaction proportional to $c_4$. Differently to the previous case, the charged resonance will therefore be more strongly coupled to $X_{2\over 3}\bar{b}$ and $X_{5/3}\bar{t}$, since it can interact only to the $SU(2)_L$ singlet current $(X_{2\over 3} \bar{b} + X_{5\over 3}\bar{t})$ at zeroth order in $\xi$. For the neutral vector, on the other hand, the decays to $T\bar{t}$ and $B\bar{b}$ will still be the most important one among the heavy-light channels, analogously to the $\rho_\mu^L$ heavy vector. Finally, as regards the couplings to two top partners, the situation is similar to the previous one: the relevant interactions of the neutral resonance are the same as the ones listed for the left-handed case, whereas the charged $\rho_R^+$ will couple preferably to $X_{2\over 3}\bar{B}$ and $X_{5\over 3}\bar{T}$, again because of the different quantum numbers of the left-handed and right-handed vectors.

We have summarized all the relevant couplings for this second model in Table 2, where their scaling with $g_{\rho_R}$ is given neglecting corrections arising after EWSB.
\begin{table}[ht!]\label{Table2}
\begin{tabular}{|l|l|}
  \hline
   Couplings & Scaling \\[0,3cm]
  \hline
   $g_{\rho_R^+ W_L Z_L }$, $g_{\rho_R^+ W_L H}$, $g_{\rho_R^0 W_L W_L}$, $g_{\rho_R^0 Z_L H}$ & $a_{\rho_R}^2 g_{\rho_R}$ \\ [0,3cm]
 
  $g_{\rho_R^0 ffY}$ &  $\displaystyle {g^{\prime 2} \over g_{\rho_R}}$ \\ [0,3cm]
   
  $g_{\rho_R^0 t_L t_L}$, $g_{\rho_R^0 b_L b_L}$ & ${\displaystyle g^{\prime 2} \over \displaystyle g_{\rho_R}} + c_4 g_{\rho_R} \displaystyle{y_L^2 f^2 \over y_L^2 f^2 + M_\Psi^2}$ \\ [0,3cm]
  
  $g_{\rho_R^+  X_{{2\over 3}L} b_L}$, $g_{\rho_R^+  X_{{5\over 3}L} t_L}$  & $\displaystyle c_4 g_{\rho_R} \frac{y_L f }{\sqrt{y_L^2 f^2 + M_\Psi^2}}$ \\ [0,3cm]
  
   $g_{\rho_R^0 T_L t_L}$, $g_{\rho_R^0 B_L b_L}$ & $\displaystyle c_4 g_{\rho_R} \frac{y_L f M_\Psi}{y_L^2 f^2 + M_\Psi^2}$ \\ [0,3cm]
   
    $g_{\rho_R^+ X_{{2\over 3}L} B_L}$ & $\displaystyle c_4 g_{\rho_R}\frac{M_\Psi}{\sqrt{y_L^2f^2+M_\Psi^2}}$ \\ [0,3cm]
 
   $g_{\rho_R^0 T_L T_L}$, $g_{\rho_R^0 B_L B_L}$ & $\displaystyle {g^{\prime 2} \over g_{\rho_R}} + c_4 g_{\rho_R}\frac{M_\Psi^2}{y_L^2f^2+M_\Psi^2}$ \\ [0,3cm]

  $g_{\rho_R^+ X_{{5\over 3}L}T_L }$, $g_{\rho_R^+ X_{{5\over 3}R}T_R }$, $g_{\rho_R^+  X_{{2\over 3}R}B_R}$, $g_{\rho_R^0 X_{{5\over 3}} X_{{5\over 3}}}$, $g_{\rho_R^0 X_{{2\over 3}} X_{{2\over 3}}}$, $g_{\rho_R^0 T_R T_R}$, $g_{\rho_R^0 B_R B_R}$ & $ \displaystyle{g^{\prime 2} \over g_{\rho_R}} + c_4 g_{\rho_R}$\\ [0,3cm]
  \hline
\end{tabular}
\caption{\small List of the couplings arising before EWSB and their scaling with the strong coupling constant $g_{\rho_R}$ in the mass eigenstate basis, for the $\rho^\mu_R$ resonance coupled to top partners.}
\end{table}         

\subsection{Two Lagrangians for $\rho_\mu^X$}

We consider now the phenomenology of a spin-1 resonance transforming only under the abelian $U(1)_X$ as a gauge field,
\begin{equation}
\rho_\mu^X \rightarrow \rho_\mu + \partial_\mu \alpha^X,
\end{equation}
with $\alpha^X \in U(1)_X$, and interacting with top partners in the singlet of $SO(4)$, $\widetilde{T}$. This vector has very different properties with respect to the left-handed and right-handed cases; we expect it to be more strongly coupled to particles which do not transform under $SO(4)$, $t_R$ and $\widetilde{T}$, so that its phenomenology can be significantly different if the $t_R$ belongs to the composite sector or if it is an elementary state linearly coupled to the new dynamics. We explore both these possibilities building two models, $\textbf{M}_{\textbf{X}}^{\textbf{1}}$ for the first situation and $\textbf{M}_{\textbf{X}}^{\textbf{2}}$ for the second. The Lagrangians for the two models read, respectively,
\begin{equation}\label{LagsX}
\mathcal{L}_{\textbf{M}_{\textbf{X}}^{\textbf{1}}} = \mathcal{L}_{light} + \mathcal{L}_{\widetilde{T}^1}+\mathcal{L}_{\rho_X^1}, \qquad \mathcal{L}_{\textbf{M}_{\textbf{X}}^{\textbf{2}}} = \mathcal{L}_{light} + \mathcal{L}_{\widetilde{T}^2}+\mathcal{L}_{\rho_X^2},
\end{equation}
with 
\begin{equation}\label{X1}
\begin{array}{ll}
\mathcal{L}_{\widetilde{T}^1} =& \bar{\widetilde{T}} i \slashed{D} \widetilde{T} - M_{\Psi} \bar{\widetilde{T}} \widetilde{T} + \left[y_L f (\bar{Q}_L^5)^I U_{I 5}\widetilde{T}_R+ y_L c_2 f(\bar{Q}_L^5)^IU_{I5} t_R + \text{h.c.} \right], \\

\\

\mathcal{L}_{\rho_X^1}= & \displaystyle -{1\over 4}\rho_{\mu \nu}^X \rho^{X \mu \nu } + {m_{\rho_X}^2 \over 2 g_{\rho_X}^2} (g_{\rho_X}\rho_\mu^X - g_{el}^\prime B_\mu )^2 + c_5 \bar{t}_R \gamma^\mu  (g_{\rho_X}\rho_\mu^X - g_{el}^\prime B_\mu )t_R  \\
& + c_6 \bar{\widetilde{T}}\gamma^\mu (g_{\rho_X}\rho_\mu^X - g_{el}^\prime B_\mu )\widetilde{T},

\end{array}
\end{equation}
and
\begin{equation}\label{X1}
\begin{array}{ll}
\mathcal{L}_{\widetilde{T}^2} = & \bar{\widetilde{T}} i \slashed{D} \widetilde{T} - M_{\Psi} \bar{\widetilde{T}} \widetilde{T} + \left[y_L f (\bar{Q}_L^5)^I U_{I 5}\widetilde{T}_R+ y_R f (\bar{Q}_R^5)^I U_{I 5}\widetilde{T}_L +\text{h.c.} \right], \\

\\

\mathcal{L}_{\rho_X^2}=& \displaystyle -{1\over 4}\rho_{\mu \nu}^X \rho^{X \mu \nu } + {m_{\rho_X}^2 \over 2 g_{\rho_X}^2} (g_{\rho_X}\rho_\mu^X - g_{el}^\prime B_\mu )^2 + c_6 \bar{\widetilde{T}}\gamma^\mu (g_{\rho_X}\rho_\mu^X - g_{el}^\prime B_\mu )\widetilde{T}.
\end{array}
\end{equation}
The Lagrangians $\mathcal{L}_{\widetilde{T}^1}$ and $\mathcal{L}_{\widetilde{T}^2}$ contain the kinetic term of the top partner and its interaction with the $t_R$ allowed by the symmetries; the fermion mass matrix is in general non-diagonal and must be diagonalised in both cases. The Lagrangians $\mathcal{L}_{\rho_X^1}$ and $\mathcal{L}_{\rho_X^2}$ describe the kinetic term of the vector singlet, with the field strength $\rho_{\mu \nu}^X$ obviously defined as
\begin{displaymath}
\rho_{\mu\nu}^X = \partial_\mu \rho^X_\nu - \partial_\nu \rho^X_\mu,
\end{displaymath}
and its direct coupling with $\widetilde{T}$. In model $\textbf{M}_{\textbf{X}}^{\textbf{1}}$ also a direct coupling with $t_R$ is present whereas the same interaction is forbidden for a partially composite $t_R$. The $\rho_\mu^X$ mixes in every case with the abelian gauge field $B_\mu$, which is needed to preserve invariance under $U(1)_X$, so that the mass matrix of the neutral spin-1 sector must be diagonalised by a field rotation. The two models have nine parameters in common, $g$, $g_{el}^\prime$ and $f$, that will be fixed to reproduce the experimental values of $\alpha$, $G_F$ and $m_Z$ according to Eq.~(\ref{AlphaGF}), $\xi$, $y_L$, the mass scales $M_{\Psi}$ and $m_{\rho_X}$, the strong coupling $g_{\rho_X}$ and the $O(1)$ parameter $c_6$. Model $\textbf{M}_{\textbf{X}}^{\textbf{1}}$ has two additional parameters, $c_2$, which must be fixed in order to reproduce the top mass, and $c_5$; apart from $\xi$ which is an observable, the six unfixed parameters could be traded for the mass of the heavy fermion, $m_{\widetilde{T}}$, and its coupling to a gauge boson and top quark, the mass of the heavy vector, its coupling to leptons, to the top quark and to the $\widetilde{T}$ particle. Model $\textbf{M}_{\textbf{X}}^{\textbf{2}}$, on the other hand, has one additional parameter, $y_R$; in this case we will fix $y_L$ to reproduce the top mass and the remaining free parameters can be expressed in terms of physical quantities similarly to the $\textbf{M}_{\textbf{X}}^{\textbf{1}}$ case.

We discuss now the rotation to the mass eigenstate basis and the spectrum of the models. As regards model $\textbf{M}_{\textbf{X}}^{\textbf{1}}$, the mass matrix of the fermionic sector has already been analysed in \cite{Hunters}, which we refer for the details. We just report here the expressions for the masses of the top quark and $\widetilde{T}$ at leading order in $\xi$,
\begin{equation}\label{tTMass1}
m_{top} = \frac{{c_2} y_Lf}{\sqrt{2}}\sqrt{\xi }, \qquad m_{\widetilde{T}} ={M_\Psi}+ \frac{y_L^2 f^2}{4 {M_\Psi}}\xi ,
\end{equation}  
and we notice that the two fields do not mix before EWSB, because the mass matrix is diagonal when $\xi = 0$. On the other hand, the mass matrix in model $\textbf{M}_{\textbf{X}}^{\textbf{2}}$ is 
\begin{equation}\label{FermMassMatX}
\left(
\begin{array}{ll}
\bar{t}_L \\
\bar{\widetilde{T}_L}
\end{array}
\right)
\left(
\begin{array}{cc}
 0 & \displaystyle -\frac{y_L f }{\sqrt{2}}\sqrt{\xi } \\
 f \sqrt{1-\xi } y_R & -{M}_{\Psi } \\
\end{array}
\right)
\left(
\begin{array}{ll}
{t}_R \\
{\widetilde{T}_R}
\end{array}
\right),
\end{equation} 
with eigenvalues 
\begin{equation}\label{TopTMassX2}
m_{top} = \frac{{y_L} {y_R}f^2 \sqrt{\xi }}{\sqrt{2} \sqrt{{y_R}^2 f^2+{M_\Psi}^2}}, \quad m_{\widetilde{T}} = \sqrt{f^2 {y_R}^2+{M_\Psi}^2}-\frac{f^2 \left(2 f^2 {y_R}^4-{M_\Psi}^2 \left({y_L}^2-2 {y_R}^2\right)\right)}{4 \left(f^2 {y_R}^2+{M_\Psi}^2\right)^{3/2}}\xi,
\end{equation}
which receive further corrections from higher orders in an expansion in $\xi$. In this case, the field rotation needed to diagonalise the mass matrix before EWSB is
\begin{equation}\label{FieldRotPartX}
t_R \rightarrow \frac{{M}_{\Psi } }{\sqrt{ y_R^2 f^2+{M}_{\Psi }^2}}t_R-\frac{ y_R f}{\sqrt{ y_R^2 f^2+{M}_{\Psi }^2}}\widetilde{T}_R,
\end{equation}
with the orthogonal transformation for the $\widetilde{T}_R$ field. Considering, on the other hand, the spin-1 sector, the mass matrix is the same for both models and, in the basis of Eq. (\ref{MassTerm}), it is given by:
\begin{equation}\label{BosMassX}
M_0^2= \left(
\begin{array}{ccc}
 \displaystyle  \frac{1}{4}  g_{\text{el}}^2 f^2 \xi   & 0 &  \displaystyle -\frac{1}{4}g_{\text{el}} g_{\text{el}}' f^2 \xi \\
 0 &\displaystyle {m}_{\rho _X}^2 & \displaystyle -\frac{g_{\text{el}}'}{g_{\rho _X}}{m}_{\rho _X}^2 \\
 \displaystyle -\frac{1}{4} g_{\text{el}} g_{\text{el}}' f^2 \xi  & \displaystyle -\frac{ g_{\text{el}}'}{g_{\rho _X}}{m}_{\rho _X}^2 &   \displaystyle \frac{\left(g_{\text{el}}'\right){}^2}{4} \left(\frac{4 {m}_{\rho _X}^2}{g_{\rho _X}^2}+f^2 \xi \right)\\
\end{array}
\right),
\end{equation}
where we notice that the zero entries are due to the absence of mixing of the $\rho_\mu^X$ singlet with $W_\mu^3$. The spectrum of the neutral sector contains the massless photon, the $W$ and $Z$ boson, whose masses have the same expressions as in Eq.~(\ref{Spin-1Masses}) at linear order in $\xi$, and the vector singlet, with mass
\begin{equation}\label{XMass}
M_{\rho_X}^2 =\frac{g_{\rho _X}^2}{g_{\rho _X}^2-\left(g'\right)^2}{m}_{\rho _X}^2+ \frac{ \left(g'\right)^4}{ g_{\rho _X}^2- \left(g'\right)^2}\frac{f^2 \xi}{4}+O(\xi^2),
\end{equation} 
where we have defined the SM coupling $g^\prime$ as in Eq.~(\ref{SMCoupR}), with $g_{\rho_R}$ replaced by $g_{\rho_X}$. 

Once the rotation is performed, it is straightforward to derive the couplings of the vector singlet to the heavy fermions and SM particles in the mass eigenstate basis; we discuss here their basic phenomenological features, stressing the differences with respect to the left-handed and right-handed cases. First of all, the couplings to gauge bosons and fully elementary fermions are the same in both models. Since $\rho_\mu^X$ is not charged under $SO(4)$, it cannot couple directly with the longitudinally polarized $W$ and $Z$ bosons, so that the functions $g_{\rho_X WW}$ and $g_{\rho_X ZH}$ arise only because of the mixing with the $B_\mu$ gauge field and must be generated after EWSB. They scale like $g^{\prime 2}/g_{\rho_X} \xi$ and are therefore strongly suppressed, contrary to what happens for $\rho_\mu^L$ and $\rho_\mu^R$. The couplings to elementary fermions, on the other hand, behave similarly to the previous cases: they are generated only because of the universal composite-elementary mixing and scale like $g^{\prime 2}/g_{\rho_X}$. In particular, the function $g_{\rho_X ffY}$ is produced before EWSB, because the mixing with $B_\mu$ arises at zeroth order in $\xi$, whereas $g_{\rho_X ffL}$ must be proportional to $\xi$, since the singlet does not mix with $W_\mu^3$.   

The two models differ in the couplings of the vector singlet to the top quark and $\widetilde{T}$, as it can be seen from Table 3, where we have summarized the scaling of the relevant couplings arising before EWSB. In both models, the function $g_{\rho_X t_R t_R}$, besides the universal contribution from the elementary-composite mixing, receives an additional enhancement which in model $\textbf{M}_{\textbf{X}}^{\textbf{1}}$ is due to the direct interaction proportional to $c_5$ and in model $\textbf{M}_{\textbf{X}}^{\textbf{2}}$ results from the interaction proportional to $c_6$ as a consequence of the fermionic rotation. The coupling $g_{\rho_X \widetilde{T}_L t_L}$ must be generated in both cases at linear order in $\xi$, since $t_L$ and $\widetilde{T}_L$ do not mix before EWSB, whereas the function $g_{\rho_X \widetilde{T}_R t_R}$ arises after EWSB in model $\textbf{M}_{\textbf{X}}^{\textbf{1}}$, because in this case $t_R$ and $\widetilde{T}_R$ mix when $\xi \neq 0$, and before EWSB in model $\textbf{M}_{\textbf{X}}^{\textbf{2}}$, since now the two fields mix even before EWSB and the coupling is proportional to the rotation angle. Finally, as regards the interaction between the vector singlet and two top partners, following the same reasoning, it is clear that the function $g_{\rho_X \widetilde{T}_L \widetilde{T}_L}$ must be the same for both models, whereas the coupling $g_{\rho_X \widetilde{T}_R \widetilde{T}_R}$ receives the contribution of the rotation angle before EWSB in model $\textbf{M}_{\textbf{X}}^{\textbf{2}}$, which is instead absent if the $t_R$ is a full singlet of the strong dynamics.

As a result of the previous analysis, we expect a relevant decay channel of the vector singlet to be $t\bar{t}$ in both models; among the channels involving the top partners, $\widetilde{T}\bar{\widetilde{T}}$ has great importance in both cases, whereas $\widetilde{T}\bar{t}$ is suppressed by the small value of $\xi$ in model $\textbf{M}_{\textbf{X}}^{\textbf{1}}$ and is instead enhanced in model $\textbf{M}_{\textbf{X}}^{\textbf{2}}$. This features will lead to a different phenomenology for the two models, so that the vector singlet is particularly sensitive to the degree of compositeness of the $t_R$ quark.    
\begin{table}[ht!]\label{Table3}
\begin{center}
\begin{tabular}{|l|l|l|}
  \hline
Couplings &  Scaling $\textbf{M}_{\textbf{X}}^{\textbf{1}}$ & Scaling $\textbf{M}_{\textbf{X}}^{\textbf{2}}$ \\[0,3cm]
  \hline
  $g_{\rho_X ffY}$ & $ \displaystyle {g^{\prime 2} \over g_{\rho_X}}$ & $ \displaystyle {g^{\prime 2} \over g_{\rho_X}}$\\ [0,3cm]   
   
  $g_{\rho_X t_R t_R}$ & ${\displaystyle g^{\prime 2} \over \displaystyle g_{\rho_X}} + c_5 g_{\rho_X}$ & ${\displaystyle g^{\prime 2} \over \displaystyle g_{\rho_X}} + \displaystyle c_6 g_{\rho_X}\frac{y_R ^2 f^2}{{y_R^2f^2+M_\Psi^2}}$ \\ [0,3cm]
   
   $g_{\rho_X \widetilde{T}_R t_R}$ &  & $\displaystyle {c_6 g_{\rho_X} \frac{y_R f M_{\Psi}}{y_R^2f^2 + M_{\Psi}^2}}$ \\ [0,3cm]
   
    $g_{\rho_X \widetilde{T}_L \widetilde{T}_L}$ & $\displaystyle{{g^{\prime 2} \over g_{\rho_X}} +  c_6 g_{\rho_X}}$ & $\displaystyle{{g^{\prime 2} \over g_{\rho_X}} +  c_6 g_{\rho_X}}$ \\ [0,3cm]
 
  $g_{\rho_X \widetilde{T}_R \widetilde{T}_R}$ & $\displaystyle{{g^{\prime 2} \over g_{\rho_X}} +  c_6 g_{\rho_X}}$ & $\displaystyle{{g^{\prime 2} \over g_{\rho_X}} +  c_6 g_{\rho_X}\frac{ M_{\Psi}^2}{y_R^2f^2 + M_{\Psi}^2}}$ \\ [0,3cm]
  \hline
\end{tabular}
\caption{\small List of the couplings arising before EWSB and their scaling with the strong coupling constant $g_{\rho_X}$ in the mass eigenstate basis, for the $\rho^\mu_X$ resonance in models $\textbf{M}_{\textbf{X}}^{\textbf{1}}$ and $\textbf{M}_{\textbf{X}}^{\textbf{2}}$.}
\end{center}
\end{table} 


\section{Production and decay of vector resonances at the LHC} 
\label{sec:ProdDec}

We discuss in this section the main LHC production mechanisms and the decay channels of the vector resonances under consideration. We will parametrize the production cross section in terms of some fundamental functions that can be computed with a Monte Carlo code, like $\mathtt{MadGraph5}$ \cite{MadGraph}, and some universal couplings, whose expressions can be derived either analytically or numerically once the rotation to the mass eigenstate basis has been performed. This procedure is very useful to scan the parameter space of the theories, as we shall see when discussing the bounds from LHC direct searches. We will then study the most relevant decay channels and introduce an efficient analytical computation of the branching ratios with the $\mathtt{FeynRules}$ package, \cite{FeynRules}, as functions of the couplings in Appendix \ref{AppCoup}. 

\subsection{Production cross section}

The main production mechanisms of the vector resonances at the LHC, at a center of mass energy of $\sqrt{s}= 8 \ \text{TeV}$, are Drell-Yan processes and VBF. Under the validity of the Narrow Width Approximation (NWA), each production rate can be factorized into an on-shell cross section times a decay branching fraction. For the Drell-Yan case, the on-shell cross sections are controlled by the universal couplings $g_{\rho^+ffL}$, $g_{\rho^0 ffL}$, $g_{\rho^0 ffY}$ and can be written as
\begin{equation}\label{CrossSection}
\begin{array}{ll}
\sigma(pp \rightarrow \rho^+ + X ) = g_{\rho^+ ffL}^2 \cdot \sigma_{u \bar{d}}, \\
\sigma(pp \rightarrow \rho^- + X ) = g_{\rho^+ ffL}^2 \cdot \sigma_{d \bar{u}}, \\
\sigma(pp \rightarrow \rho^0 + X ) = g_{\rho^0 uu}^2 \cdot \sigma_{u \bar{u}} + g_{\rho^0 dd}^2 \cdot \sigma_{d \bar{d}},
\end{array}
\end{equation} 
where $\rho$ stands for $\rho_L$, $\rho_R$ or $\rho_X$ and $g_{\rho^0 uu}$ and $g_{\rho^0 dd}$ are the coupling strength of respectively up- and down-type fermions to the resonance,
\begin{equation}\label{UnivCoup}
\begin{array}{ll}
\displaystyle {g_{\rho^0 uu} = \left[\left( {1\over 2} \left(g_{\rho^0 ffL}-g_{\rho^0 ffY} \right) + {2\over 3}g_{\rho^0 ffY} \right) ^2 + \left( {2\over 3} g_{\rho^0 ffY}\right)^2 \right]^{1/2}}, \\
\displaystyle {g_{\rho^0 dd} = \left[\left( -{1\over 2} \left(g_{\rho^0 ffL}-g_{\rho^0 ffY} \right) - {1\over 3}g_{\rho^0 ffY} \right) ^2 + \left( {-1\over 3} g_{\rho^0 ffY}\right)^2 \right]^{1/2}}. 
\end{array}
\end{equation}
We have furthermore defined the partonic cross sections as
\begin{equation}
\begin{array}{ll}
\sigma_{u \bar{d}}= \sum\limits_{\psi_u, \psi_d}^{} \sigma(pp \rightarrow \psi_u \bar{\psi}_d \rightarrow \rho^+ + X )\left. \right |_{g_{\rho^+ ffL}=1}, \\
\sigma_{d \bar{u}}= \sum\limits_{\psi_u, \psi_d}^{} \sigma(pp \rightarrow \psi_d \bar{\psi}_u \rightarrow \rho^0 + X )\left. \right |_{g_{\rho^+ ffL}=1}, \\
\sigma_{u \bar{u}}= \sum\limits_{\psi_u}^{} \sigma(pp \rightarrow \psi_u \bar{\psi}_u \rightarrow \rho^0 + X )\left. \right |_{g_{\rho^0 uu}=1}, \\
\sigma_{d \bar{d}}= \sum\limits_{\psi_d}^{} \sigma(pp \rightarrow \psi_d \bar{\psi}_d \rightarrow \rho^0 + X )\left. \right |_{g_{\rho^0 dd}=1},
\end{array}
\end{equation}
where we have schematically indicated $\psi_u = u,c$ and $\psi_d=d,s$. The total production rates (\ref{CrossSection}) are thus simply given in terms of the fundamental cross sections, which include the contributions of all the initial partons and can be computed with a Monte Carlo code, appropriately rescaled by the couplings $g_{\rho^+ ffL}$, $g_{\rho^0 uu}$ and $g_{\rho^0 dd}$.

Analogously, the VBF production cross sections are controlled by the couplings $g_{\rho^+ WZ}$, $g_{\rho^0 WW}$ and can be parametrized as
\begin{equation}
\begin{array}{ll}
\sigma (pp \rightarrow \rho^+ + X ) = g_{\rho^+ WZ}^2 \cdot \sigma_{W^+ Z}, \\
\sigma (pp \rightarrow \rho^- + X ) = g_{\rho^+ WZ}^2 \cdot \sigma_{W^- Z}, \\
\sigma (pp \rightarrow \rho^0 + X ) = g_{\rho^0 WW}^2 \cdot \sigma_{W W},
\end{array}
\end{equation}  
with the fundamental cross sections now given by:
\begin{equation}
\begin{array}{ll}
\sigma_{W^+ Z} = \sigma (pp \rightarrow W^+ Z \rightarrow \rho^+ + X ) \left. \right | _ {g_{\rho^+ WZ }=1}, \\
\sigma_{W^- Z} = \sigma (pp \rightarrow W^- Z \rightarrow \rho^- + X ) \left. \right | _ {g_{\rho^+ WZ }=1},\\
\sigma_{W^+ W^-} = \sigma (pp \rightarrow W^+ W^- \rightarrow \rho^0 + X ) \left. \right | _{ g_{\rho^0 WW }=1}.
\end{array}
\end{equation}
Again, once these cross sections are computed numerically at the partonic level, we can get the total production rates by simply rescaling with the couplings of the vectors to gauge bosons which are easily computed in the mass eigenstate basis. Finally, since both the couplings of the resonance to lighter quarks and to gauge bosons depend on $\xi$, $g_\rho$ and $M_\rho$, the production cross section for Drell-Yan and VBF processes is a function of only these three parameters.
\begin{figure}[t!]
\begin{center}
\includegraphics[width=0.49\textwidth]{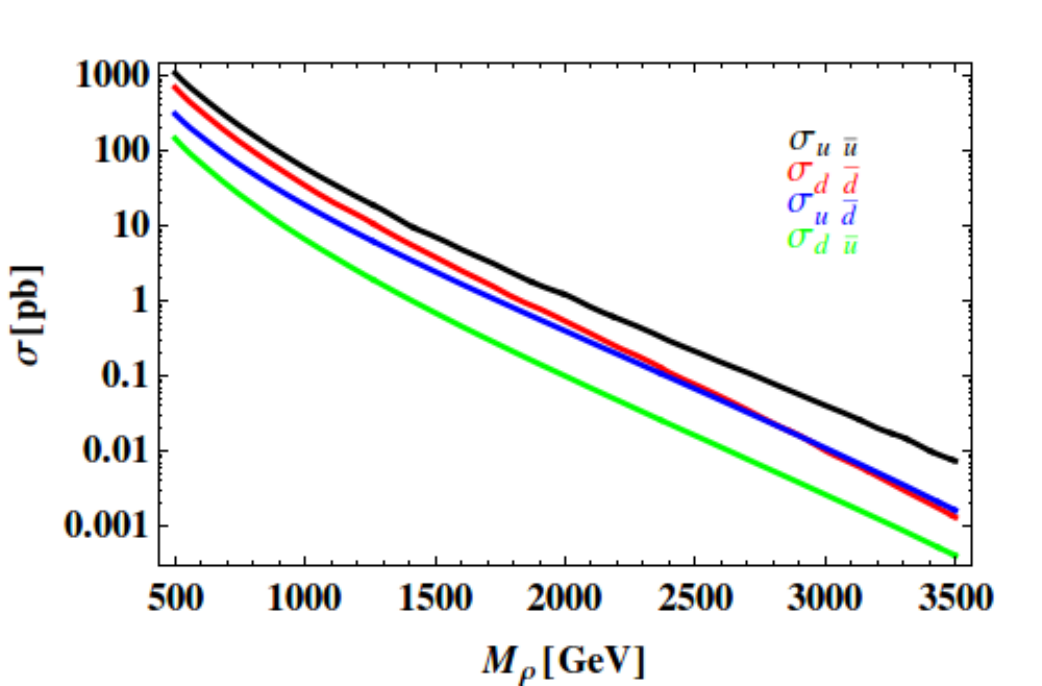}
\includegraphics[width=0.49\textwidth]{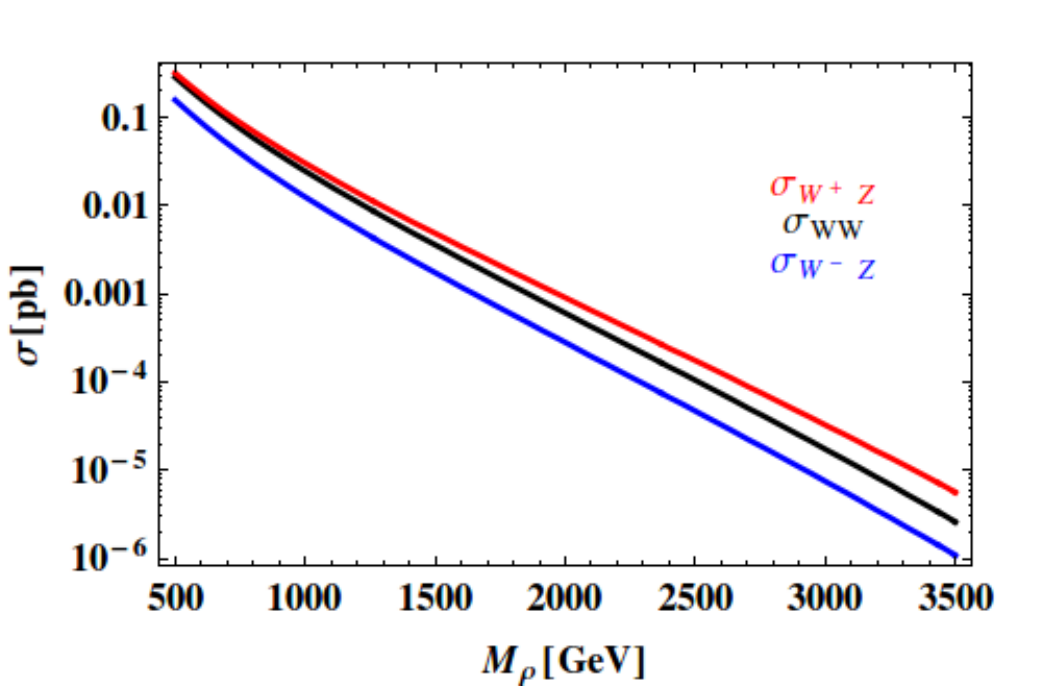}
\end{center}
\caption{\small Fundamental cross sections as functions of the physical mass of the resonance at $\sqrt{s}= 8 \ \text{TeV}$. Left panel: fundamental cross sections for the DY process. Right panel: fundamental cross sections for the VBF process.}
\label{fig:FundCross}
\end{figure}  

\begin{figure}[t!]
\begin{center}
\includegraphics[width=0.49\textwidth]{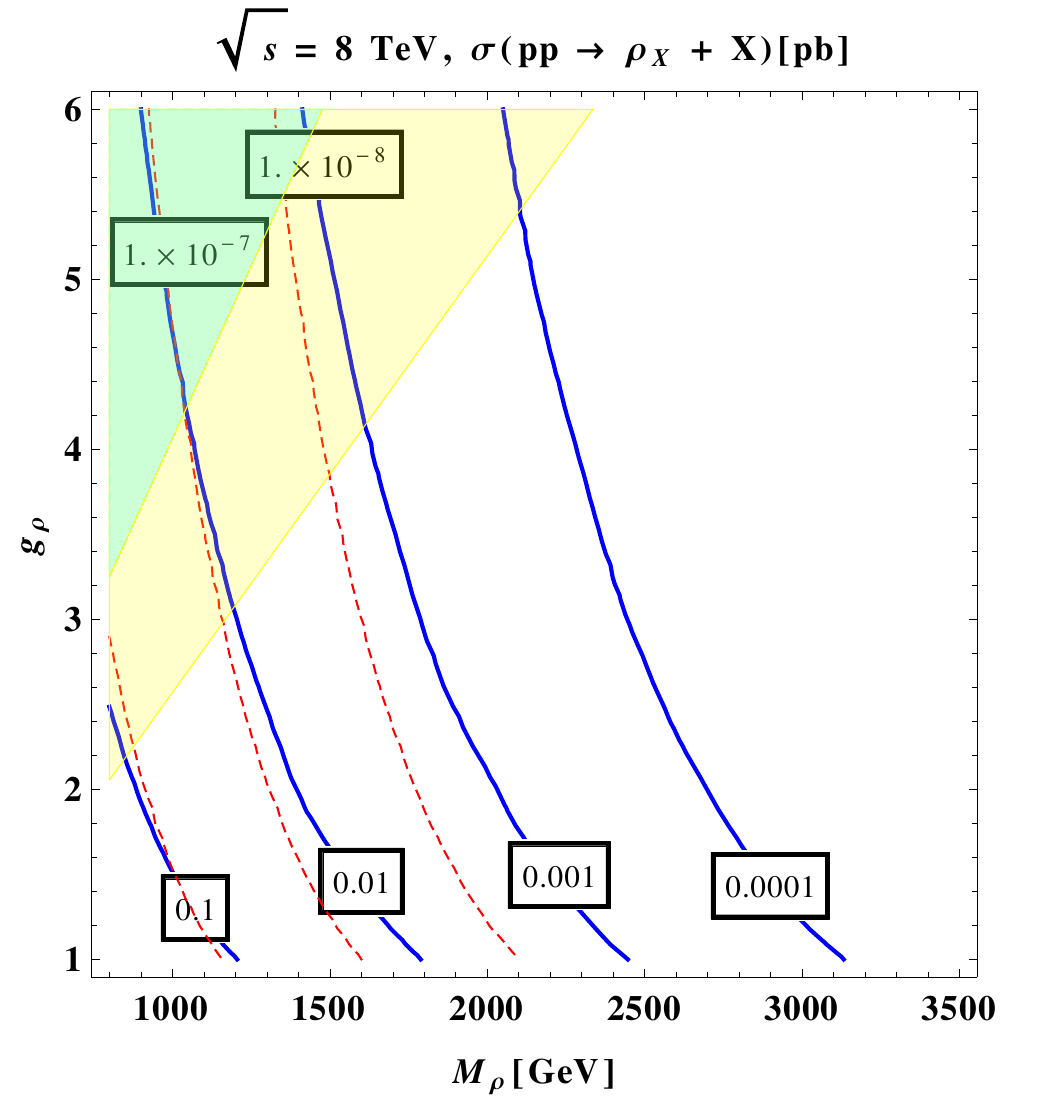}
\end{center}
\caption{\small Contours of constant cross section (blue lines for the DY process, red dashed lines for the VBF process) in the plane ($M_{\rho_X}$, $g_{\rho_X}$) for the production of the vector singlet. The yellow region corresponds to $\xi > 0.4$, the light blue one to $\xi > 1$.}
\label{fig:CrossX}
\end{figure}

We now discuss the relevance of these two production mechanisms for the three vectors in our models. In general, we expect the fundamental cross sections for the VBF process to be much smaller than the corresponding ones for the DY process. In fact, DY is a one-body process and the corresponding cross section goes like $\sim  g^4/g_{\rho}^2$, whereas VBF is a three-body process, so that the cross section is further suppressed by a phase space factor and scales like $\sim g^4/((16 \pi^2)^2 g_{\rho}^2)$. This is confirmed by a quantitative estimation of the two mechanisms, as it can be seen in Fig.~(\ref{fig:FundCross}), where the various fundamental cross sections are plotted as a function of the resonance mass. The relative importance of the two complete production rates depends however on the coupling strengths that rescale the partonic cross sections. Since the couplings of the resonances to elementary fermions decrease with increasing $g_{\rho}$, the Drell-Yan process is smaller for larger values of the strong coupling constant. On the other hand, the couplings to longitudinally polarized gauge bosons increase with $g_\rho$, so that the VBF mechanism can have a chance to compete with the DY one for more strongly coupled scenarios. 
\begin{figure}[t!]
\begin{center}
\includegraphics[width=0.49\textwidth]{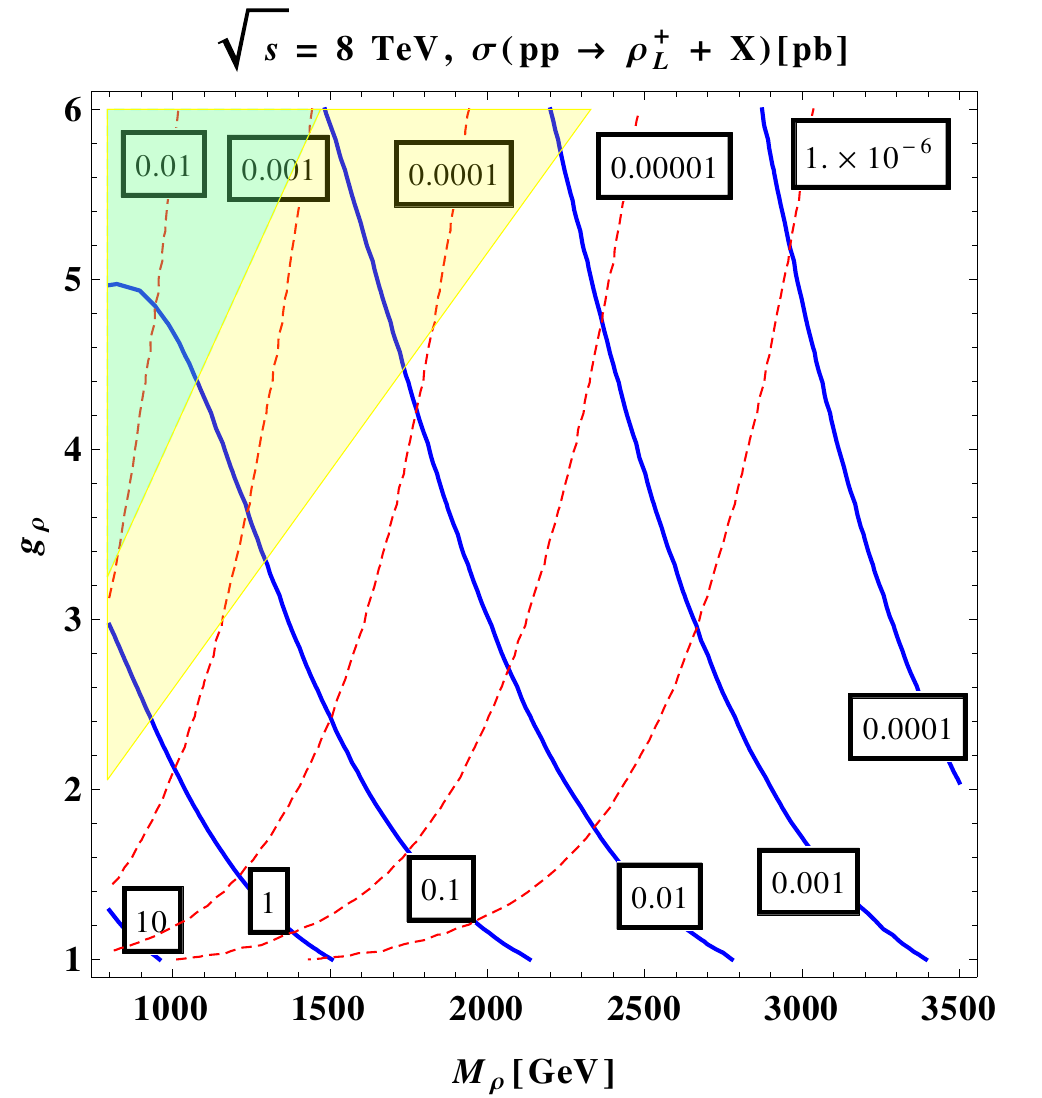}
\includegraphics[width=0.49\textwidth]{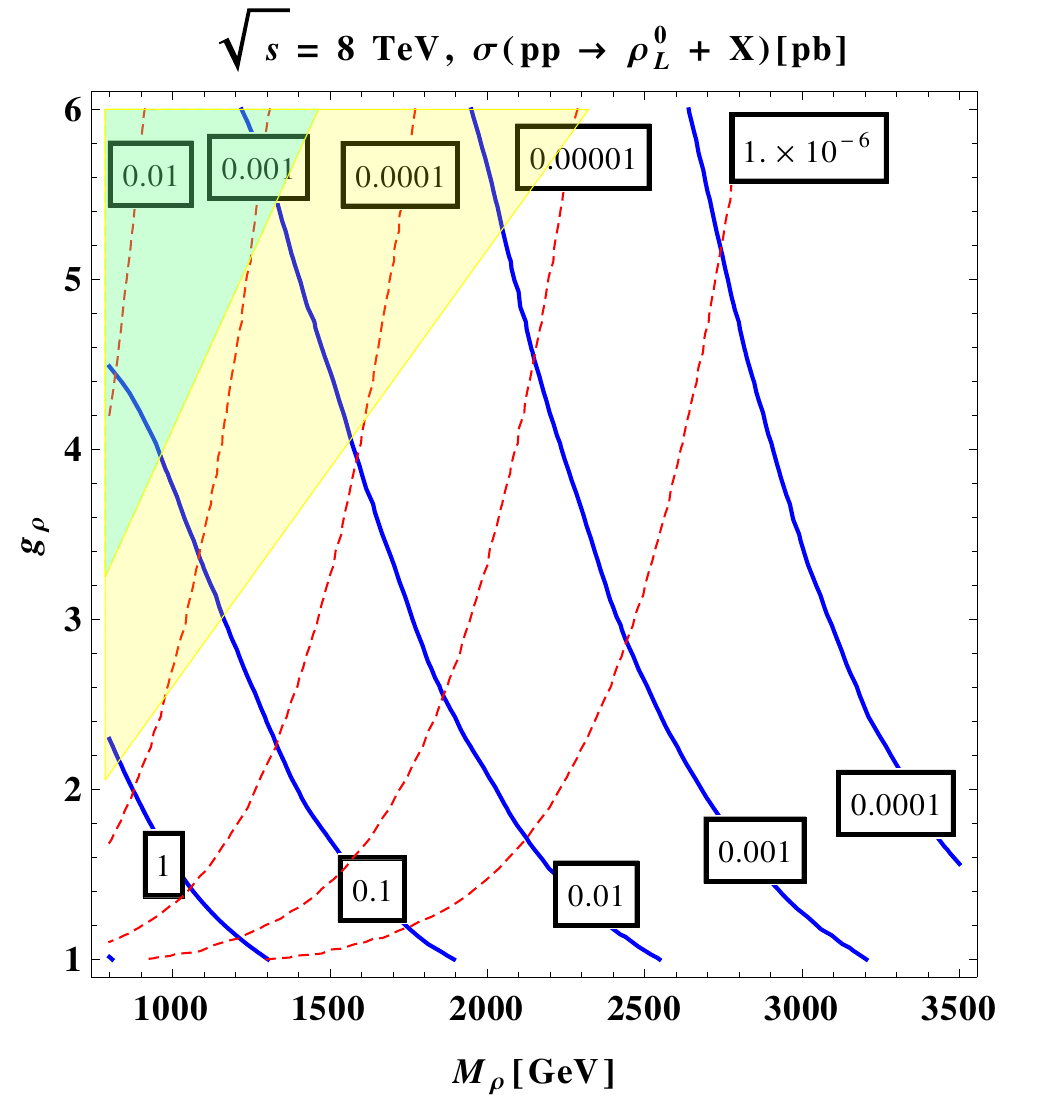}
\end{center}
\begin{center}
\includegraphics[width=0.49\textwidth]{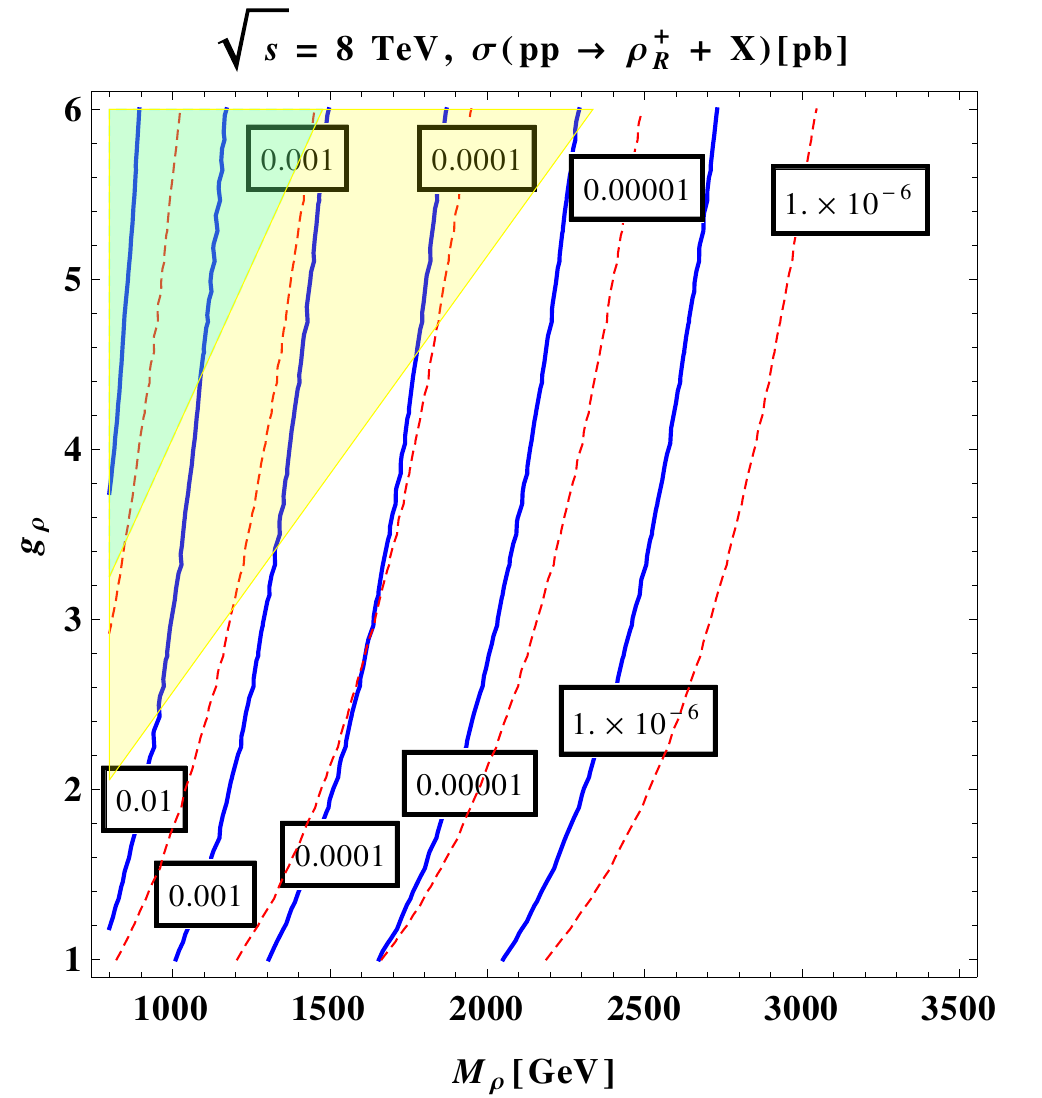}
\includegraphics[width=0.49\textwidth]{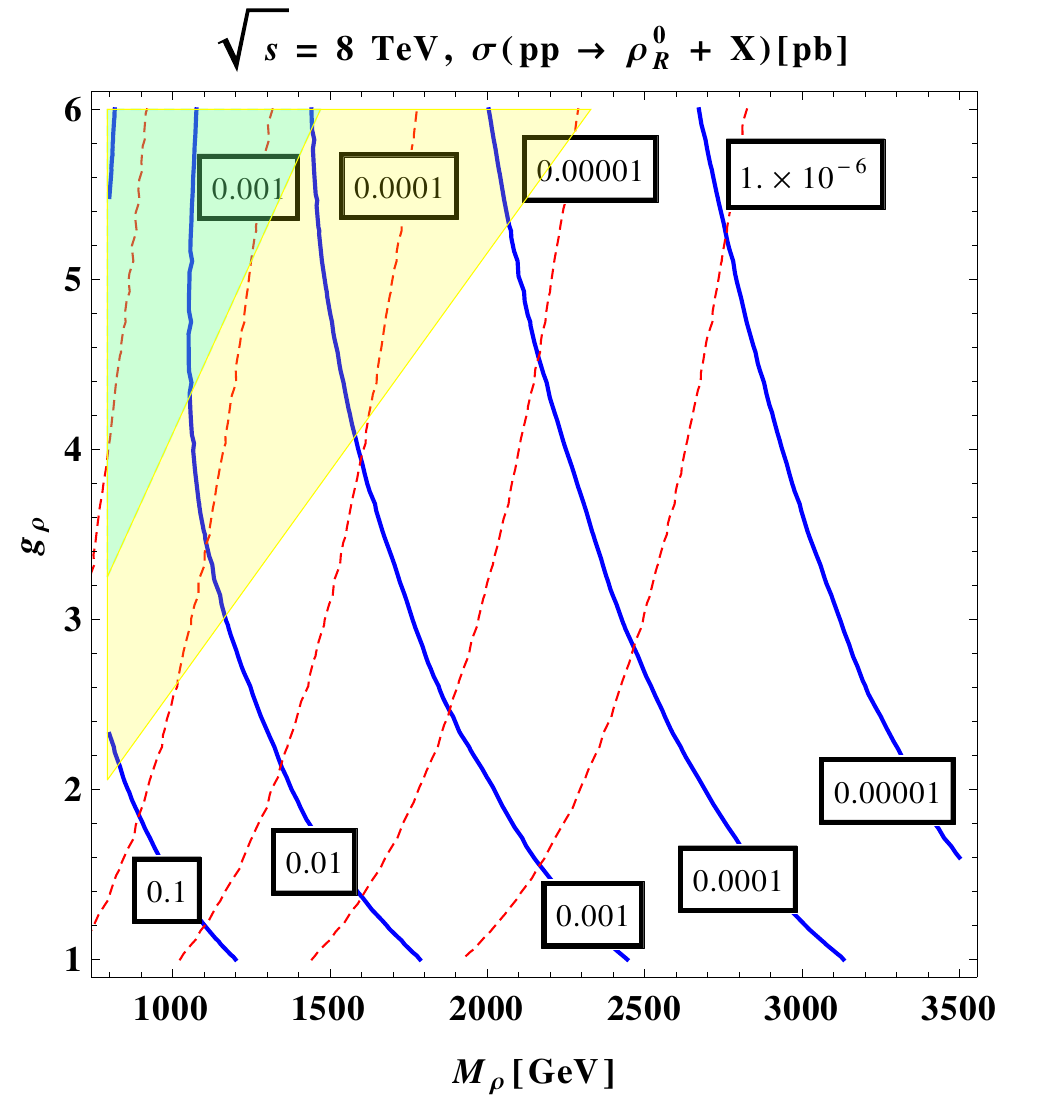}
\end{center}
\caption{\small Contours of constant cross section (blue lines for the DY process, red dashed lines for the VBF process) in the plane ($M_{\rho}$, $g_{\rho}$) for the production of the charged (left panel) and neutral (right panel) left-handed (top) and right-handed (bottom) vector triplets. The yellow region corresponds to $\xi > 0.4$, the light blue one to $\xi > 1$.}
\label{fig:CrossLeft}
\end{figure}
The total production cross sections for the two processes are illustrated in Figs.~(\ref{fig:CrossLeft}) and (\ref{fig:CrossX}), where we plot the contours of constant cross sections, both for DY and VBF processes, for the three heavy vectors in the $(M_\rho, g_\rho)$ plane. In every case, in order to enforce the NDA relation (\ref{VectorNDA}) between the coupling and the mass, we have rescaled $\xi$ as
\begin{equation}\label{XiRescale}
\xi = a_\rho^2 {1 \over \sqrt{2} G_F } \left({g_\rho \over M_\rho} \right) ^2,
\end{equation}   
and we have fixed $a_\rho = 1$, for illustration. We have also indicated the region of the parameter space where the value of $\xi$ exceeds $1$, and is therefore not allowed, and the region where $\xi$ exceeds $0.4$, which corresponds to the experimentally disfavoured limit where our analytical expressions for the couplings at leading order in $\xi$ start losing their validity. From Fig.~(\ref{fig:CrossLeft}), we see that, despite the suppressed couplings of the resonances to elementary fermions, the DY cross section for both the charged and neutral $\rho_\mu^L$ vector dominates over the VBF one even for large $g_{\rho}$ and increases for smaller values of the strong coupling, since in that limit the couplings to SM fermions get larger as a result of the larger elementary-composite mixing. The VBF cross section increases for higher values of $g_\rho$, but remains nevertheless sub-dominant in all regions of the parameter space where $\xi<0.4$. Analogous considerations are valid also for the production cross section of the neutral $\rho_\mu^R$; the shapes of the contours are similar, but the overall size of the cross section is smaller by a factor $\sim (g^\prime / g_\rho )^2$. As regards the charged $\rho^R_\mu$ vector, the couplings to the SM fermions are weaker than the previous cases, since they arise after EWSB; as a result, the two production rates are both very small and comparable, so that in this case the VBF mechanism competes with the DY in every region of the parameter space. Since for both mechanisms the production cross section is extremely small, however, this resonance is produced at low rate at the LHC and is much more difficult to discover. Finally, the vector singlet will be mostly produced by DY process, as shown in Fig.~(\ref{fig:CrossX}), since it does not interact with longitudinally polarized gauge bosons before EWSB and the VBF cross section is therefore further suppressed. These results on the behaviour of the production cross sections for the various kinds of vector resonances are in agreement with those obtained in a similar context in \cite{Prod1, Prod2, Prod3, Prod4, Prod5, Prod6}.

\subsection{Branching ratios}

We now turn to the study of the vector resonances decays. Following our natural assumptions on the dynamics of the strong sector, we consider the top partners to be the lightest heavy states and we fix for illustration $M_\Psi = 800 \ \text{GeV}$. This value for the masses of the $X_{5\over 3}$ and $X_{2\over 3}$ fields is in agreement with the bounds coming from the LHC direct searches of new exotic quarks of charge $5/3$, \cite{ExpTPMass}, and automatically satisfies the bounds from searches of other top-like fermions, which are generally weaker. Under these conditions, we will study the most relevant decay channels of the heavy bosons and how the presence of the lighter top partners affects their branching ratios. All the partial decay widths described in this section can be computed analytically by using the $\texttt{Feynrules}$ package once the couplings in Appendix \ref{AppCoup} are derived at leading order in $\xi$.
\begin{figure}[t!]
\begin{center}
\includegraphics[width=0.49\textwidth]{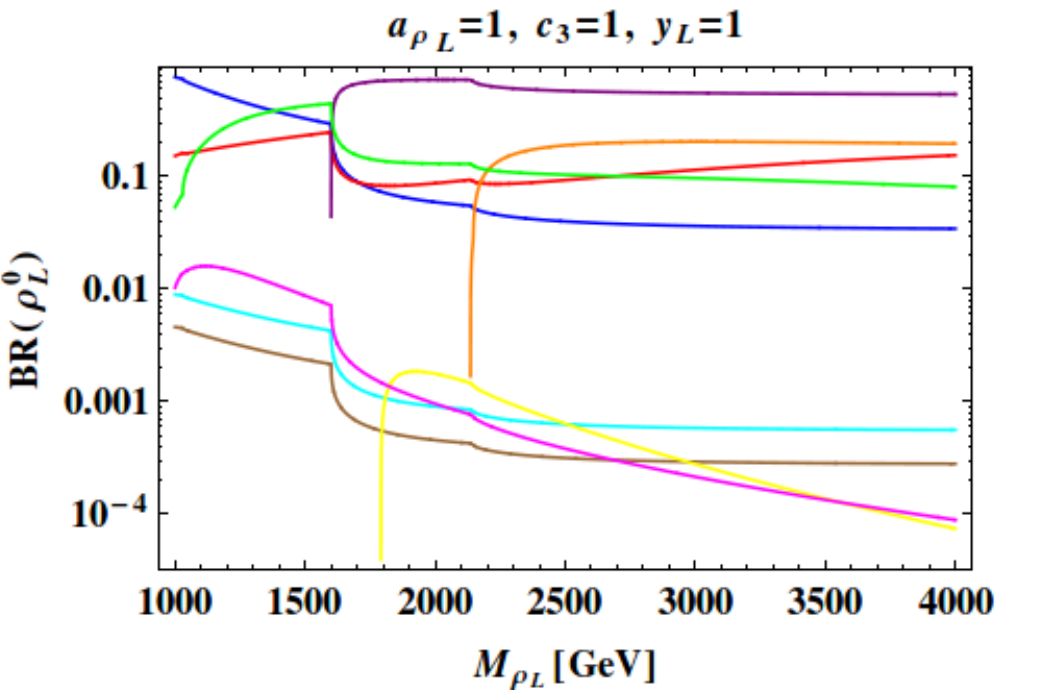}
\includegraphics[width=0.49\textwidth]{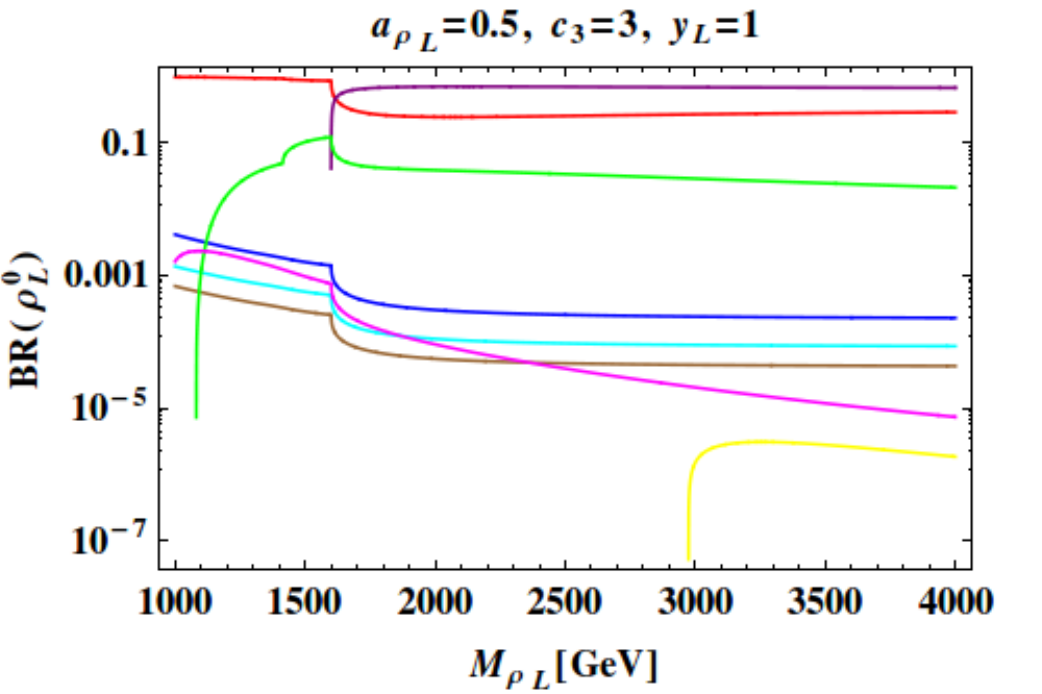}
\end{center}
\caption{\small Decay branching ratios of the neutral left-handed vector as a function of the resonance mass for $g_{\rho_L}=3$, $M_\Psi = 800 \ \text{GeV}$ and two different sets of the free parameters. The various curves correspond to the following decay channels: $WW+Zh$ (blue), $t\bar{t}+b\bar{b}$ (red), $l^+l^-$ (brown), $u\bar{u}+d \bar{d}$ (cyan), $X_{5\over 3}\bar{X}_{5\over 3}+X_{2\over 3}\bar{X}_{2\over 3}$ (purple), $T \bar{T}+B\bar{B}$ (orange), $X_{2\over 3}\bar{T}$ (yellow), $X_{2\over 3}\bar{t}$ (magenta), $T\bar{t}+B\bar{b}$ (green).}
\label{fig:BRLeftNeut}
\end{figure}

We start considering the case of the neutral right-handed and left-handed vector resonances; their decay widths are very similar, since they couple to the same top partners fields before EWSB and their couplings to gauge bosons and SM fermions are comparable. We have therefore shown in Fig.~(\ref{fig:BRLeftNeut}) the different branching ratios as a function of the resonance mass only for $\rho_L^0$, omitting the analogous case of $\rho_R^0$, for the benchmark value of the strong coupling constant $g_{\rho_L} = 3$ and varying $\xi$ as in Eq.~(\ref{XiRescale}). The importance of the different decay channels depends obviously on the  choice of the various free parameters of the theory; in particular, $a_{\rho_L}$, $c_3$ and $y_L$ play a dominant role in setting the strength of the interaction with gauge bosons, third family quarks and heavy fermions, whereas we do not expect $c_1$ to give a relevant contribution to the different decays. We have thus set $c_1=1$ and shown the branching ratios for two different choices of the remaining parameters that change the behaviours of the branching ratios as a function of $M_{\rho_L}$. In the first case, the three relevant parameters are all set to one, according to the most natural expectations dictated by NDA. We see that in the lower mass region, $M_{\rho_L}<2 M_\Psi$, the dominant decays are $WW/Zh$, $t\bar{t}/b\bar{b}$ and $T\bar{t}/B\bar{b}$,\footnote{For the importance of heavy-light decay channels in a similar context, see for example \cite{HeavyLight}.} whereas above threshold, $M_{\rho_L}>2M_\Psi$, the vector resonance will mainly decay to pairs of heavy fermions, in particular $X_{2\over 3}$ and $X_{5\over3}$. The relevance of the light decay channels below threshold, when the free parameters are chosen so as to perfectly match their NDA estimate, has also been pointed out in \cite{WarpedComposite}. The situation can be considerably changed with a slight violation of NDA, as shown for the second choice of free parameters, $a_\rho=0.5$ and $c_3=3$. In this case, the decay width to gauge bosons and Higgs is extremely reduced in the lower mass region, since their couplings now get smaller, and the heavy vector mainly decays to two tops or two bottoms, whereas above threshold the decays to two 5/3 charged exotic states and to two top-like $X_{2\over 3}$ particles remain still the dominant ones. We notice that for this particular choice of parameters the fermionic elementary-composite mixing is stronger, so that the couplings of the vector resonance to a heavy fermion and a third family quark are weaker than the corresponding couplings to two tops or bottoms. The branching ratio for the heavy-light decay channels is therefore reduced, whereas the $t\bar{t}$ and $b\bar{b}$ decays are considerably enhanced. In both cases, the branching ratios for decays to leptons and first two quark families are instead strongly suppressed, as expected, as well as the decays to the top partners whose couplings to the heavy vectors are not allowed by isospin conservation before EWSB. We note finally that the branching fractions to $WW$ and $Zh$ are equal to a very good approximation, as implied by the Equivalence Theorem, which works well since $M_{\rho_L} \gg m_{W/Z}$ for the chosen values of parameters. The approximate custodial symmetry also implies that $BR(t\bar{t}) \sim BR(b \bar{b})$ and $BR(u \bar{u}) \sim BR(d \bar{d}) \sim 3 BR(l^+ l^-)$.  
\begin{figure}[t!]
\begin{center}
\includegraphics[width=0.49\textwidth]{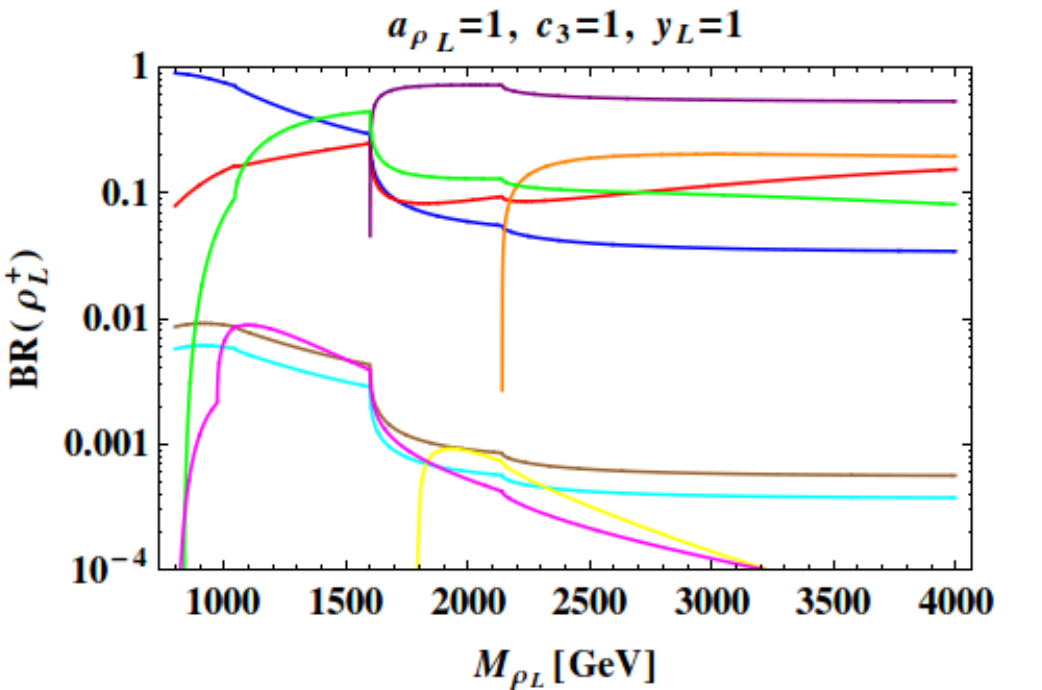}
\includegraphics[width=0.49\textwidth]{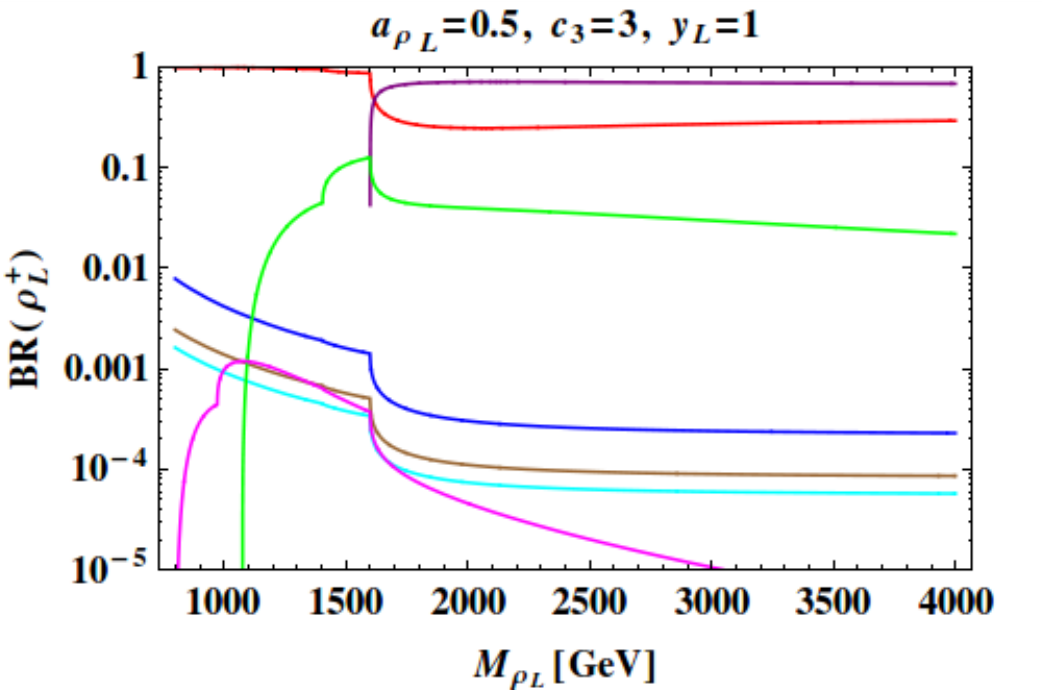}
\end{center}
\begin{center}
\includegraphics[width=0.49\textwidth]{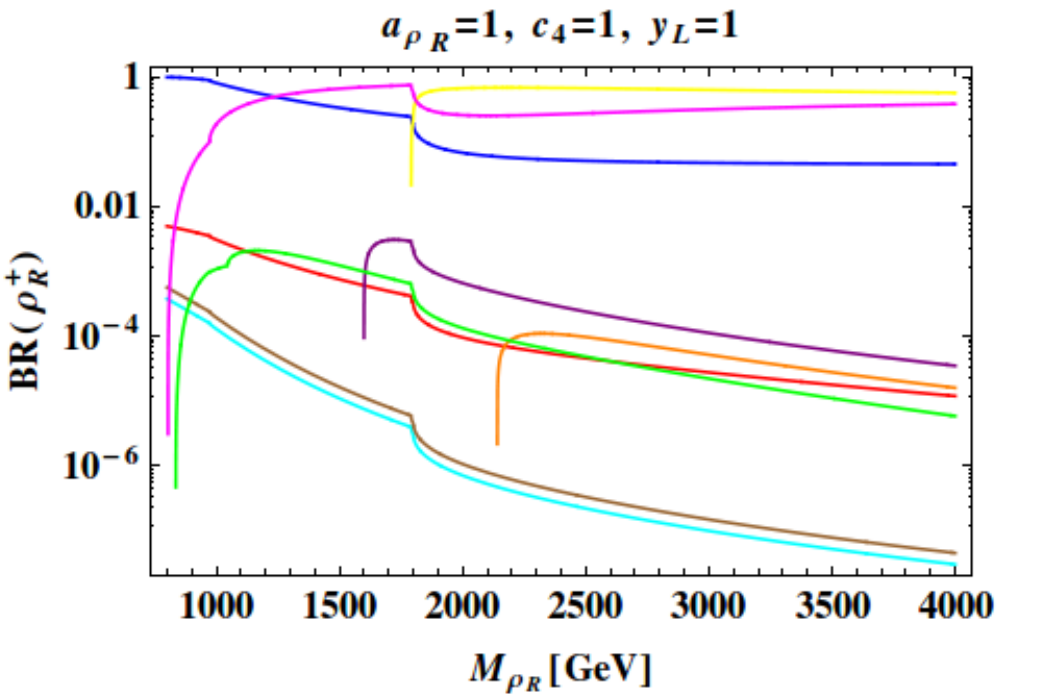}
\includegraphics[width=0.49\textwidth]{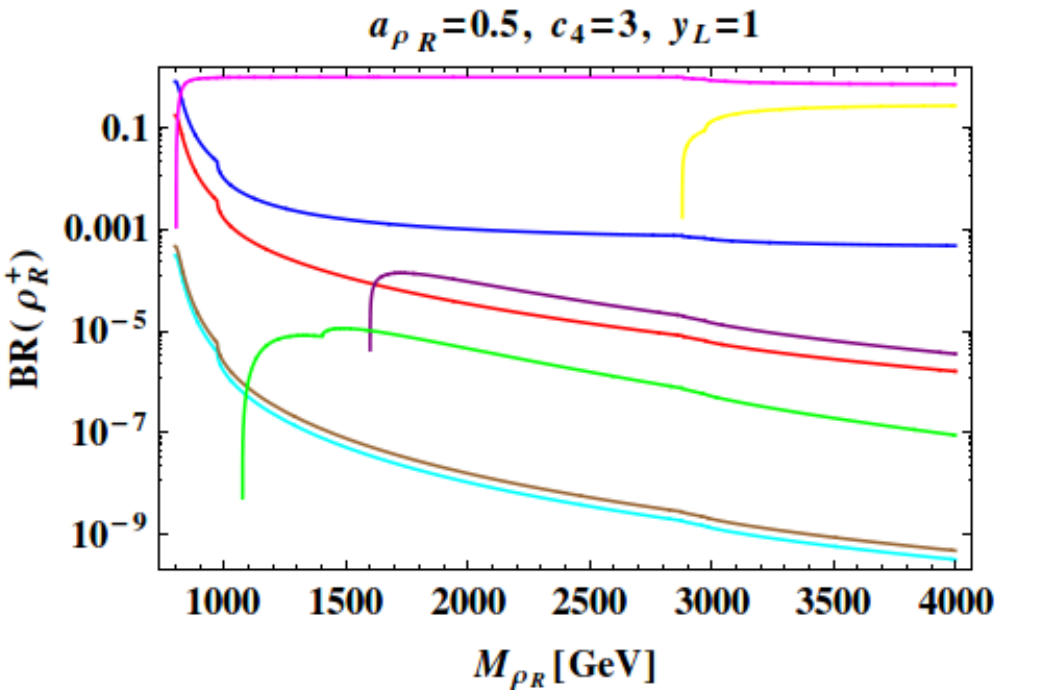}
\end{center}
\caption{\small Decay branching ratios of the charged left-handed (top) and right-handed (bottom) vectors as a function of the resonance mass for $g_{\rho_{L/R}}=3$, $M_\Psi = 800 \ \text{GeV}$ and two different sets of the free parameters. The various curves correspond to the following decay channels: $WZ+Wh$ (blue), $t\bar{b}$ (red), $l \nu$ (cyan), $u\bar{d}$ (brown), $X_{5\over 3}\bar{X}_{2\over 3}$ (purple), $T \bar{B}$ (orange), $X_{5\over 3}\bar{T}+X_{2\over 3}\bar{B}$ (yellow), $X_{5\over 3}\bar{t}+X_{2\over 3}\bar{b}$ (magenta), $T\bar{b}+B\bar{t}$ (green).}
\label{fig:BRLeftRightCharged}
\end{figure}

As concerns the decay channels of the charged left-handed and right-handed vector resonances, their behaviour is now completely different, as implied by their different quantum numbers. The branching ratios for both cases are shown in Fig.~(\ref{fig:BRLeftRightCharged}), for the same value of the strong coupling as before and the same two sets of free parameters, the first one fully matching the NDA estimate, the second one slightly departing from the natural expectations. The decay to two gauge bosons, $WZ$, and to $Wh$ is dominant in the low mass region for both resonances when $a_\rho=1$, but a soon as $a_\rho$ gets smaller and $c_{3/4}$ is increased this channel is strongly suppressed. The $t \bar{b}$ decay becomes the most important one in the low mass region when $a_\rho = 0.5$ and $c_3=3$, for the $\rho_L^+$ particle, as implied by partial compositeness, whereas it is always sub-dominant for the $\rho_R^+$ case, because of its suppressed couplings to third family quarks. The heavy-light decay channel for the charged left-handed vector is again reduced for the second choice of parameters because, analogously to its neutral counterpart, for smaller values of $a_{\rho_L}$ the couplings to one heavy fermion and a third family quark are weaker. Above threshold, the most relevant decay channel of the left-handed vector is that involving two top partners, for every choice of the free parameters. This latter charged vector will in fact mainly decay to $X_{5\over 3}\bar{X}_{2\over 3}$, with almost unit branching ratio. Among the $\rho_R^+$ decays involving top partners, on the other hand, the dominant ones are the channels $X_{5\over 3}\bar{t}/X_{2\over 3}\bar{b}$, which is kinematically favoured since it opens up as soon as $M_{\rho_R} > M_\Psi$, and $X_{5\over 3}\bar{T}/X_{2\over 3}\bar{B}$. They are both dominant above the threshold for the first choice of parameters, whereas in the second case the decay to $X_{5\over 3}\bar{t}/X_{2\over 3}\bar{b}$ is the most relevant one among all the others for every value of the resonance mass. Finally, the decay to leptons and first two quark families are again suppressed, but the branching ratios for the $\rho_R^+$ are much smaller, since its couplings to fully elementary fermions are further suppressed by a factor of $\xi$.   
\begin{figure}[t!]
\begin{center}
\includegraphics[width=0.49\textwidth]{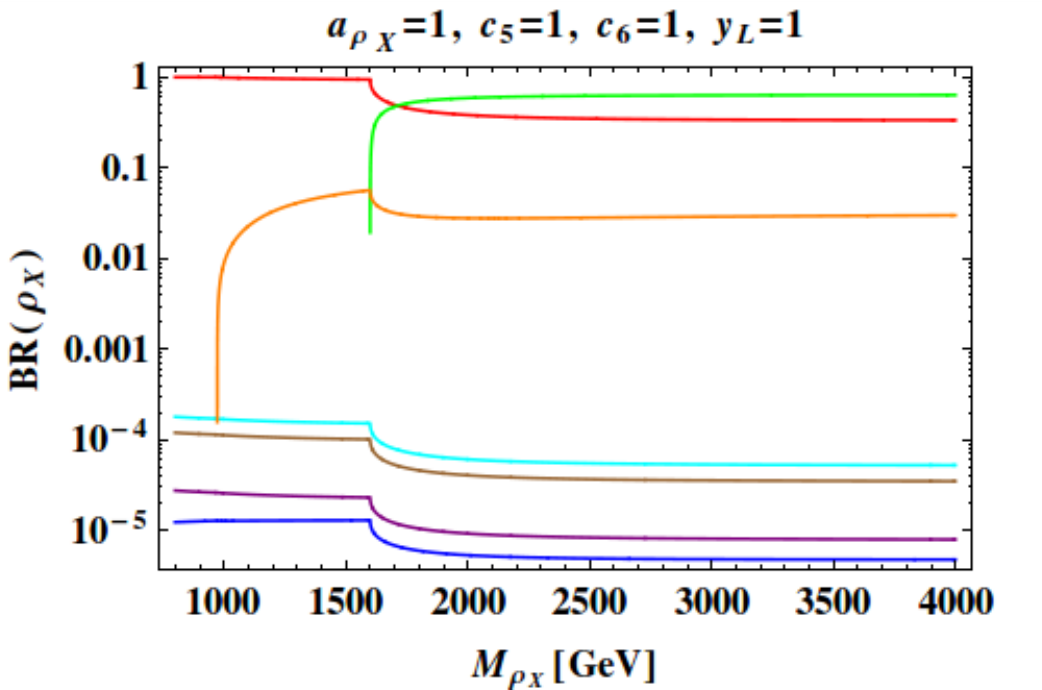}
\includegraphics[width=0.49\textwidth]{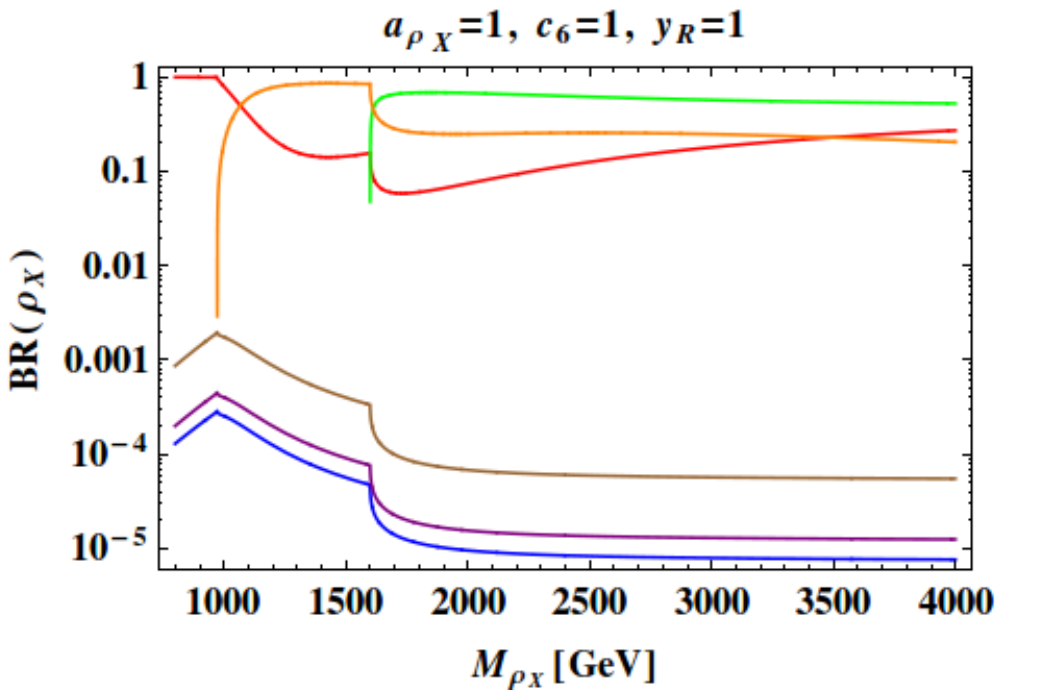}
\end{center}
\caption{\small Decay branching ratios of the vector singlet as a function of the resonance mass for $g_{\rho_X}=3$ and $M_\Psi = 800 \ \text{GeV}$ in models $\textbf{M}_{\textbf{X}}^{\textbf{1}}$ (left panel) and $\textbf{M}_{\textbf{X}}^{\textbf{2}}$ (right panel). The various curves correspond to the following decay channels: $WW+Zh$ (blue), $t\bar{t}$ (red), $l^+l^-$ (cyan), $u\bar{u}+d \bar{d}$ (brown), $b\bar{b}$ (purple), $\widetilde{T} \bar{t}$ (orange), $\widetilde{T}\bar{\widetilde{T}}$ (green).}
\label{fig:BRX}
\end{figure}

We finally discuss the most important decay channels of the singlet in the two models $\textbf{M}_{\textbf{X}}^{\textbf{1}}$ and $\textbf{M}_{\textbf{X}}^{\textbf{2}}$; the branching ratios are shown in Fig.~(\ref{fig:BRX}), for $g_{\rho_X}=3$. In both models, the decays to lighter SM fermions, gauge bosons and Higgs are always suppressed, due to their extremely weak couplings to the vector resonance; the parameter $a_{\rho_X}$ therefore does not play any major role in improving the relevance of the $WW$ and $Zh$ channels. The most important decays are thus $t\bar{t}$, $\widetilde{T}\bar{t}$ and $\widetilde{T}\bar{\widetilde{T}}$, as expected. In the $\textbf{M}_{\textbf{X}}^{\textbf{1}}$ case, the two important parameters are $c_5$ and $c_6$; setting them to one, as illustration, shows that, below the threshold for the production of two heavy fermions, the singlet mainly decays to two tops, whereas above the threshold the channel to two top partners becomes the dominant one. The decay width to one top partner and the top quark, on the other hand, is smaller since it is generated only after EWSB. The situation is different in model $\textbf{M}_{\textbf{X}}^{\textbf{2}}$; after setting the relevant parameter $c_6$ to one, we see that the channel $\widetilde{T}\bar{t}$ is the most important one below the threshold, because it now arises before EWSB. When $M_{\rho_X}>2 M_\Psi$, on the other hand, the decay to two top partners is still the most relevant, even if now the channel involving the top and $\widetilde{T}$ is stronger than in the previous model.

\section{Bounds from LHC direct searches} 
\label{sec:LHCBounds}

Many searches of spin-1 resonances have been performed by the ATLAS and CMS collaborations, with the data collected at the 8 TeV LHC, both for neutral and charged heavy vector particles. The main decay channels that have been considered for the charged resonance can be summarized as follows:
\begin{itemize}
\item the decay to third family quarks, $\rho^+ \rightarrow t \bar{b}$, both by ATLAS in \cite{ATLASChargedTBChannel} and CMS in \cite{CMSChargedTBChannel},
\item the leptonic decay $\rho^+ \rightarrow l \bar{\nu}$, by ATLAS in \cite{ATLASChargedLeptonNeutrinoChannel} and by CMS in \cite{CMSChargedLeptonNeutrinoChannel},
\item the fully hadronic decay to gauge bosons, $\rho^+ \rightarrow WZ \rightarrow jj $, by CMS in \cite{CMSCharged/NeutralWZChannels} and in \cite{CMSCharged/NeutralWZJet}, 
\item the fully leptonic decay to gauge bosons, $\rho^+ \rightarrow WZ \rightarrow 3l\nu$, by ATLAS in \cite{ATLASWZLeptonChargedChannel} and by CMS in \cite{CMSChargedWZLeptChannel}.
\end{itemize} 
As regards the searches of new neutral resonant states, the decay channels which have been extensively analysed by the two experiments are:
\begin{itemize}
\item the leptonic decay, $\rho^0 \rightarrow l^+\bar{l}^-$, by ATLAS in \cite{ATLASNeutrallLLChannel} and by CMS in \cite{CMSDileptonChannelNeutral},
\item the decay to two tops, $\rho^0 \rightarrow t \bar{t}$, by ATLAS in \cite{ATLASNeutralttChannel} and by CMS in \cite{CMSNeutralttChannel},
\item the decay channels to two $\tau$ leptons, $\rho^0 \rightarrow \tau \bar{\tau}$, bt ATLAS in \cite{ATLASTauNeutralChannel},
\item the semi-leptonic decay to two gauge bosons, $\rho^0 \rightarrow WW \rightarrow l \bar{\nu} jj $, by CMS in \cite{CMSNeutralWWChannel},
\item the fully hadronic decay to two gauge bosons, $\rho^0 \rightarrow WW \rightarrow jj$, by CMS in \cite{CMSCharged/NeutralWZChannels}.
\end{itemize} 
The results of these searches are all presented as limits on the production cross section times branching ratio, $\sigma \times BR$, as a function of the resonant mass. This allows us to recast very easily these analyses as exclusion regions in the parameter space of our models: once the cross section is computed semi-analytically with the method described in the previous section and the branching ratios are derived as a function of the couplings, we can immediately compare the theoretical predictions with the experimental data. Similar exclusion contours on the parameters of a vector resonance, charged under $SU(2)_L$, have already been presented in \cite{Bridge}, without considering the effects of partial compositeness or lighter heavy fermions. We will show how these bounds are altered by the stronger coupling of third family quarks to the resonance and by the presence of lighter top partners, for which we will conveniently choose again $M_\Psi = 800 \ \text{GeV}$, and compare them with the indirect information coming from the resonances contribution to Electroweak Precision Observables, derived in Appendix \ref{AppSTU}. In deriving the exclusion bounds on the parameters of our models, we will finally take into account only the DY production mechanism and compute the total production cross section without considering the contribution of the VBF process, this latter being much smaller than the DY one.

We finally stress that the results presented in this section are based on the validity of the Narrow Width Approximation. This latter assumes that the production rate can be factorized into an on-shell cross section times a decay branching ratio and neglects the interference with the SM background. Experimental analyses performed by following this approach must be carried out consistently with its underlying assumptions, namely that the limits on the production rate of the new particles should be set by focussing on the on-shell signal region; for a detailed discussion of these aspects see Ref.~\cite{Bridge}. We will take into account the limitations of the NWA approach by showing in the exclusion plots the contours of constant $\Gamma / M_\rho$ in the parameter space of our models. In the region where this ratio is less the $10\%$, the resonance is narrow enough for the Narrow Width Approximation to be a reliable estimate of the production rate, otherwise a more refined description must be considered in order to analyse the results of the experimental searches.

\subsection{Bounds on $\rho_\mu^L$}

\begin{figure}[ht!]
\begin{center}
\includegraphics[width=0.49\textwidth]{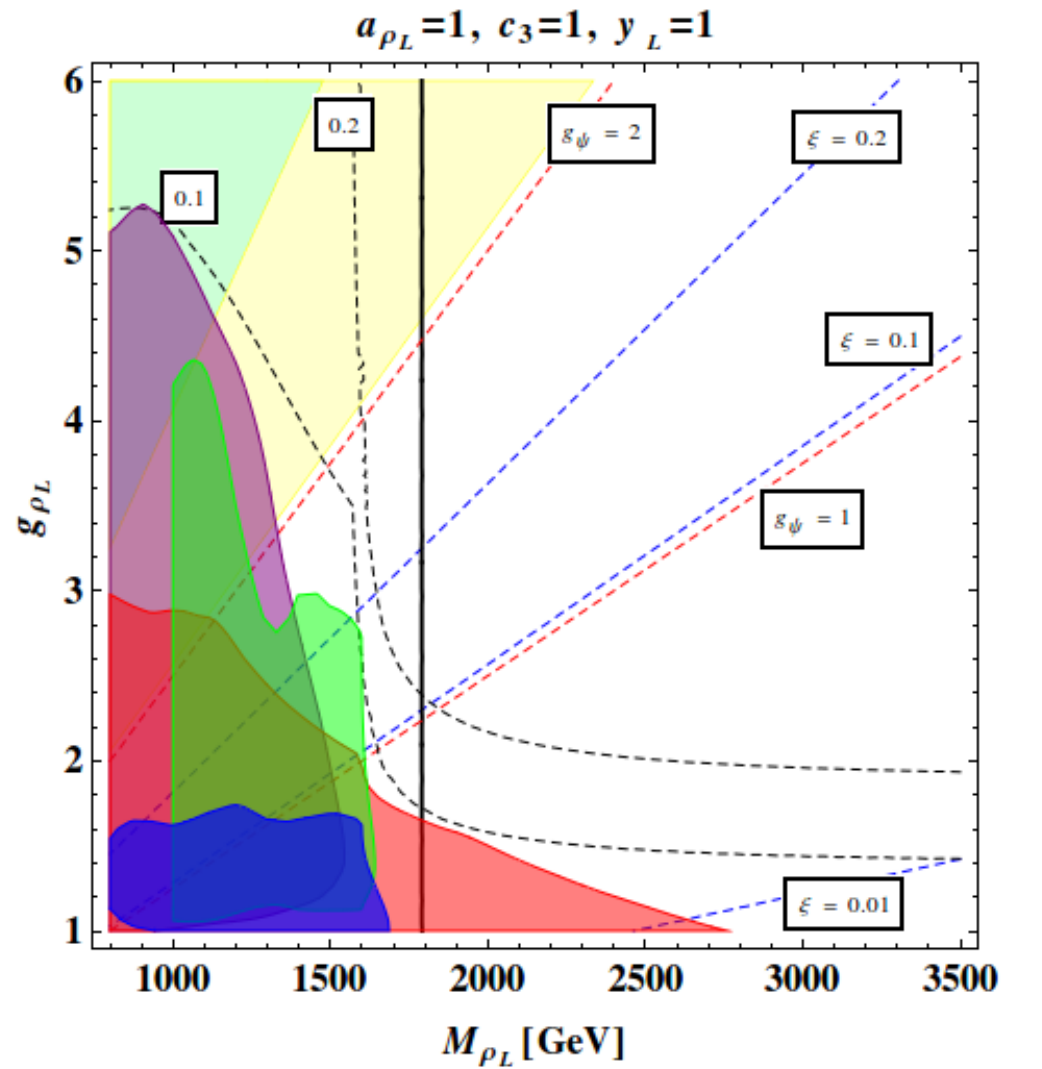}
\includegraphics[width=0.49\textwidth]{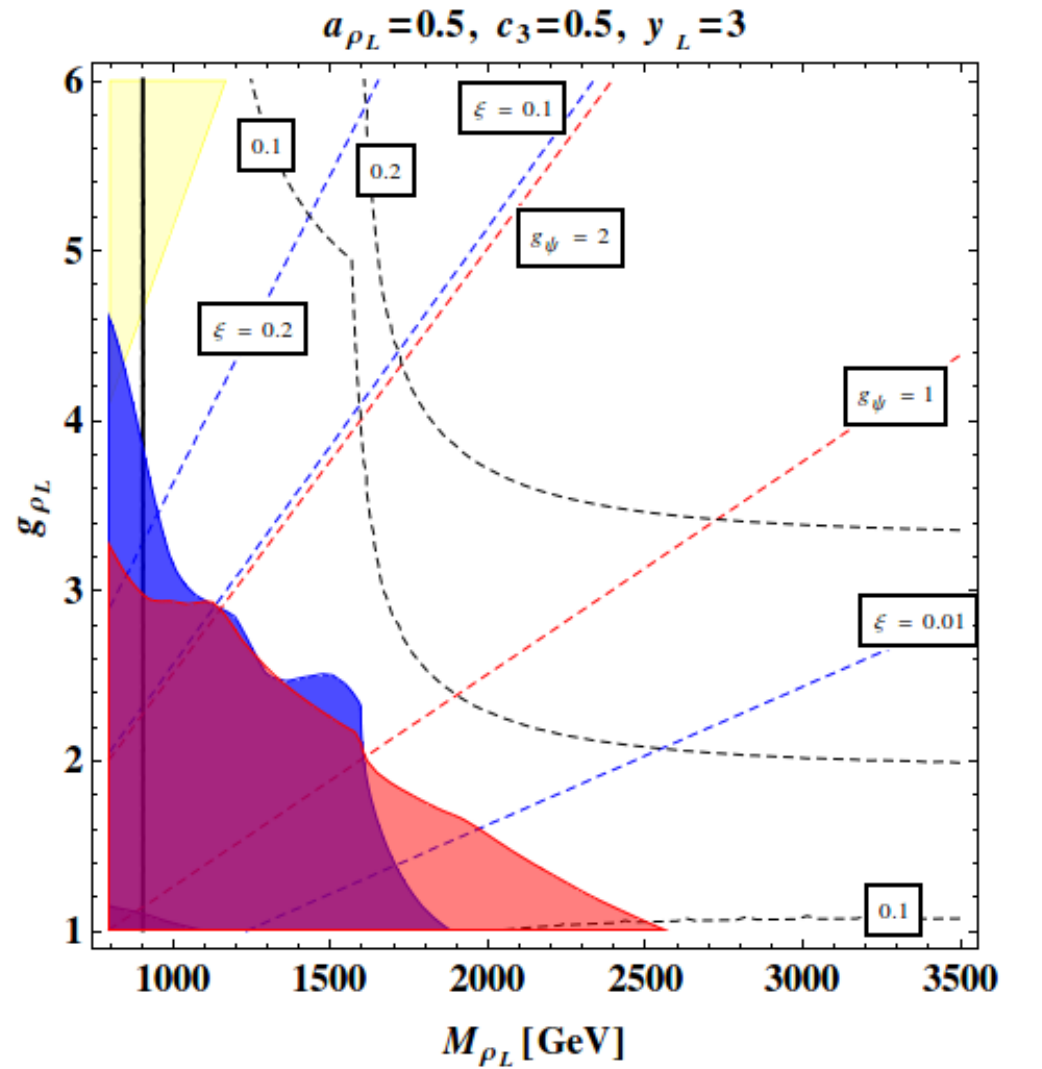}
\end{center}
\caption{\small Excluded regions in the $(M_{\rho_L}, g_{\rho_L})$ plane for the charged left-handed vector resonance for two different sets of the free parameters and for $M_\Psi=800 \ \text{GeV}$. The exclusions are derived from the $\rho^+ \rightarrow t\bar{b}$ searches in \cite{CMSChargedTBChannel} (blue), the $\rho^+ \rightarrow l \bar{\nu}$ searches in \cite{CMSChargedLeptonNeutrinoChannel} (red), the $\rho^+ \rightarrow WZ \rightarrow jj $ searches in \cite{CMSCharged/NeutralWZChannels} (purple) and the $\rho^+ \rightarrow WZ \rightarrow 3l\nu $ searches in \cite{ATLASWZLeptonChargedChannel} (green). The plot also shows the contours of constant $\Gamma/M_{\rho_L}$ (dashed black lines), of constant $\xi$ (dashed blue lines) and of constant $g_\Psi$ (dashed red lines). The region on the left of the thick black line is excluded by experimental constraints on the $\hat{S}$ parameter. The yellow region corresponds to $\xi > 0.4$, the light blue one to $\xi > 1$.}
\label{fig:RegionPlotLeftMg}
\end{figure}

We start the study of the experimental constraints on the parameters of our models by considering the case of the left-handed heavy vector. The tree-level exchange of this particle contributes to the $\hat{S}$ and $W$ parameters \cite{STU, EWFit1, EWFit2}, among which the most stringent bounds come from the first one, since $W$ is smaller by a factor of $g^2/g_{\rho_L}^2$. In Fig.~(\ref{fig:RegionPlotLeftMg}) we show the excluded regions in the $(M_{\rho_L}, g_{\rho_L})$ plane from four different direct searches, one for each of the main decay channels considered by the experimental groups, and we compare them with the limits coming from the $\hat{S}$ variable. We also show how the bounds change for two different choices of the free parameters: in one case, we fix $a_{\rho_L} = c_3 = y_L =1$; in the second case we have analysed the set $a_{\rho_L} = c_3 = 0.5$, $y_L = 3$. The variable $\xi$ always scales as in Eq. (\ref{XiRescale}). Only the bounds for the charged heavy vector case are presented, for illustration; the exclusion limits for the neutral resonance are similar and are not reported here.   

Let us discuss the results for the first choice of parameters. The searches of a heavy vector decaying to gauge bosons, which subsequently decay fully leptonically or fully hadronically, give the most important constraints in the low mass region, $M_{\rho_L} < 2 M_{\Psi}$, since for the chosen value of $a_{\rho_L}$ the branching ratio of the $WZ$ channel is still dominant below the threshold. These searches do not give any information in the high mass region, $m_{\rho_L} > 2 M_{\Psi}$, however, due to the opening of the $X_{5\over 3}\bar{X}_{2\over 3}$ channel, which significantly reduce the branching ratio to gauge bosons. On the other hand, despite the suppressed couplings to the vector resonance of SM leptons, the searches in the $l\bar{\nu}$ channel are competitive with the previous ones and can also provide exclusion limits above the threshold for small values of the strong coupling constant. From Fig.~(\ref{fig:RegionPlotLeftMg}), we also see how the direct results compete with the indirect bounds from the $\hat{S}$ parameter; this latter excludes the mass of the heavy resonance up to $\sim 1.8 \ \text{TeV}$ and still gives the most powerful information on the parameter space of the model.

These bounds derived for the charged left-handed heavy vector, for $a_{\rho_L} = 1$, agree with the results obtained in analogous contexts; the relevance of the experimental searches in the gauge bosons and leptonic channels was for instance already discussed in \cite{Bridge}. However, taking into account the enhanced coupling of third family quarks to the resonance, we see that exclusion limits can be obtained below threshold and for small values of $g_{\rho_L}$ also from the $t \bar{b}$ search, which does not give any constraint when treating the top-bottom doublet as fully elementary.   

Fig~(\ref{fig:RegionPlotLeftMg}) also shows different contours in the plane $(M_{\rho_L}, g_{\rho_L})$ which provides information on the validity of the NWA approach and of our theoretical assumptions based on naturalness requirements. The curves corresponding to the contours of constant $\Gamma/M_{\rho_L}$ show that the experimental constraints are always confined in the region when this ratio is smaller than $10\%$, so that the NWA works well for all the four main searches. The dashed blue lines, on the other hand, correspond to contours of constant $\xi$ and give thus information on the amount of tuning required for different combination of the mass and coupling of the heavy resonance. The most natural region compatible with the experimental constraints on $\xi$ is the window between $\xi \sim 0.1$ and $\xi \sim 0.2$, a portion of which is already excluded by the direct searches below the threshold; below the $\xi \sim 0.1$ line, more tuning is required to accommodate a reasonably light Higgs in the spectrum, so that these regions correspond to the more unnatural ones where our hypothesis of lighter top partners is no longer justified. Contours of constant $g_{\Psi}$ are also shown; the fermionic coupling constant can be in fact derived, using both Eq.~(\ref{FermionNDA}) and Eq.~(\ref{VectorNDA}), as
\begin{equation}
g_\Psi = {a_{\rho_L} \over a_\Psi} {M_\Psi \over M_{\rho_L}} g_{\rho_L};
\end{equation}     
we have shown the lines corresponding to the naturally favoured values $g_\Psi = 1$ and $g_\Psi = 2$ fixing $a_\Psi = a_{\rho_L}$ for illustration. We see that the preferred natural window corresponds also to the portion of parameter space where the fermionic coupling is in its theoretically expected range; the region where $g_\Psi \lesssim 1$, on the other hand, coincides with the unnatural one, where $\xi$ assumes very small values and the lightness of top partner can no longer be justified by naturalness arguments.

We focus now on the exclusion limits for the second set of parameters. In this case, the values of $a_{\rho_L}$ and $c_3$ are reduced and $y_L$ is instead incremented in order to show the effects on the bounds of the reduced interaction strength between gauge bosons and heavy vectors, on one side, and of a higher top quark degree of compositeness, on the other side. Since now the branching ratio to gauge bosons is suppressed even in the low mass region, no excluded region can be extracted from any of the searches involving the $WZ$ decay channel. On the other hand, the experimental analyses in $t\bar{b}$ channel provide a bigger exclusion limit with respect to the previous case, due to the bigger value of $y_L$ which now increases the strength of the interaction between the charged resonance and the $q_L$ doublet despite the reduced value of $c_3$. The constraints coming from the $l\bar{\nu}$ searches are still competitive and important above the threshold, so that this decay channel is extremely powerful in providing information on the physics of new heavy states or for a potential discovery. Another main difference with respect to the previous study is that, choosing $a_{\rho_L} = 0.5$, the limit coming from the $\hat{S}$ parameter is reduced by a factor of two, excluding the mass of the heavy vector up to $\sim 1 \ \text{TeV}$. When the $a_{\rho_L}$ parameter is lower than one, we therefore find that the direct searches are much more competitive and can exclude portions of the parameter space beyond the reach of indirect information.

As regards the NWA approach, also in this case the bounds are well constrained in the region where this approximation is reliable and valid. The natural window $0.1 \lesssim \xi \lesssim 0.2$ is now achieved in more strongly coupled scenarios, due to the reduced value of $a_{\rho_L}$, and still part of it is excluded by the two shown searches. The contours of constant $g_\Psi$ are derived again for $a_\Psi = a_{\rho_L}$ and, as before, the less fine-tuned region coincides with higher values of the fermionic coupling.

\subsection{Bounds on $\rho_\mu^R$}

\begin{figure}[ht!]
\begin{center}
\includegraphics[width=0.49\textwidth]{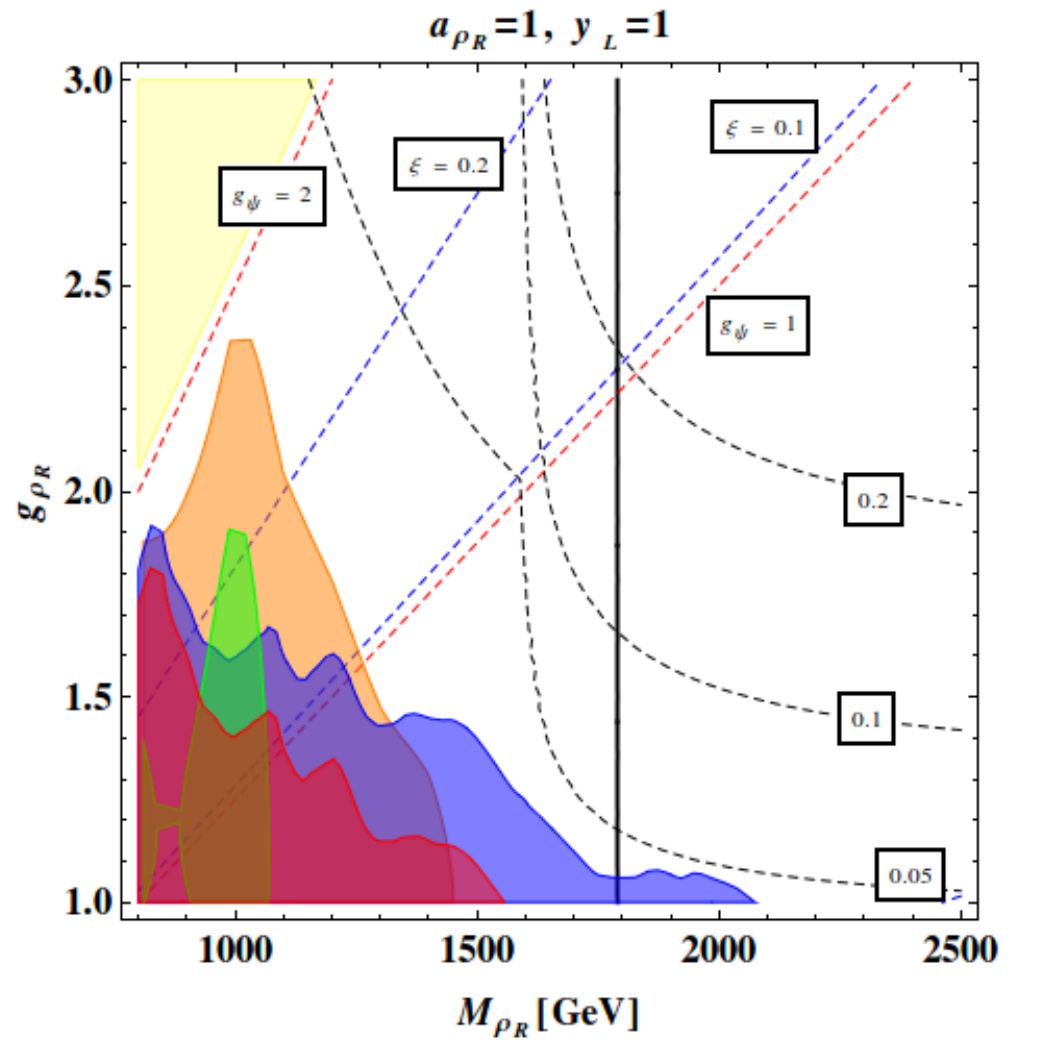}
\includegraphics[width=0.49\textwidth]{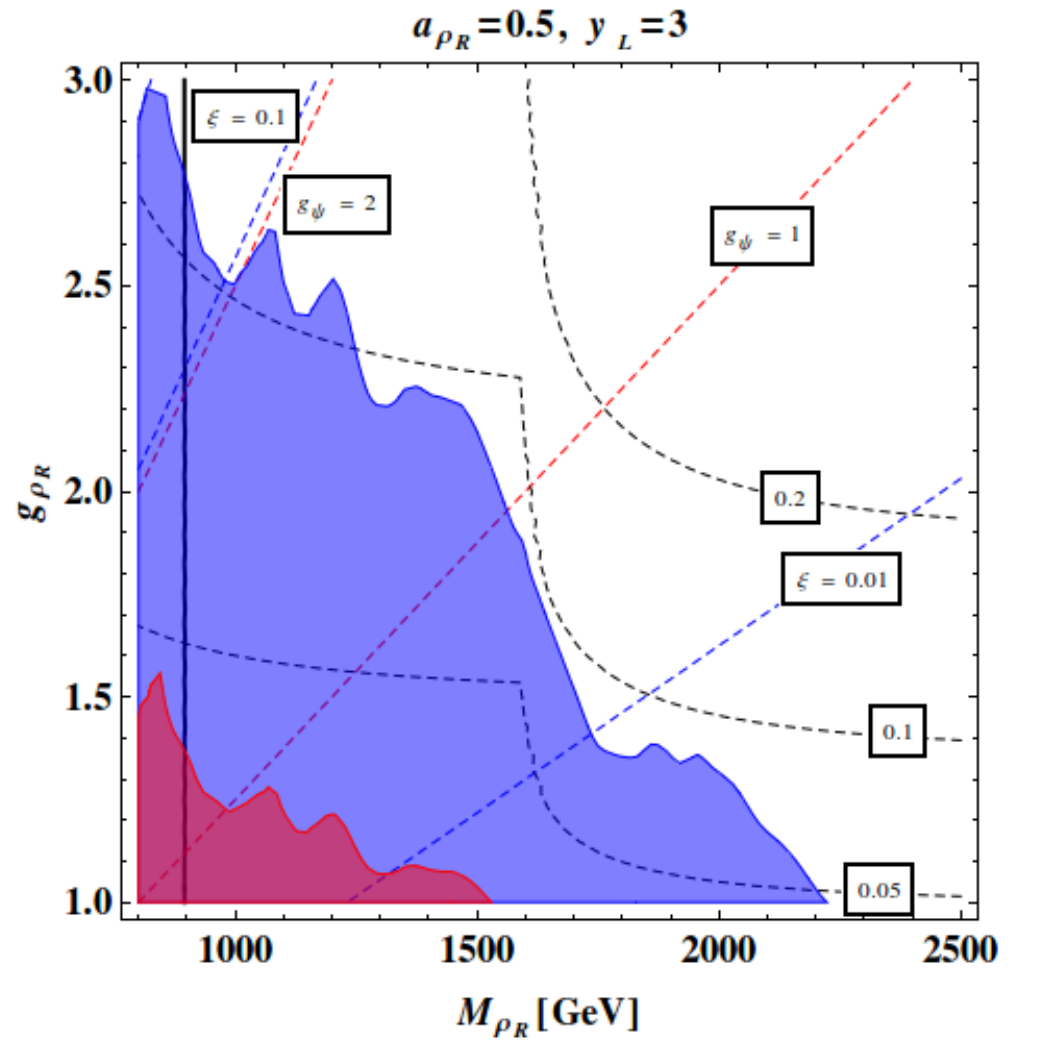}
\end{center}
\caption{\small Excluded regions in the $(M_{\rho_R}, g_{\rho_R})$ plane for the neutral right-handed vector resonance for two different sets of the free parameters and for $M_\Psi=800 \ \text{GeV}$. The exclusions are derived from the $\rho^0 \rightarrow l\bar{l} $ searches in \cite{CMSDileptonChannelNeutral} (in red for $c_4 = 1$, in blue for $c_4 = 0$) and the $\rho^0 \rightarrow WW \rightarrow l\nu jj $ searches in \cite{CMSNeutralWWChannel} (in green for $c_4 = 1$, in orange for $c_4 = 0$). The plot also shows the contours of constant $\Gamma/M_{\rho_R}$ (dashed black lines), of constant $\xi$ (dashed blue lines) and of constant $g_\Psi$ (dashed red lines). The region on the left of the thick black line is excluded by experimental constraints on the $\hat{S}$ parameter. The yellow region corresponds to $\xi > 0.4$.}
\label{fig:RegionPlotRightMg}
\end{figure}

We consider now the bounds on the parameter space of the right-handed resonance. This heavy particle contributes at tree level to the $\hat{S}$ and $Y$ parameters; this latter being suppressed by a factor of $g^{\prime 2}/g_{\rho_R}^2$, we again expect the most stringent limit on the mass of the new state to come from the $\hat{S}$ variable. Since the total production cross section of the charged right-handed vector is very small, for both VBF and DY mechanisms at the LHC, we can only extract bounds on the model parameters for the neutral $\rho_R^0$; these are shown in Fig. (\ref{fig:RegionPlotRightMg}), as excluded regions in the $(M_{\rho_R}, g_{\rho_R})$ plane for two different sets of the free parameters and recasting the results of the searches in the lepton channel and in the semi-leptonic $WW$ channel. We have presented the different exclusion contours for two values of $c_4$, when it is vanishing and when it is 1, in order to clearly analyse the effects of the lighter top partners on the bounds from direct searches. 

Let us start briefly considering the case in which $a_{\rho_R} = 1$ and $y_L = 1$. For these values of the free parameters, the $WW$ channel provides constraints in the low mass region, analogously to the left-handed resonance, and it is not sensitive to the portion of parameter space above the threshold $2 M_\Psi$. In the extreme situation where $c_4 = 0$ and the direct coupling to top partners is completely eliminated, the constraints are obviously much stronger and they gradually reduce as $c_4$ is increased and the branching ratios for the top partners channels become important. As regards the experimental search in the leptonic channel, the bounds can give exclusions above the threshold and again they are stronger for small $c_4$, as expected. We note also the main difference between the right-handed and the left-handed case: the production cross section for the $\rho_R$ resonance being smaller by a factor $(g^{\prime}/g_{\rho})^2$, the bounds in the parameter space of the right-handed vector are in general much weaker than those of the left-handed counterpart. Finally, the NWA approach works well also in this situation, the excluded regions being confined in the portion of the $(M_{\rho_R}, g_{\rho_R})$ plane where $\Gamma/M_{\rho_R} < 0.1$. The discussion on the natural window and the comparison with the limits from the $\hat{S}$ variable are similar to the $\rho_L$ case.

We discuss now how the bounds change for $a_{\rho_R} = 0.5$ and $y_L = 3$. As expected, no exclusion contours can be derived from the $WW$ search channel, since the branching ratios to gauge bosons are now suppressed. The only bounds come from the analysis performed with the $ll$ decay channel; for $c_4 = 0$, they are much stronger, whereas, when the decay to top partners and third family quarks are enhanced with $c_4 = 1$, a very tiny region of parameter space is excluded. This is again due to the smaller production cross section that makes this resonance in general much harder to constrain and to discover with respect to the previous one. The NWA is again well satisfied and the region where our natural assumptions are well justified has the same behaviour as the analogous left-handed case.

We finally notice that no exclusion regions can be derived from the experimental search of neutral resonances in the $t\bar{t}$ channel. The experiments performed using this particular decay are indeed much less sensitive than the others, so that, despite the enhanced coupling strength of the top quark to the neutral vector, we find no bounds even for high degrees of top compositeness and for larger values of $c_4$. For this reasons, we do not expect this final state to be enough powerful for the discovery of a neutral spin-1 particle.    

\subsection{Bounds on $\rho_\mu^X$}

\begin{figure}[t!]
\begin{center}
\includegraphics[width=0.49\textwidth]{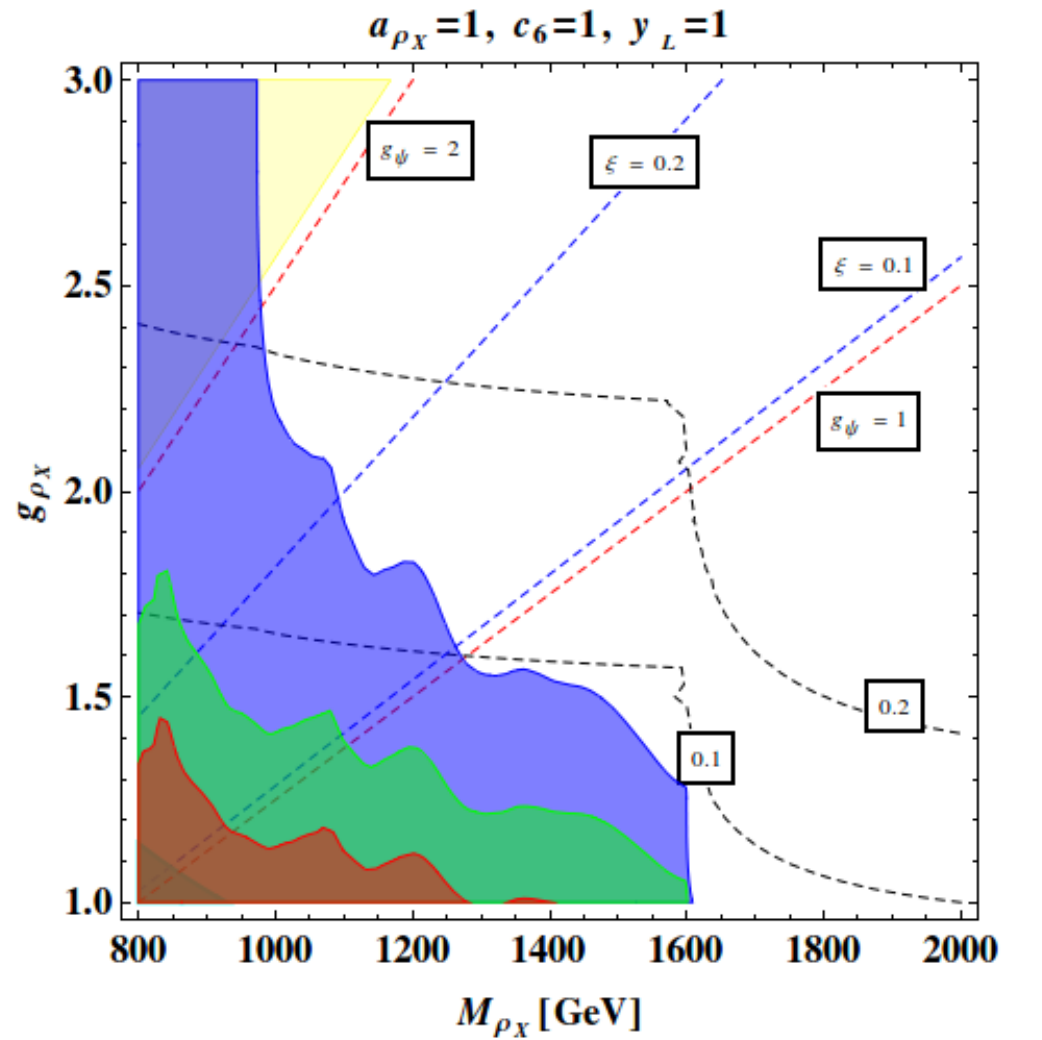}
\includegraphics[width=0.49\textwidth]{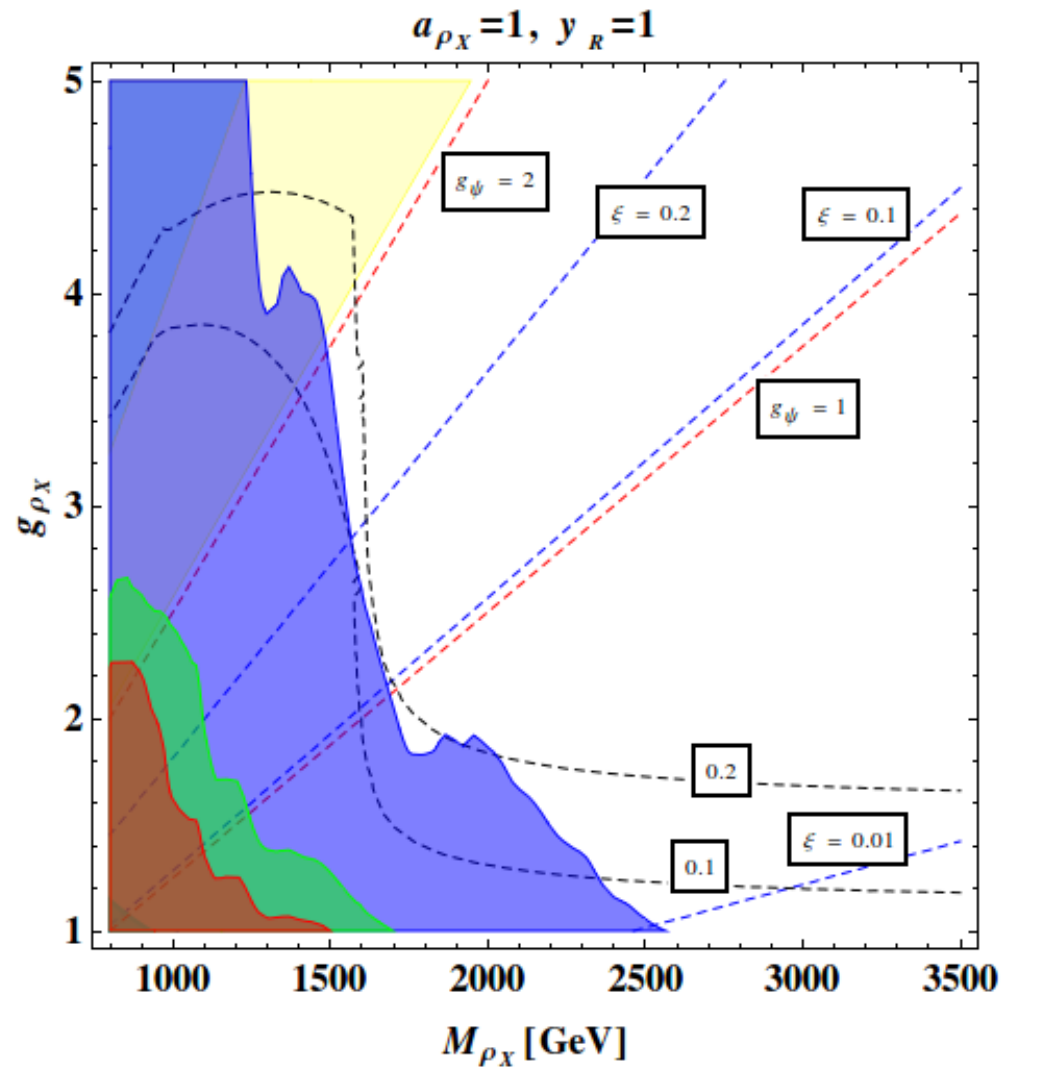}
\end{center}
\caption{\small Excluded regions in the $(M_{\rho_X}, g_{\rho_X})$ plane for the vector singlet in models $\textbf{M}_{\textbf{X}}^{\textbf{1}}$ (left) and $\textbf{M}_{\textbf{X}}^{\textbf{2}}$ (right), fixing $M_\Psi = 800 \ \text{GeV}$. The exclusions are derived from the $\rho^0 \rightarrow l\bar{l} $ searches in \cite{CMSDileptonChannelNeutral}. Left panel: in red the excluded region for $c_5 = 1$, in green for $c_5 = 0.5$, in blue for $c_5 = 0$. Right panel: in red the excluded region for $c_6 = 1$, in green for $c_6 = 0.5$, in blue for $c_6 = 0$.  The plot also shows the contours of constant $\Gamma/M_{\rho_X}$ (dashed black lines), of constant $\xi$ (dashed blue lines) and of constant $g_\Psi$ (dashed red lines). The yellow region corresponds to $\xi > 0.4$.}
\label{fig:RegionPlotXMg}
\end{figure}

The experimental searches for a neutral heavy resonance can also be recast as a bound on the parameter space of the vector singlet. This heavy particle contributes only to the $Y$ parameter, which however always gives very weak constraints; in this case, the exclusion limits from direct searches are therefore the most relevant ones and electroweak precision measurements have very little exclusion power.\footnote{Since the vector singlet does not contribute to the $\hat{S}$ parameter, our theoretical picture of heavier spin-1 resonances and lighter top partners could be not so well justified for this particle, allowing the possible existence of a vector which is as light as or lighter than the spin-1/2 resonances. Consistency with the idea that the new strong sector should be characterised by only two mass scales and that all spin-1 heavy states should behave similarly, however, leads us to consider also the singlet to belong to the tower of heavier resonances at the $m_\rho$ scale.} The excluded regions in the $(M_{\rho_X}, g_{\rho_X})$ plane are presented in Fig.~(\ref{fig:RegionPlotXMg}), both for model $\textbf{M}_{\textbf{X}}^{\textbf{1}}$ and $\textbf{M}_{\textbf{X}}^{\textbf{2}}$ and for different values of the free parameters. In both cases, the most relevant experimental search is always the decay channel to the $ll$ final state, since the searches involving the decay to $WW$ do not obviously give any constraint, due to the extremely weak coupling strength of the singlet to the $W$ boson. We will therefore fix $a_{\rho_X}=1$ in all the cases considered, since different values of this parameter will only alter the shape of the contours of constant $\xi$ and $g_\Psi$, but will not significantly change the exclusion contours. Despite the enhanced coupling strength to top quarks, finally, the searches with the $t\bar{t}$ final state produce no limits on the parameter space of the two models, similarly to the right-handed neutral resonance.

Considering now the specific results for model $\textbf{M}_{\textbf{X}}^{\textbf{1}}$, we have fixed $y_L = 1$ and shown the bounds for three different values of $c_5$. The most stringent constraints on the parameter space of the singlet are obviously obtained when $c_5 = 0$; in this extreme case, the direct coupling to the $t_R$ quark is suppressed and the branching ratio to leptons increases, so that the experimental search under consideration gives stronger bounds. Increasing $c_5$, on the other hand, makes the bounds much weaker and for $c_5=1$ only a very tiny portion of parameter space is excluded. This is due again to the $g^\prime$ suppression in the coupling of the vector singlet to lighter quarks, which makes the total production cross section smaller than the left-handed case. All the exclusion regions are concentrated in the low mass region, $M_{\rho_X} < M_{\Psi}$, and abruptly end when $M_{\rho_X} =2 M_{\Psi}$, due to the opening of the decay channel to two top partners. 

The situation is similar for model $\textbf{M}_{\textbf{X}}^{\textbf{2}}$; we have shown the exclusion regions for $a_{\rho_X} = y_R = 1$ and for three values of the free parameter $c_6$, ranging from 0 to 1. When $c_6$ is vanishing, the bounds are much stronger and they can extend above the threshold due to the absence of a direct interaction with the $\widetilde{T}$ heavy fermion. Increasing $c_6$ makes the exclusion limits weaker; the bounds are now confined in the low mass region and are less stringent than the neutral left-handed case due to the hypercharge suppression.

Finally, the NWA approach is reliable for both models. In Fig.~(\ref{fig:RegionPlotXMg}), we have in fact shown the contours of constant $\Gamma/M_{\rho_X}$ only for $c_5=1$ and $c_6=1$, corresponding to the excluded region in red. The contours for the other two smaller values of these parameters, corresponding to the excluded regions in blue and green, lie outside the portion of the $(M_{\rho_X}, g_{\rho_X})$ plane which is presented. Therefore, the bounds corresponding to $c_5= 0 , 0.5$ and to $c_6 = 0 , 0.5$ automatically satisfy the requirements of a narrow resonance, whereas the bound for $c_5 = 1$ and $c_6 = 1$ lie completely in the portion of parameter space where the total decay width in units of $M_{\rho_X}$ is less than $10\%$. Also in this final case the NWA is therefore a valid prescription for analysing the experimental results. For both models, the natural window where our theoretical assumptions are well justified is excluded in the low mass region, but still allowed for larger values of the resonant mass and for more strongly coupled scenarios.

\section{Conclusions} 
\label{sec:Concl}

In this paper we have introduced a simplified description based on an effective low-energy Lagrangian of the phenomenology of heavy vector resonances in the minimal composite Higgs model, studying their interaction with lighter top partners. Our approach is based on two classes of assumptions, one regarding the symmetry structure of the theory and one regarding its dynamical features. As concerns the symmetries, we considered the minimal case of a new confining dynamics with an approximate global $G = SO(5)\times U(1)_X$ symmetry spontaneously broken to $H = SO(4)\times U(1)_X$. The Higgs boson emerges as pNGB and the electroweak scale is dynamically generated via loop effects. In this framework, we focussed on heavy vector triplets, transforming as a $(\textbf{3}, \textbf{1})$ and $(\textbf{1}, \textbf{3})$, and on heavy vector singlets, transforming as a $(\textbf{1}, \textbf{1})$ of $SO(4)$. Following the paradigm of partial compositeness, we introduced a linear coupling to the strong sector for the top-bottom doublet and we considered the $t_R$ to be a bound state of the strong dynamics, except in one case in which we studied the implications of a partially composite $t_R$ quark. In this scenario, we characterised the couplings of heavy vectors to top partners in the singlet and in the fourplet of $SO(4)$. In the most natural realizations of the composite Higgs idea these are indeed the lightest fermionic resonances that must be present in the spectrum. We constructed four simplified models which are suitable for studying the phenomenology of heavy vectors, capturing the most important features of the underlying symmetry structure.

As concerns the dynamics, we parametrised the new strong sector with two mass scales, a heavier one for vector resonances, $m_\rho$, and a lighter one for fermionic resonances, $m_\psi$. We have clarified under which conditions our effective Lagrangian description is a good approximation of the full underlying dynamics and what its regime of validity is. Our simplified approach is in fact reliable whenever the mass of the heavy vector satisfies the relation $m_\psi < M_\rho \ll m_\rho$, in which case, using the criterion of partial UV completion \cite{EffectRes}, the tower of the remaining and unknown resonances can be integrated out of the spectrum. Our approximate models provide therefore a systematic simplified description of the phenomenology of spin-1 heavy states in an expansion of $M_\rho / m_\rho$. These constructions loose their validity as soon as $M_\rho \sim m_\rho$, in which case using an effective Lagrangian is formally inappropriate. However, we expect our approach to provide a good interpretation of the experimental results, at least qualitatively, also in this second case. We have assessed this issue considering the particular situation in which two vector resonances of the composite tower are present in the spectrum. We show in Appendix \ref{AppIntEff} that neglecting the spectrum degeneracy is a reliable approximation for a basic quantitative description of their phenomenology.

One of the most important simplification of our procedure is to describe the phenomenology of heavy vectors in terms of a manageable set of free parameters. Once the basic electroweak observables and the top mass are fixed, we are left with one mass and one coupling for each resonance, the misalignment angle and some additional $O(1)$ parameters controlling the interaction with top partners and SM fermions. Of these, $c_1$ has no role in the production and decay of the vector resonances, so that their phenomenology can be significantly affected only by the remaining $(c_3, y_L)$ for $\rho_\mu^L$, $(c_4, y_L)$ for $\rho_\mu^R$, $(c_5, c_6, y_L)$ in model $\textbf{M}_{\textbf{X}}^{\textbf{1}}$ and $(c_6, y_R)$ in model $\textbf{M}_{\textbf{X}}^{\textbf{2}}$. In this sense, the effective Lagrangian approach based on specific underlying assumptions on the symmetry structure of the theory has the virtue of expressing all the couplings of the vectors to top partners and SM particles in terms of only these quantities. This reduces considerably the degrees of freedom that one would have in a complete model-independent procedure, like in \cite{Bridge, ModelIndip}, and allows us to formulate a consistent description of the interaction with lighter fermions, which necessarily requires some knowledge of the underlying symmetries, \cite{Hunters}. Our model-dependent approach is therefore essential in order to capture the most important features of the interplay between heavy vectors and top partners, that would be impossible to analyse without any robust assumption on the symmetry structure of the theory. 

For each resonance, we studied the main phenomenological features, analysing the mixing angles, the spectrum and the most important couplings arising before EWSB. We have shown that the left-handed and right-handed vectors couple strongly to the longitudinally polarized $W$ and $Z$ bosons and Higgs, thanks to the Equivalence Theorem, and that they both couple very weakly to fully elementary SM fermions. Concerning their interaction with top partners and third family quarks, conservation of isospin gives the most important rationale to extract the relevant couplings: only those conserving isospin without any Higgs vev insertion can arise before EWSB and the corresponding decay channels give a dominant contribution to the decay width. We have also considered the very different case of the singlet, which has peculiar properties with respect to the other resonances. It couples very weakly both to SM fermions and to gauge bosons, whereas it interacts strongly with the $t_R$ and the top partner $\widetilde{T}$, with interaction strength depending on whether the $t_R$ is partially composite or not. This vector is also special since it does not give any contribution to the $\hat{S}$ parameter, so that direct searches are the most important mean to constrain its parameter space. We have finally studied the decay branching ratios of all the three vectors, noticing the dominance of the top partner decay channel above the threshold $M_\rho = 2M_\psi$ and studying the relevance of the decays to SM particles below the threshold for different values of the free parameters.

Using our effective Lagrangian description, we have devised an efficient semi-analytical method to compare the theoretical predictions of our models with the LHC data on direct searches of vector resonances. These latter are given as exclusion limits of $\sigma \times BR$ as a function of the resonance mass, under the validity of the Narrow Width Approximation. In order to compute the total production cross section, we have numerically calculated the parton level contribution once for all, setting the relevant trilinear couplings to unity, and we have then rescaled with the analytical expression of the couplings at linear order in $\xi$. We have also studied the main production mechanisms, DY and VBF, noticing that the former is the most relevant one in all cases of interest. Following this method, it is very fast to analytically recast the experimental searches as bounds on the parameter space of the resonances, once the LHC data are rescaled with the BRs that can be computed analytically in our models. The calculation of the cross sections as well as the numerical diagonalization of the vector mass matrices, at every order in $\xi$, have been implemented in a $\texttt{Mathematica}$ notebook that is available on a dedicated website, \cite{Website}.

We have applied this methodology to extract exclusion limits on the parameter space of our models using the presently available 8 TeV LHC data. The results can be found in Figs.~(\ref{fig:RegionPlotLeftMg}), (\ref{fig:RegionPlotRightMg}) and (\ref{fig:RegionPlotXMg}), where exclusion regions are shown for some relevant direct searches of heavy vectors. We have analysed what information can be obtained from the decay channels considered by the experimental groups for different values of the free parameters of the theories. For the left-handed vector, we concluded that the most constraining decay channels at the LHC are $WZ$ and $l\bar{\nu}$, when the free parameters are chosen so as to respect the NDA estimate. A slight violation of NDA, obtained by reducing $a_{\rho_L}$, shows, however, that the decay channels to gauge bosons can give no bound at all and that a very important decay channel that can be extensively studied in the future is the $t\bar{b}$, since partially composite quarks are more strongly coupled to the heavy vectors than to the other SM fermions. The situation is similar for the neutral right-handed resonance; again, for values of the free parameters respecting the NDA expectations, the $WW$ and the $l\bar{l}$ channels give the most stringent bounds, whereas reducing the value of $a_{\rho_R}$ shows that exclusion regions can be drawn only from the leptonic decay channel. As regards the searches with a $t\bar{t}$ final state, in this case they do not provide any constraint, since the production cross section for $\rho_{\mu}^R$ is smaller than the corresponding one for the left-handed vector by a factor $(g^\prime / g)^2$. This suppression is the reason why the enhanced coupling to top quarks does not improve the sensitivity of this channel. Finally, considering the $\rho_\mu^X$ case, the most constraining decay channel is the $l\bar{l}$, since the couplings of the singlet to $W$ bosons are very weak. Also in this case, the $t\bar{t}$ channel does not give any significant bound, the production cross section being again reduced by a factor $(g^\prime / g)^2$. The suppression in the production cross sections of the right-handed vector and of the singlet is in general the reason why the bounds for the $\rho_\mu^R$ and $\rho_\mu^X$ resonances are much weaker than the bounds on $\rho_\mu^L$, making them more difficult to constrain or discover at the LHC. Finally, all these results can be readily interpreted as a test of our notion of naturalness and of our dynamical assumptions on the nature of the strong dynamics. We have shown the most natural expected window of parameter space and considered how the data already exclude part of it in the low-mass and small coupling region. But for bigger values of the mass and for more strongly coupled scenarios, there is still room for a natural realization of the composite Higgs idea with heavier vectors decaying to lighter top partners.

\section*{Acknowledgements}
We thank Riccardo Rattazzi, Roberto Contino and Francesco Riva, who carefully read the manuscript; Andrea Thamm, whose constant support and useful suggestions during the various stages of completion of this work are acknowledged by D.G.; Kohei Kamada, who helped D.L. with the Mathematica plotting functions. D.G. is supported by the Swiss National Science Foundation under contract 200020-150060. D.L. acknowledges support by the China Scholarship Foundation.

\appendix

\section{CCWZ variables}\label{AppCCWZ}

We report in this appendix some important formulae related to the CCWZ formalism that are used in the main text. We indicate with $T^{\widehat{a}}$ ($\widehat{a} = 1, \cdots , 4$) the broken generators parametrizing the coset $SO(5)/SO(4)$ and with $T^{a_L/a_R}$ ($a_L/a_R = 1, 2, 3$) the $SO(4)$ unbroken generators, whose expressions can be found in \cite{Hunters}. The $5\times 5$ Goldstone boson matrix, $U(\Pi) = e^{i \sqrt{2}/f \Pi^{\hat{a}}T^{\widehat{a}}}$, has the following form in the unitary gauge:
\begin{equation}\label{U}
U= \left(
\begin{array}{l|ll}
\mathbb{I}_3 &  &   \\
\hline 
 &  \cos \left( \theta+{ h \over f}\right) &\sin\left( \theta + { h \over f}\right) \\
 &  -\sin\left( \theta + { h \over f}\right) & \cos \left( \theta+{ h \over f}\right)
\end{array}
\right),
\end{equation}
with the $d^{\widehat{a}}_\mu$, $E^{a_L}_{\mu}$ and $E^{a_R}_{\mu}$ variables defined by the relation:
\begin{equation}\label{UDE}
-i U^\dagger D_\mu U = d_\mu^{\widehat{a}}T^{\widehat{a}}+E^{a_L}_{\mu} T^{a_L}_L+E^{a_R}_{\mu} T^{a_R}.
\end{equation}   
$D_\mu$ is the SM covariant derivative containing the elementary gauge fields,
\begin{equation}\label{CovDer}
D_\mu = \partial_\mu - i g_{el} {W^i_\mu \over 2} \sigma^i - i g_{el}^{\prime} Y B_\mu,
\end{equation}
where $i=1, 2, 3$ and $\sigma^i$ are the Pauli matrices. 
 
The $d$ and $E$ symbols, on the other hand, can be easily computed once $U(\Pi)$ is known; up to quadratic order in the unitary gauge their expression is given by:
\begin{equation}\label{d}
\begin{array}{l}
\displaystyle d_\mu^{\widehat{a}} = A_\mu^{\widehat{a}}+{\sqrt{2} \over f}\partial_\mu h +{\sqrt{2} \over 2 f} h (\delta^{a_L \widehat{a}} A_{\mu }^{a_L}- \delta^{a_R \widehat{a}} A_{\mu} ^{a_R}),\\[0,2cm]
\displaystyle E_\mu^{a_L} = A_\mu^{{a_L}}- \delta^{a_L \widehat{a}}{\sqrt{2}\over 2 f} h A_\mu^{\widehat{a}},\\[0,2cm]
\displaystyle E_\mu^{a_R} = A_\mu^{{a_R}}-\delta^{a_R \widehat{a}}{\sqrt{2}\over 2 f} h A_\mu^{\widehat{a}},
\end{array}
\end{equation}
where we have defined the Kronecker $\delta^{\widehat{a}i}$, for a generic index $i=1, 2, 3$, as:
\begin{displaymath}
\delta^{i\widehat{a}} = \left \{
\begin{array}{ll}
1 \qquad \text{if} \qquad \widehat{a} = i \\
0 \qquad \text{if} \qquad \widehat{a} \neq i \qquad \text{or} \qquad \widehat{a} = 4
\end{array}
\right. .
\end{displaymath}
We notice that in this work we always use the expression of the connection truncated at quadratic order, as in equations (\ref{d}), since we are mainly interested in trilinear couplings and we are neglecting the contribution of dimension-6 operators.

The external gauge fields appearing in the formulae for the $d$ and $E$ symbols, for a given value of the angle $\theta$, have the following forms:
\begin{equation}
\begin{array}{llll}
\displaystyle A_\mu ^{\widehat{a}} = { \sin  \theta \over \sqrt{2}} (\delta^{\widehat{a}i} g_{el}W_\mu^i-\delta^{\widehat{a}3}g^{\prime}_{el}B_\mu ), \qquad A_\mu^{\widehat{4}}= 0,\\ [0,2cm]

\displaystyle A_\mu^{a_L}= \delta^{a_L i} \left({1+\cos \theta} \over 2 \right)g_{el} W^i_\mu + \delta^{{a_L}3}\left( {1-\cos \theta }\over 2\right)g_{el}^{\prime}B_\mu,\\[0,2cm]

\displaystyle A_\mu^{a_R}= \delta^{a_R i} \left({1-\cos \theta} \over 2 \right) g_{el}W^i_\mu + \delta^{{a_R}3}\left( {1+\cos \theta }\over 2\right)g_{el}^{\prime}B_\mu,\\[0,2cm]
\end{array}
\end{equation} 
where $g_{el}$ and $g_{el}^{\prime}$ are the weak coupling of the elementary sector.

Under a global transformation $g \in SO(5)$, the Goldstone boson matrix transforms as:
\begin{equation}\label{UTransf}
U(\Pi) \rightarrow g U(\Pi) h^\dagger(\Pi,g),
\end{equation}
where $h(\Pi,g) \in SO(4)$. As a consequence of Eq.~(\ref{UDE}), the previous relation implies the following transformation rules for $d$ and $E$:
\begin{equation}\label{dETransf}
\begin{array}{ll}
\displaystyle d^{\widehat{a}}_\mu \rightarrow h(\Pi,g) d^{\widehat{a}}_\mu h^\dagger (\Pi, g)\\[0,2cm]

\displaystyle E^{a_{L/R}}_\mu \rightarrow h( \Pi, g) E_\mu^{a_{L/R}} h^\dagger (\Pi, g)- i h(\Pi, g) \partial_\mu h^\dagger (\Pi, g),
\end{array}
\end{equation}
showing that both these variables transform under a local $SO(4)$ symmetry when acted upon with $g$. Since in particular $E^{a_{L/R}}_\mu$ behaves like a gauge field under $h(\Pi,g)$, we can introduce the covariant derivative 
\begin{equation}\label{CovE}
\nabla_\mu = \partial_\mu - i E^{a_{L}}_\mu T^{a_{L}} - i E^{a_{R}}_\mu T^{a_{R}}
\end{equation}
and a field strength 
\begin{equation}\label{FieldStrengthE}
\begin{array}{ll}
E_{\mu \nu }^{L/R}= \partial_\mu E_\nu^{L/R} - \partial_\nu E_\mu^{L/R} + i [E_\mu^{L/R}, E_\nu^{L/R}]\\[0,2cm]

E_{\mu \nu}^{L/R} \rightarrow h(\Pi, g) E_{\mu \nu}^{L/R} h^\dagger (\Pi, g),
\end{array}
\end{equation}
where $E_\mu^{L/R} = E_\mu^{a_{L/R}}T^{a_{L/R}}$.

\section{Contribution to the Electroweak Precision Observables}\label{AppSTU}

In this appendix, we briefly study the contribution to the Electroweak Precision Observables generated by integrating out at tree level the vectors in our models. In general, the deviations from the SM in the vector boson vacuum polarization amplitudes can be described by four effective form factors: $\hat{S}, \hat{T}, W$ and $ Y$. New physics contributions to the four parameters can be expressed as a function of the Wilson coefficients of the leading dimension-6 operators  obtained by integrating out the BSM sector. If the BSM sector respects the custodial symmetry, as in the case of the minimal composite Higgs model, $\hat{T}$ is vanishing and we are left with the remaining three oblique parameters. In the SILH basis, \cite{SILH}, $\hat{S}$ comes from the linear combination of $O_W + O_B$, $W$ and $Y$ on the other hand are generated by $O_{2W}$ and $O_{2B}$ respectively. In order to get the Wilson coefficients of these dimension-6 operators, we integrate out the $\rho$ resonances using the EOM at $O(p^3)$:
\begin{equation}\label{IntegOut}
\rho^{a_L/a_R}_{\mu} = E^{a_L/a_R}_\mu - \frac{1}{M_{\rho_{L/R}}^2} \nabla_\mu E^{ a_L/a_R \ \mu \nu}  + {O}(p^5) , \qquad \rho_\mu^X = B_\mu - {\partial_\mu B^{\mu \nu} \over M_{\rho_X}^2}+O(p^5);
\end{equation}
we have to keep up to three derivative terms in the EOM, because the operators $O_{2W}$ and $O_{2B}$ include six derivatives according to the SILH power counting (gauge fields count as one derivative). Once evaluated on the equation of motions, we obtain from the $\mathcal{L}_{\rho}$ term in Eqs. (\ref{LagLeft}), (\ref{LagRight}), (\ref{LagsX}), the following low-energy Lagrangian:
\begin{equation}\label{LowEnerLag}
\begin{split}
\mathcal{L}_6 =& - \frac{1}{4 g_{\rho_L}^2} (E_{\mu\nu}^{a_L})^2- \frac{1}{4 g_{\rho_R}^2} (E_{\mu\nu}^{a_R})^2- {1 \over 4 g_{\rho_X}^2}B^{\mu \nu}B_{\mu \nu} - \frac12 \frac{1}{M_{\rho_L}^2g_{\rho_L}^2} \nabla_\mu E^{a_L \mu \nu} \nabla_\rho E^{a_L \rho}_{\ \ \ \ \nu} \\
&- \frac12 \frac{1}{M_{\rho_R}^2g_{\rho_R}^2} \nabla_\mu E^{a_R \mu \nu} \nabla_\rho E^{a_R \rho}_{\ \ \ \ \nu} - {1\over 2}{1\over M_{\rho_X}^2 g_{\rho_X}^2 }\partial_\mu B^{\mu \nu} \partial_\rho B^{\rho}_{\ \nu} + \cdots \, ,
\end{split}
\end{equation}
where  the dots imply terms more than quadratic in the field strength and with at least four partial derivatives. The first two terms will give rise to $O_W$ and $O_B$ and the last two terms will instead lead to $O_{2W}, O_{2B}$. To see this explicitly, we rewrite the formulae for the $E_\mu$ connections in terms of the  Higgs current; the relevant terms are 
\begin{equation}\label{HiggsCurrent}
\begin{split}
E_\mu^{a_L} &= \delta^{a_L i} g_{el} W^i_\mu + \frac{i}{f^2} H^\dagger \frac{\sigma^a}{2} \overleftrightarrow{D_\mu} H + \cdots\, , \\
E_\mu^{3_R} &= g_{el}^\prime B_\mu+ \frac{i}{f^2} H^\dagger \frac{1}{2} \overleftrightarrow{D_\mu} H + \cdots. ,
\end{split}
\end{equation}
and, after substituting in \ref{LowEnerLag}, we get:
\begin{equation}
\begin{split}
\mathcal{L}_6 =& \frac{i g}{g_{\rho_L}^2 f^2} H^\dagger \frac{\sigma^a}{2} \overleftrightarrow{D}^\mu H D^\nu W^a_{\mu\nu}  + \frac{i g^\prime}{g_{\rho_R}^2 f^2} H^\dagger \frac{1}{2} \overleftrightarrow{D}_\mu H \partial_\nu B^{\mu\nu}-\frac12 \frac{g^2}{g_{\rho_L}^2 M_{\rho_L}^2} D^\mu W_{\mu \nu}^a D_\rho W^{a\rho\nu} \\
 & - \frac12 \frac{g^{\prime2}}{g_{\rho_R}^2 M_{\rho_R}^2} \partial^\mu B_{\mu \nu} \partial_\rho B^{\rho\nu}- \frac12 \frac{g^{\prime2}}{g_{\rho_X}^2 M_{\rho_X}^2} \partial^\mu B_{\mu \nu} \partial_\rho B^{\rho\nu}.
\end{split}
\label{L6}
\end{equation}
From the previous formulae, we can immediately find the expression of the three oblique parameters:
\begin{equation}
\hat{S} = c_W + c_B = a_{\rho_L}^2\frac{m_W^2}{M_{\rho_L}^2}  +  a_{\rho_R}^2\frac{m_W^2}{M_{\rho_R}^2}, \qquad W = \frac{g^2m_W^2}{g_{\rho_L}^2 M_{\rho_L}^2}, \qquad Y = \frac{g^{\prime2} m_W^2}{g_{\rho_R}^2 M_{\rho_R}^2}+\frac{g^{\prime2} m_W^2}{g_{\rho_X}^2 M_{\rho_X}^2}.
\end{equation}

\section{Couplings}\label{AppCoup}

In this appendix, we give some technical details on the structure of the Lagrangian in the mass eigenstate basis, for the case of a heavy vector triplet and a heavy vector singlet. We will focus on trilinear interactions, neglecting for simplicity the quartic vertices. 

We start considering the Lagrangian of a vector triplet with top partners in the fourplet, $\mathcal{L}^{T}_{\rho}$. Without making explicit reference to the representation under which the spin-1 resonances fall, we can rewrite in full generality the Lagrangian after rotation to the mass eigenstate basis as a set of three fields, the charged $\rho_\mu^{\pm}$ and the neutral $\rho_\mu^0$, interacting with the SM particles and the top partners. The couplings between the heavy vectors and the other bosons and fermions are in general a function of all the free parameters of the theory and they explicitly depend on the model under consideration; we will name them $g_{\rho^+i j}$, for the couplings of the charged pair, and $g_{\rho^0 i j}$, for the couplings of the neutral state, where $i$ and $j$ generically stand for two particles the resonance interacts with. We can therefore introduce the following decomposition for $\mathcal{L}^T_{\rho}$:
\begin{equation}\label{LTripDecomp}
\mathcal{L}^T_{\rho} = \mathcal{L}^T_{gbh}+\mathcal{L}^T_{ef}+\mathcal{L}^T_{tb}
+\mathcal{L}^T_{TPtb}+\mathcal{L}^T_{TP},
\end{equation}
where $\mathcal{L}^T_{gbh}$ contains the interactions between the $\rho$'s and the gauge bosons and between the $\rho$'s, the Higgs and a gauge boson, whereas $\mathcal{L}^T_{ef}$, $\mathcal{L}^T_{tb}$, $\mathcal{L}^T_{TPtb}$ and $\mathcal{L}^T_{TP}$ comprise, respectively, the couplings of the spin-1 heavy states to fully elementary fermions, to top and bottom quarks, to one top partner and one heavy quarks and finally to two top partners. It is straightforward to derive the form of the different contributions in the mass eigenstate basis and in the unitary gauge; we find:\footnote{All interaction terms between SM fermions and spin-1 resonances in this Lagrangian are flavor diagonal. This follows from assuming that all the lightest fermions are fully elementary: in absence of elementary-composite fermion mixings one can always make fields rotations to diagonalize the fermionic kinetic terms in flavor space. By allowing for some degrees of compositeness for leptons and the first two quark families and thus for non-vanishing elementary-composite couplings $\lambda$, the Lagrangian \ref{LTripDecomp} is valid up to $O(\lambda)$ in the weak interaction eigenbasis for the fermions. In this basis the fermion masses are not diagonal in flavor space. After rotating the fermion fields to diagonalize the mass matrices, a $V_{CKM}$ matrix appear in the vertex $\rho_\mu^+ \bar{\psi}_u \psi_d$, while the interactions of $\rho^0$ remain diagonal.}
\begin{equation}\label{gbhInter}
\begin{array}{ll}
\displaystyle \mathcal{L}_{gbh}^T = & \displaystyle i g_{\rho^0 WW} \left[ (\partial_\mu W_\nu^+ - \partial_\nu W_\mu^+ ) W^{\mu -}\rho^{0 \nu}+{1\over 2}(\partial_\mu \rho_\nu^0 - \partial_\nu \rho_\mu ^0)W^{\mu +}W^{\nu -} +\text{h.c.} \right] \\

\displaystyle & \displaystyle + i g_{\rho^{+}WZ} \left[ (\partial_\mu \rho_\nu ^{+}- \partial_\nu \rho_\mu ^{+}) W^{\mu {-}}Z^\nu- (\partial_\mu W_\nu ^- - \partial_\nu W_\mu ^-)\rho^{\mu + }Z^\nu \right.\\

\displaystyle  & \displaystyle \left.  +(\partial_\mu Z_\nu - \partial_\nu Z_\mu )\rho^{\mu +}W^{\nu -}  + \text{h.c.} \right]+g_{\rho^0 ZH } h \rho_\mu^0 Z^\mu + g_{\rho^+ W H }(h \rho_\mu^+ W_\mu^- + \text{h.c.}), \\
\end{array}
\end{equation}
\begin{equation}\label{efInter}
\begin{array}{ll}
\displaystyle \mathcal{L}^T_{ef} = & \displaystyle {1\over \sqrt{2}}g_{\rho^+ ffL} (\rho_\mu^+ \bar{\psi}_u \gamma^\mu P_L \psi_d + \text{h.c.}) \\

&  \displaystyle  + \rho_\mu^0 \bar{\psi}_u \gamma^\mu \left[ {1\over 2}(g_{\rho^0 ffL}- g_{\rho^0 ffY}) P_L + g_{\rho^0 ffY} Q[\psi_u] \right]\psi_u \\

& \displaystyle +  \rho_\mu^0 \bar{\psi}_d \gamma^\mu \left[- {1\over 2}(g_{\rho^0 ffL}- g_{\rho^0 ffY}) P_L + g_{\rho^0 ffY} Q[\psi_d] \right]\psi_d, 

\end{array}
\end{equation}
\begin{equation}\label{tbInter}
\begin{array}{ll}
\displaystyle \mathcal{L}^T_{tb} = & \displaystyle {1\over \sqrt{2}}g_{\rho^+ tb} (\rho_\mu^+ \bar{t}_L \gamma^\mu  b_L + \text{h.c.})\\

& \displaystyle   + g_{\rho^0 t_L t_L} \rho_\mu^0 \bar{t}_L \gamma^\mu t_L +g_{\rho^0 t_R t_R}\rho_\mu^0 \bar{t}_R \gamma^\mu t_R +g_{\rho^0 b_L b_L} \rho_\mu^0 \bar{b}_L \gamma^\mu b_L, 

\end{array}
\end{equation}
\begin{equation}\label{TPtbInter}
\begin{array}{ll}
\displaystyle \mathcal{L}^T_{TPtb} = & \displaystyle {1\over \sqrt{2}}\left[\rho_\mu^+ \left (g_{\rho^+ T_L b_L}  \bar{T}_L \gamma^\mu  b_L +g_{\rho^+ X_{{2\over 3} L} b_L} \bar{X}_{{2 \over 3}L} \gamma^\mu  b_L+g_{\rho^+ B_L t_L}  \bar{t}_L \gamma^\mu  B_L \right. \right. \\

& \left. \left. \displaystyle +g_{\rho^+ X_{{5\over 3} L} t_L} \bar{X}_{{5 \over 3}L} \gamma^\mu  t_L+ g_{\rho^+ B_R t_R}  \bar{t}_R \gamma^\mu  B_R +g_{\rho^+ X_{{5\over 3} R} t_R} \bar{X}_{{5 \over 3}R} \gamma^\mu  t_R\right) + \text{h.c.}\right]\\

& \displaystyle + \rho_\mu^0 \left( g_{\rho^0 T_L t_L}\bar{T}_L \gamma^\mu t_L +g_{\rho^0 X_{{2\over 3} L} t_L} \bar{X}_{{2 \over 3}L}  \gamma^\mu t_L +g_{\rho^0 B_L b_L} \bar{B}_L \gamma^\mu b_L \right.  \\

& \displaystyle \left. + g_{\rho^0 T_R t_R}\bar{T}_R \gamma^\mu t_R +g_{\rho^0 X_{{2\over 3} R} t_R} \bar{X}_{{2 \over 3}R}  \gamma^\mu t_R+ \text{h.c.} \right), 

\end{array}
\end{equation}
\begin{equation}\label{TPInter}
\begin{array}{ll}
\displaystyle \mathcal{L}^T_{TP} = & \displaystyle {1\over \sqrt{2}}\left[\rho_\mu^+ \left (g_{\rho^+ T_L B_L}  \bar{T}_L \gamma^\mu  B_L +g_{\rho^+ X_{{2\over 3} L} B_L} \bar{X}_{{2 \over 3}L} \gamma^\mu  B_L+g_{\rho^+ X_{{5\over 3} L} T_L} \bar{X}_{{5 \over 3}L} \gamma^\mu  T_L \right. \right. \\

& \left. \left. \displaystyle +(L \leftrightarrow R)+g_{\rho^+ X_{{5\over 3} } X_{{2\over 3} } } \bar{X}_{{5 \over 3}} \gamma^\mu  X_{{2\over 3} }  \right) + \text{h.c.}\right]\\

& \displaystyle + \rho_\mu^0 \left( g_{\rho^0 T_L T_L}\bar{T}_L \gamma^\mu T_L +g_{\rho^0 X_{{2\over 3} L} T_L}( \bar{X}_{{2 \over 3}L}  \gamma^\mu T_L  + \text{h.c.}) +g_{\rho^0 B_L B_L} \bar{B}_L \gamma^\mu B_L + (L \leftrightarrow R) \right.  \\

& \displaystyle \left. +  g_{\rho^0 X_{2\over 3}X_{2\over 3}} \bar{X}_{2 \over 3} \gamma^\mu X_{2 \over 3}  +  g_{\rho^0 X_{5\over 3}X_{5\over 3}} \bar{X}_{5 \over 3} \gamma^\mu X_{5 \over 3} \right).

\end{array}
\end{equation}
We make some comments on the parametrization chosen in the previous formulae. As regards the couplings to fully elementary fermions, we have collectively indicated with $\psi_u$ ($\psi_d$) any of the SM up-type quarks and neutrinos (down-type quarks and charged leptons) and we have introduced their charge through the function $Q[\psi_u]$ ($Q[\psi_d]$). The form chosen for $\mathcal{L}_{ef}^T$ is convenient for the implementation of the models in a $\mathtt{Mathematica}$ code, since the couplings to different kinds of leptons and quarks can be easily and unambiguously derived from the universal functions $g_{\rho^{+/0}ffL}$ and $g_{\rho^{+/0}ffY}$. The top-bottom doublet and the $t_R$ are instead treated differently, as seen in equation \ref{tbInter}; we introduce specific couplings for every vertex between the heaviest quarks and the spin-1 resonances, in order to take into account the enhancement in the interactions due to partial compositeness. Finally, in the last term of the Lagrangian, $\mathcal{L}_{TP}^T$, we have differentiated the couplings of the heavy vectors to left-handed and right-handed top partners, because they are in general expected to be different. The only exceptions are the interactions involving only the exotic $X_{5\over 3}$ and the top-like $X_{2\over 3}$, namely $g_{\rho^0 X_{2\over 3}X_{2\over 3}}$, $g_{\rho^0 X_{5\over 3}X_{5\over 3}}$ and $g_{\rho^+ X_{5\over 3}X_{2\over 3}}$; in this case the couplings to states of different chirality are equal since these $X_{5/3}$ top partner is left invariant by the rotation in the fermionic sector, whereas the $X_{2/3L}$ and $X_{2/3R}$ fields transforms in the same way under the fermionic rotation, \cite{Hunters}.

We finally consider the Lagrangian for the singlets: a neutral vector resonance interacting with a fermionic heavy state, both being invariant under the unbroken $SO(4)$. The Lagrangian can be decomposed analogously to the previous formulae as:
\begin{equation}\label{LSingDecomp}
\mathcal{L}^S_\rho = \mathcal{L}^S_{gbh}+\mathcal{L}^S_{ef}+\mathcal{L}^S_{tb}
+\mathcal{L}^S_{TPtb}+\mathcal{L}^S_{TP}.
\end{equation}
The first three terms have the same expressions as the Lagrangian for the neutral heavy state, $\rho_\mu^0$, in $\mathcal{L}_\rho^T$. The last two contributions can be instead easily rewritten after rotations to the mass eigenstate basis and specifically depend on the choice of the representation for the top partner; we find:
\begin{equation}\label{TPtbInterS}
\begin{array}{ll}
\displaystyle \mathcal{L}^S_{TPtb} = & \displaystyle  \rho_\mu^0 \left( g_{\rho^0 \widetilde{T}_L t_L}\bar{\widetilde{T}}_L \gamma^\mu t_L  + g_{\rho^0 \widetilde{T}_R t_R}\bar{\widetilde{T}}_R \gamma^\mu t_R +\text{h.c.}\right),
\end{array}
\end{equation}
\begin{equation}\label{TPInterS}
\begin{array}{ll}
\displaystyle \mathcal{L}^S_{TP} = & \displaystyle  \rho_\mu^0 \left( g_{\rho^0 \widetilde{T}_L \widetilde{T}_L}\bar{\widetilde{T}}_L \gamma^\mu \widetilde{T}_L  + g_{\rho^0 \widetilde{T}_R \widetilde{T}_R}\bar{\widetilde{T}}_R \gamma^\mu \widetilde{T}_R \right).
\end{array}
\end{equation}
As before, the couplings are a function of all the free input parameters of the theory and we find different expressions if the $t_R$ is fully composite or only partially composite.

\section{Effects of a degenerate spectrum}\label{AppIntEff}

In this appendix, we clarify the phenomenological effects of relaxing the assumption that one vector resonance is much lighter and the other two belong to the tower of states that are integrated out. We want to analyse the possible consequences of having an almost degenerate spectrum and, for simplicity, we will not consider the most complicated case in which all the three heavy states are present together. We will only analyse, instead, the simpler situation in which two resonances are degenerate and the other one is heavier and is thus integrated out. We therefore introduce the three following cases,  
\begin{equation}\label{Cases}
\begin{array}{ll}
\text{(I)}\, \  (\rho_L,\,\rho_R )\, \ \text{with Lagrangian} \ \mathcal{L}_{L+R}= \mathcal{L}_{light}+\mathcal{L}_\Psi+\mathcal{L}_{\rho_L}+\mathcal{L}_{\rho_R}, \\
\text{(II)}\, \ (\rho_L,\,\rho_X )\, \ \text{with Lagrangian} \ \mathcal{L}_{L+X}= \mathcal{L}_{light}+\mathcal{L}_\Psi+\mathcal{L}_{\widetilde{T}^1}+\mathcal{L}_{\rho_L}+
\mathcal{L}_{\rho_X^1}, \\
\text{(III)}\, \ (\rho_R,\,\rho_X )\, \ \text{with Lagrangian} \ \mathcal{L}_{R+X}= \mathcal{L}_{light}+\mathcal{L}_\Psi+\mathcal{L}_{\widetilde{T}^1}+\mathcal{L}_{\rho_R}+
\mathcal{L}_{\rho_X^1};
\end{array}
\end{equation}
in all combinations the $t_R$ quark arises as a singlet of the composite dynamics, so that we have considered only the interference with model $\textbf{M}_{\textbf{X}}^{\textbf{1}}$ in (II) and (III). 

When considering the degeneracy of the particle spectrum, there are different effects on our analysis of direct searches that we must take into account with respect to the situations studied in the main text. First of all, we expect that the expressions of the couplings in the mass eigenstate basis will be corrected and that the more degenerate the spectrum is, the stronger these corrections will be. Secondly, the branching ratios will change as well, due to the opening of new decay channels, a heavy-light one, with a vector resonance decaying to a second heavy vector and a gauge boson, and a heavy-heavy one, which involves a vector state decaying to other two heavy spin-1 resonances. These two classes of modifications could significantly alter the results concerning the bounds on the free parameters of our models; we will analyse them in the following, showing that considering only one resonance at a time and integrating out the other two is a good basic approximation for interpreting the experimental data.

Let us start considering how the couplings change in case (I). The spectrum now contains two charged and two neutral heavy vector particles. The mass matrix is given by a $3\times 3$ charged block and a $4\times 4$ neutral block, whose expressions is not reported here, but can be found in \cite{Proceeding}, where also some of the modified couplings in the mass eigenstate basis are given. Since the $\rho_\mu^R$ and $\rho_\mu^L$ resonances belong to different representations of the unbroken $SO(4)$, all the corrections to the couplings in Appendix \ref{AppCoup} must arise after EWSB and are therefore suppressed. As a consequence, we do not expect that the degeneracy of the resonances masses will induce important differences on the branching ratios that have already been analysed in this work, so that no relevant modifications on the bounds can be induced by the changes in the couplings.

In case (II) and (III), on the other hand, one charged and two neutral vector resonances are present. The charged block of the mass matrix is not affected by the interference with the singlet, which mixes only with the $B_\mu$ boson, so that no modification is induced on the couplings of the charged vector. The neutral block, on the other hand, becomes now a $4\times 4$ matrix and, after rotation to the mass eigenstate basis, the couplings of the neutral resonances will be indeed modified with respect to the situation considered in the main text. In particular, in model (II) these corrections must be suppressed by $\xi$, since $\rho_\mu^L$ mixes with $B_\mu$ only after EWSB, whereas in model (III) both $\rho_R^3$ and $\rho^X$ mix with $B_\mu$ before EWSB, therefore inducing interference effects that can have important consequences on their phenomenology. We conclude that the approximate description adopted in the main text works well for case (II), even with a degenerate spectrum, whereas in case (III) the bounds and branching ratios should be corrected if the two resonances have comparable masses.  

\begin{figure}[!tb]
\begin{center}
\includegraphics[width=0.49\textwidth]{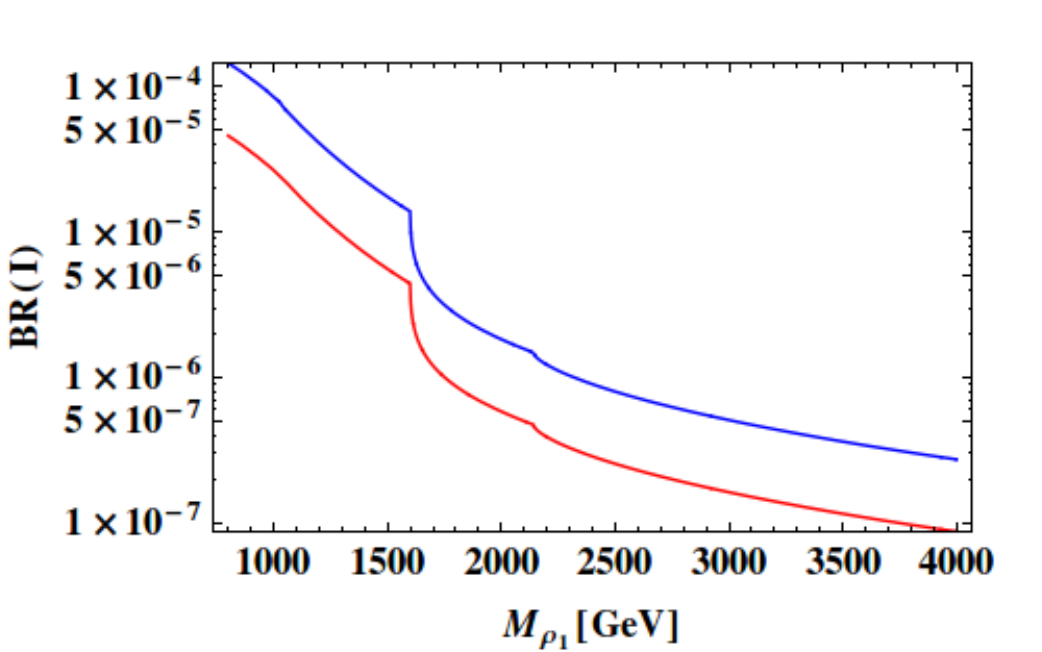}
\includegraphics[width=0.49\textwidth]{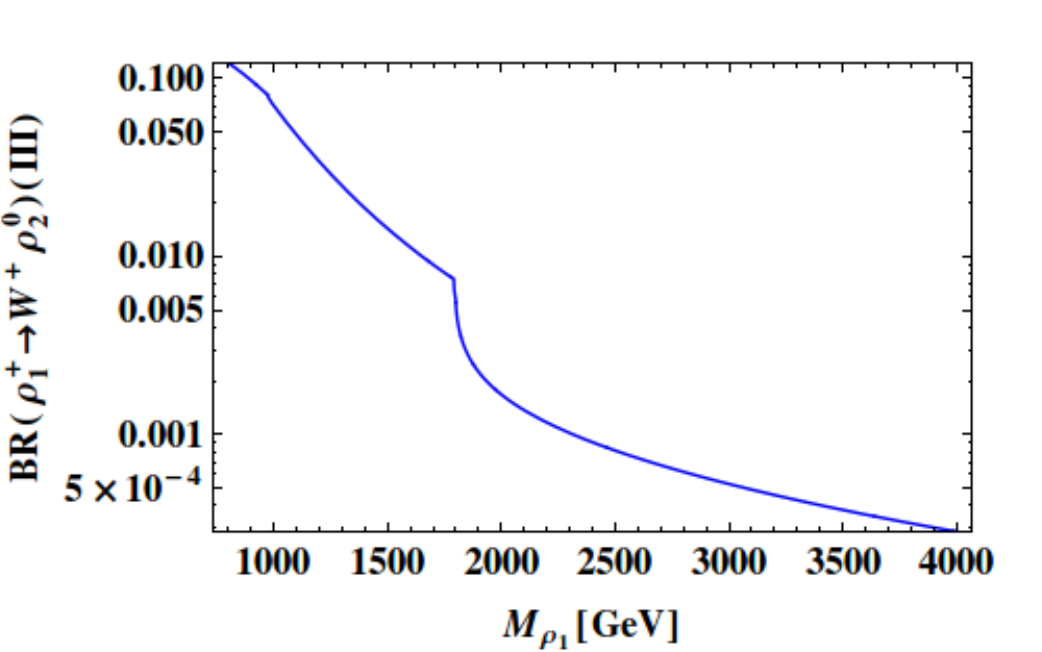}
\end{center}
\caption{\small 
Cascade decay branching ratios as a function of the heavier resonance mass, for the benchmark value $g_\rho = 3$, for case~(I) (left plot) and case~(III) (right plot) of Eq.~({\ref{Benchmarks}}). The blue line corresponds to BR($\rho_1^+ \rightarrow W^+ \rho_2^0$) and the red curve corresponds to BR($\rho_1^+ \rightarrow \rho_2^+ Z$). 
}\label{fig:CascadeDecay}
\end{figure}

We now study more quantitatively the effects of the spectrum degeneracy on the branching ratios, analysing, as illustration, the cascade decay of one heavy vector to a second spin-1 resonance and a gauge boson. We want to estimate the branching ratio of this process in the three cases, so as to understand how much the decay widths analysed in this work can be altered by the opening of this new decay channel. From triple vector couplings in the kinetic terms of the Lagrangians in (\ref{Cases}), an additional interaction between two heavy vectors is generated; we can write it as follows:
\begin{equation}\label{CascadeLagrangian1}
\begin{split}
\mathcal{L}_{XYM}   =  \,
& i  g_{X^+ Y^- M^0} \, \big[  (\partial_\mu X_{\nu}^+ - \partial_\nu X_{\mu}^+ ) Y^{\mu -}M^{0\nu}  
                            -(\partial_\mu X_{\nu}^- - \partial_\nu X_{\mu}^- ) Y^{\mu +} M^{0\nu }  \\[0.1cm]
& \phantom{ig_{X^+ Y^-  M^0}\,\big[}    
                           + (\partial_\mu Y_\nu^+ - \partial_\nu Y_\mu^+ ) X^{\mu -} M^{0\nu}  -(\partial_\mu Y_\nu^- - \partial_\nu Y_\mu^- ) X^{\mu +} M^{0\nu }  \\[0.1cm]
& \phantom{ig_{X^+ Y^-  M^0}\,\big[}           
                           + (\partial_\mu M^0_{\nu} - \partial_\nu M^0_{\mu} ) (X^{\mu +}Y^{\nu -} - X^{\mu -}Y^{\nu +} ) \big],  \\
\end{split}
\end{equation}
when $X$ is different from $Y$, and
\begin{equation}\label{CascadeLagrangian2}
\begin{split}
\mathcal{L}_{XXM}   =  \,
& i  g_{X^+ X^- M^0} \, \big[  (\partial_\mu X_{\nu}^+ - \partial_\nu X_{\mu}^+ ) X^{\mu -}M^{0\nu}  
                            -(\partial_\mu X_{\nu}^- - \partial_\nu X_{\mu}^- ) X^{\mu +} M^{0\nu }  \\[0.1cm]
& \phantom{ig_{X^+ X^-  M^0}\,\big[}           
                           +\frac 12 (\partial_\mu M^0_{\nu} - \partial_\nu M^0_{\mu} ) (X^{\mu +}X^{\nu -} - X^{\mu -}X^{\nu +} ) \big],  \\
\end{split}
\end{equation}
when ${X}={Y}$. We have indicated with $X$, $Y$ and $M$ any of ($W/Z$, $\rho^+$, $\rho_0$). As a result, when one of the two vectors is relatively heavier than the other one, the channels $\rho_1^+ \rightarrow \rho_2^0 W^+$, $\rho_1^0 \rightarrow \rho_2^+ W^-$ and $\rho_1^+ \rightarrow \rho_2^+ Z$ open up ($\rho_1$ and $\rho_2$ stand for the vectors in different representations for each of the three cases considered). In order to illustrate the relevance of these cascade decays, we focus on the two following sets of benchmark values
\begin{equation}\label{Benchmarks}
\begin{split}
\text{(I)}  \quad m_{\rho_L} = 1.5 \, m_{\rho_R} = 1.5 \, g_{\rho_R} f \, ,\quad  g_{\rho_L} = g_{\rho_R} \equiv g_{\rho}\,, \\[0.1cm]
\text{(III)}  \quad m_{\rho_R} = 1.5 \, m_{\rho_X} = 1.5 \, g_{\rho_X} f \,\, ,\quad  g_{\rho_R} = g_{\rho_X} \equiv g_{\rho}\,, \\[0.1cm]
\end{split}
\end{equation}
and we show in Fig.~(\ref{fig:CascadeDecay}) the relative branching ratios as a function of the resonant mass, for illustration, fixing to 1 all the $O(1)$ parameters controlling the couplings to top partners. The results in case (II) are very similar to case (I) and the corresponding branching ratios are not shown. We see that the branching ratios are very tiny for cases (I), due to the fact that the mixing between a charged and a neutral state or between two charged states belonging to different representation of $H$ arises at ${O}(\xi)$ after EWSB. The situation is different for case (III); the branching ratio is now considerably bigger, even if the coupling between two different heavy vectors arises again at $O(\xi)$. This is a consequence of the small couplings of the charged right-handed resonance to SM fermions: since the branching ratios for its decay to both elementary and partially composite fermions are strongly suppressed, the decay channel to the lighter vector and a $W$ boson is much more competitive. As expected, in case (III) the corrections to the branching ratios are therefore more important. However, these corrections will not have relevant consequences on the exclusion plots we derived in the main text. These latter are in fact obtained for the neutral right-handed vector which is not affected by the presence of the relatively lighter $\rho^X_\mu$ since no couplings involving two neutral heavy vectors can be induced in our models. We thus conclude that our estimate of the branching ratios and relative bounds on the parameter space of the models is a good approximation for all the resonances, even neglecting their possible degeneracy.  

\section{The $\mathtt{MadGraph5}$ model}\label{AppNumDiag}

The four models discussed in this paper have been implemented in the parton level generator $\mathtt{MadGraph5}$ for the simulation of Monte Carlo events. All the trilinear interaction vertices involving vector resonances, SM particles and top partners have been introduced in the UFO file, following the conventions of Appendix \ref{AppCoup}. 

A $\mathtt{Mathematica}$ calculator is also provided, which performs a numerical diagonalization of the vector mass matrix and computes all the physical quantities, masses and trilinear couplings between heavy vectors and SM particles, after the input parameters are specified. This code also implements the numerical diagonalization of the fermionic mass matrices in the top partner sector and computes the trilinear couplings between heavy resonances, top partners and partially composite SM fermions to full order in $\xi$. The semi-analytical formulae for the computation of the cross sections and the partial decay widths described in the main text can be also derived with this program.

We also stress that our numerical code has been designed not only to simulate the production and decay of vector resonances, but also to study $WW$ scattering processes at the LHC and at future colliders. In order for these processes to be suitably simulated in the presence of vector resonances, also the modifications to the couplings $g_{HWW}$, $g_{HZZ}$, $g_{HHWW}$, $g_{HHZZ}$ and $g_{HHH}$ after rotation to the mass eigenstate basis must be properly taken into account. The corrections to the first four couplings are numerically calculated by the $\mathtt{Mathematica}$ file and in particular the vertices $g_{HHWW}$ and $g_{HHZZ}$ are the only four-particles interactions that are numerically derived by the calculator. On the other hand, the modification of the trilinear Higgs coupling $g_{HHH}$ for the minimal model with elementary fermions embedded in the vector representation of $SO(5)$ (MCHM5) has been derived analytically in \cite{DoubleHiggsProd} to all orders in $\xi$ and it is implemented in the code accordingly.

All the available software can be downloaded in a single package from the HEPMDB website \cite{Website} and the instruction on how to run the calculator can be found in the README file which is provided with the program.


\providecommand{\href}[2]{#2}

\end{document}